\documentclass[longauth]{aa}
\usepackage{euclid}
\usepackage{ISTL_macros}

\usepackage[english]{babel}
\usepackage{natbib}
\bibpunct{(}{)}{;}{a}{}{,} 
\usepackage{graphicx}
\usepackage{txfonts}
\usepackage{amsmath}
\usepackage{color, colortbl}
\usepackage{xcolor}
\usepackage{soul}
\usepackage{amsfonts}
\usepackage{multicol}
\usepackage{blindtext}
\usepackage[switch]{lineno}
\definecolor{linkcolor}{rgb}{0.8,0,0}
\definecolor{citecolor}{rgb}{0,0,0.7}
\definecolor{urlcolor}{rgb}{0,0,0.7}
\usepackage[breaklinks,colorlinks,citecolor=citecolor,urlcolor=urlcolor,linkcolor=linkcolor,pdfencoding=auto]{hyperref}
\definecolor{romared}{RGB}{142,0,28}
\definecolor{reddy}{rgb}{0.8,0,0}
\definecolor{greeny}{rgb}{0,0.3,0}
\definecolor{bluey}{rgb}{0.1,0.1,1}
\definecolor{bluetwo}{rgb}{0,0,0.8}
\definecolor{magentay}{rgb}{0.79,0.08,0.48}
\definecolor{linkcolorish}{rgb}{0.6,0,0}
\definecolor{citecolorish}{rgb}{0,0,0.75}
\definecolor{urlcolorish}{rgb}{0.12,0.46,0.7}
\hypersetup{linktocpage}
\usepackage{enumitem}
\usepackage{verbatim}
\usepackage[hang,small,bf]{caption}
\usepackage{rotating}
\usepackage{url}
\usepackage{times}
\usepackage{booktabs}
\usepackage{xr}
\usepackage{amsfonts,amssymb,amsmath,mathtools}
\usepackage{siunitx}
\usepackage{bm}
\usepackage{epsfig}
\usepackage[normalem]{ulem}
\usepackage{multirow}
\usepackage{tabularx}
\usepackage{fontawesome}
\usepackage[nameinlink,capitalise]{cleveref}
\usepackage{changepage}
\usepackage{listings}
\usepackage{slashed}
\usepackage{soul}

\usepackage{adjustbox}

\newcommand{\saccaddress}{\url{https://sacc.readthedocs.io}\xspace}
\newcommand{\ccladdress}{\url{https://ccl.readthedocs.io}\xspace}

\newcommand{\snakevizaddress}{\url{https://jiffyclub.github.io/snakeviz}\xspace}
\newcommand{\coffeeaddress}{\url{https://jcgoran.github.io/coffe}\xspace}

\newcommand{\cosmolssaddress}{\url{https://github.com/sjoudaki/CosmoLSS}\xspace}
\newcommand{\firecrownaddress}{\url{https://firecrown.readthedocs.io}\xspace}
\newcommand{\cobayaaddress}{\url{https://cobaya.readthedocs.io}\xspace}
\newcommand{\cosmosisaddress}{\url{https://cosmosis.readthedocs.io}\xspace}
\newcommand{\pyqtaddress}{\url{https://pypi.org/project/PyQt5}\xspace}
\newcommand{\pytestaddress}{\url{https://pypi.org/project/pytest}\xspace}
\newcommand{\jaxaddress}{\url{https://github.com/google/jax}\xspace}
\newcommand{\xlaaddress}{\url{https://github.com/openxla/xla}\xspace}
\newcommand{\euclidlibaddress}{\url{https://github.com/euclidlib/euclidlib}\xspace}
\newcommand{\fastptaddress}{\url{https://fast-pt.readthedocs.io}\xspace}
\newcommand{\getdistaddress}{\url{https://getdist.readthedocs.io}\xspace}
\newcommand{\cambaddress}{\url{https://camb.info}\xspace}
\newcommand{\classaddress}{\url{https://github.com/lesgourg/class_public}\xspace}
\newcommand{\hmcodeaddress}{\url{https://github.com/alexander-mead/HMcode}\xspace}
\newcommand{\baccoaddress}{\url{https://baccoemu.readthedocs.io}\xspace}
\newcommand{\eecodeaddress}{\url{https://github.com/miknab/EuclidEmulator2}\xspace}
\newcommand{\bcemuaddress}{\url{https://github.com/sambit-giri/BCemu}\xspace}
\newcommand{\montepythonaddress}{\url{https://brinckmann.github.io/montepython_public}\xspace}
\newcommand{\cosmomcaddress}{\url{https://cosmologist.info/cosmomc}\xspace}
\newcommand{\polychordaddress}{\url{https://github.com/PolyChord}\xspace}
\newcommand{\emceeaddress}{\url{https://emcee.readthedocs.io}\xspace}
\newcommand{\pocomcaddress}{\url{https://pocomc.readthedocs.io}\xspace}
\newcommand{\zeusaddress}{\url{https://zeus-mcmc.readthedocs.io}\xspace}
\newcommand{\nautilusaddress}{\url{https://nautilus-sampler.readthedocs.io}\xspace}
\newcommand{\multinestaddress}{\url{https://github.com/farhanferoz/MultiNest}\xspace}
\newcommand{\eftcambaddress}{\url{http://eftcamb.org/}\xspace}
\newcommand{\mgcambaddress}{\url{https://github.com/sfu-cosmo/MGCAMB}\xspace}
\newcommand{\mgclassaddress}{\url{https://gitlab.com/zizgitlab/mgclass--ii}\xspace}
\newcommand{\isitgraddress}{\url{https://github.com/mishakb/ISiTGR}\xspace}
\newcommand{\hiclassaddress}{\url{https://miguelzuma.github.io/hi_class_public}\xspace}
\newcommand{\cloe}{\textsc{CLOE}\xspace}
\newcommand{\cloelib}{\textsc{cloelib}\xspace}
\newcommand{\cloelike}{\textsc{cloelike}\xspace}
\newcommand{\cloeorg}{\textsc{cloe-org}\xspace}

\newcommand{\mac}{\textsc{Mac}\xspace}
\newcommand{\linux}{\textsc{Linux}\xspace}
\newcommand{\windows}{\textsc{Windows}\xspace}
\newcommand{\ccl}{\textsc{CCL}\xspace}
\newcommand{\pbj}{\textsc{PBJ}\xspace}
\newcommand{\coffee}{\textsc{COFFEE}\xspace}

\newcommand{\cosmopower}{\textsc{CosmoPower}\xspace}
\newcommand{\camb}{\textsc{CAMB}\xspace}
\newcommand{\class}{\textsc{CLASS}\xspace}
\newcommand{\mgcamb}{\textsc{MGCAMB}\xspace}
\newcommand{\mgclass}{\textsc{MGCLASS}\xspace}
\newcommand{\hiclass}{\textsc{hi\_class}\xspace}
\newcommand{\isitgr}{\textsc{ISiTGR}\xspace}
\newcommand{\eftcamb}{\textsc{EFTCAMB}\xspace}
\newcommand{\getdist}{\textsc{GetDist}\xspace}
\newcommand{\pyqt}{\textsc{PyQt5}\xspace}
\newcommand{\cosmolss}{\textsc{CosmoLSS}\xspace}

\newcommand{\cobaya}{\textsc{Cobaya}\xspace}
\newcommand{\cosmosis}{\textsc{CosmoSIS}\xspace}
\newcommand{\jax}{\textsc{JAX}\xspace}
\newcommand{\xla}{\textsc{XLA}\xspace}

\newcommand{\pyside}{\textsc{PySide}\xspace}

\newcommand{\cosmomc}{\textsc{CosmoMC}\xspace}
\newcommand{\nautilus}{\textsc{Nautilus}\xspace}
\newcommand{\emcee}{\textsc{emcee}\xspace}
\newcommand{\multinest}{\textsc{Multinest}\xspace}
\newcommand{\zeus}{\textsc{zeus}\xspace}
\newcommand{\pocomc}{\textsc{pocoMC}\xspace}
\newcommand{\halofit}{\textsc{Halofit}\xspace}
\newcommand{\hmcode}{\textsc{HMCODE}\xspace}
\newcommand{\hmcodetwenty}{\textsc{HMCODE-2020}\xspace}
\newcommand{\flamingo}{\textsc{Flamingo}\xspace}
\newcommand{\bacco}{\textsc{BACCO}\xspace}
\newcommand{\bcemu}{\textsc{BCemu}\xspace}
\newcommand{\eecode}{\textsc{EuclidEmulator2}\xspace}
\newcommand{\sacc}{\textsc{SACC}\xspace}
\newcommand{\firecrown}{\textsc{Firecrown}\xspace}
\newcommand{\ccode}{\textsc{C}\xspace}
\newcommand{\sphinx}{\textsc{Sphinx}\xspace}
\newcommand{\jupyternotebook}{\textsc{Jupyter Notebook}\xspace}
\newcommand{\jupyter}{\textsc{Jupyter}\xspace}
\newcommand{\jupyternotebooks}{\textsc{Jupyter Notebooks}\xspace}

\newcommand{\python}{\textsc{Python}\xspace}
\newcommand{\ubuntu}{\textsc{Ubuntu}\xspace}
\newcommand{\pip}{\textsc{pip}\xspace}
\newcommand{\fastpt}{\textsc{FAST-PT}\xspace}
\newcommand{\gsl}{\textsc{GSL}\xspace}
\newcommand{\gfortran}{\textsc{GFortran}\xspace}
\newcommand{\fortran}{\textsc{Fortran}\xspace}

\newcommand{\astropy}{\textsc{astropy}\xspace}
\newcommand{\matplotlib}{\textsc{matplotlib}\xspace}
\newcommand{\mpiforpy}{\textsc{mpi4py}\xspace}
\newcommand{\scikit}{\textsc{scikit}\xspace}
\newcommand{\seaborn}{\textsc{seaborn}\xspace}
\newcommand{\tensorflow}{\textsc{tensorflow}\xspace}
\newcommand{\pytest}{\textsc{pytest}\xspace}
\newcommand{\pytestcov}{\textsc{pytest-cov}\xspace}
\newcommand{\pytestcode}{\textsc{pytest-pycodestyle}\xspace}
\newcommand{\sphinxrtd}{\textsc{sphinx-rtd-theme}\xspace}
\newcommand{\cxxcomp}{\textsc{cxx-compiler}\xspace}
\newcommand{\fitsio}{\textsc{fitsio}\xspace}

\newcommand{\deepdish}{\textsc{deepdish}\xspace}
\newcommand{\pyhmcode}{\textsc{pyhmcode}\xspace}
\newcommand{\euclidlib}{\textsc{euclidlib}\xspace}
\newcommand{\pytestdoc}{\textsc{pytest-pydocstyle}\xspace}
\newcommand{\cosmosislib}{\textsc{cosmosis-build-standard-library}\xspace}
\newcommand{\eftoflss}{\textsc{EFTofLSS}\xspace}
\newcommand{\numpydoc}{\textsc{Numpydoc}\xspace}
\newcommand{\numpy}{\textsc{Numpy}\xspace}
\newcommand{\scipy}{\textsc{Scipy}\xspace}
\newcommand{\docker}{\textsc{Docker}\xspace}
\newcommand{\snakeviz}{\textsc{Snakeviz}\xspace}
\newcommand{\conda}{\textsc{Conda}\xspace}
\newcommand{\miniconda}{\textsc{Miniconda}\xspace}

\newcommand{\euclidemu}{\textsc{euclidemu2}\xspace}
\newcommand{\git}{\textsc{Git}\xspace}
\newcommand{\gitlab}{\textsc{GitLab}\xspace}
\newcommand{\github}{\textsc{GitHub}\xspace}
\newcommand{\polychord}{\textsc{Polychord}\xspace}
\newcommand{\montepython}{\textsc{Monte Python}\xspace}
\newcommand{\yaml}{\textsc{YAML}\xspace}
\newcommand{\html}{\textsc{HTML}\xspace}
\newcommand{\ini}{\textsc{INI}\xspace}
\newcommand{\openmp}{\textsc{OpenMP}\xspace}
\newcommand{\fftlog}{\textsc{FFTLog}\xspace}
\newcolumntype{C}{>{\centering\arraybackslash}X}

\definecolor{dkgreen}{rgb}{0,0.6,0}
\definecolor{gray}{rgb}{0.5,0.5,0.5}
\definecolor{mauve}{rgb}{0.58,0,0.82}

\definecolor{BrickRed}{RGB}{182, 50, 28}
\definecolor{forestgreen}{rgb}{0.13, 0.55, 0.13}

\usepackage{orcidlink} 
\newcommand{\orcid}[1]{\orcidlink{#1}}

\begin{document}
\title{\Euclid\/ preparation}
\subtitle{XCV. Cosmology Likelihood for Observables in Euclid (\cloe). 2. Code implementation}

\hypersetup{linkcolor=bluetwo}

\author{Euclid Collaboration: S.~Joudaki\orcid{0000-0001-8820-673X}\thanks{\email{\color{bluetwo}shahab.joudaki@ciemat.es}}\inst{\ref{aff1},\ref{aff2},\ref{aff3},\ref{aff4}}
\and V.~Pettorino\inst{\ref{aff5}}
\and L.~Blot\orcid{0000-0002-9622-7167}\inst{\ref{aff6},\ref{aff7}}
\and M.~Bonici\orcid{0000-0002-8430-126X}\inst{\ref{aff8},\ref{aff3}}
\and S.~Camera\orcid{0000-0003-3399-3574}\inst{\ref{aff9},\ref{aff10},\ref{aff11}}
\and G.~Ca\~nas-Herrera\orcid{0000-0003-2796-2149}\inst{\ref{aff5},\ref{aff12},\ref{aff13}}
\and V.~F.~Cardone\inst{\ref{aff14},\ref{aff15}}
\and P.~Carrilho\orcid{0000-0003-1339-0194}\inst{\ref{aff16}}
\and S.~Casas\orcid{0000-0002-4751-5138}\inst{\ref{aff17}}
\and S.~Davini\orcid{0000-0003-3269-1718}\inst{\ref{aff18}}
\and S.~Di~Domizio\orcid{0000-0003-2863-5895}\inst{\ref{aff19},\ref{aff18}}
\and S.~Farrens\orcid{0000-0002-9594-9387}\inst{\ref{aff20}}
\and L.~W.~K.~Goh\orcid{0000-0002-0104-8132}\inst{\ref{aff20}}
\and S.~Gouyou~Beauchamps\inst{\ref{aff21},\ref{aff22}}
\and S.~Ili\'c\orcid{0000-0003-4285-9086}\inst{\ref{aff23},\ref{aff24}}
\and F.~Keil\orcid{0000-0002-8108-1679}\inst{\ref{aff24}}
\and A.~M.~C.~Le~Brun\orcid{0000-0002-0936-4594}\inst{\ref{aff25}}
\and M.~Martinelli\orcid{0000-0002-6943-7732}\inst{\ref{aff14},\ref{aff15}}
\and C.~Moretti\orcid{0000-0003-3314-8936}\inst{\ref{aff26},\ref{aff27},\ref{aff28},\ref{aff29},\ref{aff30}}
\and A.~Pezzotta\orcid{0000-0003-0726-2268}\inst{\ref{aff31}}
\and Z.~Sakr\orcid{0000-0002-4823-3757}\inst{\ref{aff32},\ref{aff24},\ref{aff33}}
\and A.~G.~S\'anchez\orcid{0000-0003-1198-831X}\inst{\ref{aff31}}
\and D.~Sciotti\orcid{0009-0008-4519-2620}\inst{\ref{aff14},\ref{aff15}}
\and K.~Tanidis\inst{\ref{aff34}}
\and I.~Tutusaus\orcid{0000-0002-3199-0399}\inst{\ref{aff24}}
\and V.~Ajani\orcid{0000-0001-9442-2527}\inst{\ref{aff20},\ref{aff35},\ref{aff36}}
\and S.~Alvi\orcid{0000-0001-5779-8568}\inst{\ref{aff37}}
\and M.~Crocce\orcid{0000-0002-9745-6228}\inst{\ref{aff22},\ref{aff21}}
\and A.~C.~Deshpande\orcid{0000-0003-3721-4232}\inst{\ref{aff38}}
\and A.~Fumagalli\orcid{0009-0004-0300-2535}\inst{\ref{aff39},\ref{aff29}}
\and C.~Giocoli\orcid{0000-0002-9590-7961}\inst{\ref{aff40},\ref{aff41}}
\and A.~G.~Ferrari\orcid{0009-0005-5266-4110}\inst{\ref{aff41}}
\and R.~Kou\orcid{0000-0003-3408-3062}\inst{\ref{aff42},\ref{aff43}}
\and L.~Legrand\orcid{0000-0003-0610-5252}\inst{\ref{aff44},\ref{aff45}}
\and M.~Lembo\orcid{0000-0002-5271-5070}\inst{\ref{aff37},\ref{aff46}}
\and G.~F.~Lesci\orcid{0000-0002-4607-2830}\inst{\ref{aff47},\ref{aff40}}
\and D.~Navarro-Giron\'es\orcid{0000-0003-0507-372X}\inst{\ref{aff13}}
\and A.~Nouri-Zonoz\orcid{0009-0006-6164-8670}\inst{\ref{aff48}}
\and S.~Pamuk\orcid{0009-0004-0852-8624}\inst{\ref{aff49}}
\and L.~Pagano\orcid{0000-0003-1820-5998}\inst{\ref{aff37},\ref{aff46}}
\and M.~Tsedrik\orcid{0000-0002-0020-5343}\inst{\ref{aff16},\ref{aff50}}
\and S.~Arcari\orcid{0000-0002-0551-1315}\inst{\ref{aff37},\ref{aff46}}
\and E.~Artis\orcid{0009-0001-6055-8503}\inst{\ref{aff31}}
\and M.~Ballardini\orcid{0000-0003-4481-3559}\inst{\ref{aff37},\ref{aff40},\ref{aff46}}
\and J.~Bel\inst{\ref{aff51}}
\and C.~Carbone\orcid{0000-0003-0125-3563}\inst{\ref{aff8}}
\and M.~Costanzi\orcid{0000-0001-8158-1449}\inst{\ref{aff52},\ref{aff28},\ref{aff29}}
\and B.~De~Caro\inst{\ref{aff8}}
\and C.~A.~J.~Duncan\orcid{0009-0003-3573-0791}\inst{\ref{aff53}}
\and G.~Fabbian\orcid{0000-0002-3255-4695}\inst{\ref{aff54},\ref{aff45},\ref{aff55}}
\and M.~Kilbinger\orcid{0000-0001-9513-7138}\inst{\ref{aff20}}
\and T.~Kitching\orcid{0000-0002-4061-4598}\inst{\ref{aff38}}
\and F.~Lacasa\orcid{0000-0002-7268-3440}\inst{\ref{aff56},\ref{aff57}}
\and M.~Lattanzi\orcid{0000-0003-1059-2532}\inst{\ref{aff46}}
\and J.~Olivares-Miranda\orcid{0009-0005-7844-2727}\inst{\ref{aff1}}
\and L.~Salvati\inst{\ref{aff57}}
\and D.~Sapone\orcid{0000-0001-7089-4503}\inst{\ref{aff58}}
\and B.~Sartoris\orcid{0000-0003-1337-5269}\inst{\ref{aff59},\ref{aff28}}
\and E.~Sellentin\inst{\ref{aff60},\ref{aff13}}
\and P.~L.~Taylor\orcid{0000-0001-6999-4718}\inst{\ref{aff61},\ref{aff62}}
\and B.~Altieri\orcid{0000-0003-3936-0284}\inst{\ref{aff63}}
\and A.~Amara\inst{\ref{aff64}}
\and L.~Amendola\orcid{0000-0002-0835-233X}\inst{\ref{aff32}}
\and S.~Andreon\orcid{0000-0002-2041-8784}\inst{\ref{aff65}}
\and N.~Auricchio\orcid{0000-0003-4444-8651}\inst{\ref{aff40}}
\and C.~Baccigalupi\orcid{0000-0002-8211-1630}\inst{\ref{aff29},\ref{aff28},\ref{aff30},\ref{aff26}}
\and M.~Baldi\orcid{0000-0003-4145-1943}\inst{\ref{aff66},\ref{aff40},\ref{aff41}}
\and S.~Bardelli\orcid{0000-0002-8900-0298}\inst{\ref{aff40}}
\and P.~Battaglia\orcid{0000-0002-7337-5909}\inst{\ref{aff40}}
\and A.~Biviano\orcid{0000-0002-0857-0732}\inst{\ref{aff28},\ref{aff29}}
\and D.~Bonino\orcid{0000-0002-3336-9977}\inst{\ref{aff11}}
\and E.~Branchini\orcid{0000-0002-0808-6908}\inst{\ref{aff19},\ref{aff18},\ref{aff65}}
\and M.~Brescia\orcid{0000-0001-9506-5680}\inst{\ref{aff67},\ref{aff68}}
\and J.~Brinchmann\orcid{0000-0003-4359-8797}\inst{\ref{aff69},\ref{aff70}}
\and A.~Caillat\inst{\ref{aff71}}
\and V.~Capobianco\orcid{0000-0002-3309-7692}\inst{\ref{aff11}}
\and J.~Carretero\orcid{0000-0002-3130-0204}\inst{\ref{aff1},\ref{aff72}}
\and M.~Castellano\orcid{0000-0001-9875-8263}\inst{\ref{aff14}}
\and G.~Castignani\orcid{0000-0001-6831-0687}\inst{\ref{aff40}}
\and S.~Cavuoti\orcid{0000-0002-3787-4196}\inst{\ref{aff68},\ref{aff73}}
\and K.~C.~Chambers\orcid{0000-0001-6965-7789}\inst{\ref{aff74}}
\and A.~Cimatti\inst{\ref{aff75}}
\and C.~Colodro-Conde\inst{\ref{aff76}}
\and G.~Congedo\orcid{0000-0003-2508-0046}\inst{\ref{aff16}}
\and C.~J.~Conselice\orcid{0000-0003-1949-7638}\inst{\ref{aff53}}
\and L.~Conversi\orcid{0000-0002-6710-8476}\inst{\ref{aff77},\ref{aff63}}
\and Y.~Copin\orcid{0000-0002-5317-7518}\inst{\ref{aff78}}
\and F.~Courbin\orcid{0000-0003-0758-6510}\inst{\ref{aff79},\ref{aff80},\ref{aff81}}
\and H.~M.~Courtois\orcid{0000-0003-0509-1776}\inst{\ref{aff82}}
\and A.~Da~Silva\orcid{0000-0002-6385-1609}\inst{\ref{aff83},\ref{aff84}}
\and H.~Degaudenzi\orcid{0000-0002-5887-6799}\inst{\ref{aff85}}
\and S.~de~la~Torre\inst{\ref{aff71}}
\and G.~De~Lucia\orcid{0000-0002-6220-9104}\inst{\ref{aff28}}
\and A.~M.~Di~Giorgio\orcid{0000-0002-4767-2360}\inst{\ref{aff86}}
\and H.~Dole\orcid{0000-0002-9767-3839}\inst{\ref{aff57}}
\and F.~Dubath\orcid{0000-0002-6533-2810}\inst{\ref{aff85}}
\and X.~Dupac\inst{\ref{aff63}}
\and S.~Dusini\orcid{0000-0002-1128-0664}\inst{\ref{aff87}}
\and A.~Ealet\orcid{0000-0003-3070-014X}\inst{\ref{aff78}}
\and S.~Escoffier\orcid{0000-0002-2847-7498}\inst{\ref{aff88}}
\and M.~Farina\orcid{0000-0002-3089-7846}\inst{\ref{aff86}}
\and R.~Farinelli\inst{\ref{aff40}}
\and F.~Faustini\orcid{0000-0001-6274-5145}\inst{\ref{aff89},\ref{aff14}}
\and S.~Ferriol\inst{\ref{aff78}}
\and F.~Finelli\orcid{0000-0002-6694-3269}\inst{\ref{aff40},\ref{aff90}}
\and P.~Fosalba\orcid{0000-0002-1510-5214}\inst{\ref{aff21},\ref{aff22}}
\and S.~Fotopoulou\orcid{0000-0002-9686-254X}\inst{\ref{aff91}}
\and N.~Fourmanoit\orcid{0009-0005-6816-6925}\inst{\ref{aff88}}
\and M.~Frailis\orcid{0000-0002-7400-2135}\inst{\ref{aff28}}
\and E.~Franceschi\orcid{0000-0002-0585-6591}\inst{\ref{aff40}}
\and M.~Fumana\orcid{0000-0001-6787-5950}\inst{\ref{aff8}}
\and S.~Galeotta\orcid{0000-0002-3748-5115}\inst{\ref{aff28}}
\and K.~George\orcid{0000-0002-1734-8455}\inst{\ref{aff59}}
\and W.~Gillard\orcid{0000-0003-4744-9748}\inst{\ref{aff88}}
\and B.~Gillis\orcid{0000-0002-4478-1270}\inst{\ref{aff16}}
\and J.~Gracia-Carpio\inst{\ref{aff31}}
\and B.~R.~Granett\orcid{0000-0003-2694-9284}\inst{\ref{aff65}}
\and A.~Grazian\orcid{0000-0002-5688-0663}\inst{\ref{aff92}}
\and F.~Grupp\inst{\ref{aff31},\ref{aff59}}
\and L.~Guzzo\orcid{0000-0001-8264-5192}\inst{\ref{aff93},\ref{aff65}}
\and S.~V.~H.~Haugan\orcid{0000-0001-9648-7260}\inst{\ref{aff94}}
\and H.~Hoekstra\orcid{0000-0002-0641-3231}\inst{\ref{aff13}}
\and W.~Holmes\inst{\ref{aff95}}
\and I.~Hook\orcid{0000-0002-2960-978X}\inst{\ref{aff96}}
\and F.~Hormuth\inst{\ref{aff97}}
\and A.~Hornstrup\orcid{0000-0002-3363-0936}\inst{\ref{aff98},\ref{aff99}}
\and P.~Hudelot\inst{\ref{aff100}}
\and K.~Jahnke\orcid{0000-0003-3804-2137}\inst{\ref{aff101}}
\and M.~Jhabvala\inst{\ref{aff102}}
\and E.~Keih\"anen\orcid{0000-0003-1804-7715}\inst{\ref{aff103}}
\and S.~Kermiche\orcid{0000-0002-0302-5735}\inst{\ref{aff88}}
\and A.~Kiessling\orcid{0000-0002-2590-1273}\inst{\ref{aff95}}
\and B.~Kubik\orcid{0009-0006-5823-4880}\inst{\ref{aff78}}
\and K.~Kuijken\orcid{0000-0002-3827-0175}\inst{\ref{aff13}}
\and M.~K\"ummel\orcid{0000-0003-2791-2117}\inst{\ref{aff59}}
\and M.~Kunz\orcid{0000-0002-3052-7394}\inst{\ref{aff48}}
\and H.~Kurki-Suonio\orcid{0000-0002-4618-3063}\inst{\ref{aff104},\ref{aff105}}
\and S.~Ligori\orcid{0000-0003-4172-4606}\inst{\ref{aff11}}
\and P.~B.~Lilje\orcid{0000-0003-4324-7794}\inst{\ref{aff94}}
\and V.~Lindholm\orcid{0000-0003-2317-5471}\inst{\ref{aff104},\ref{aff105}}
\and I.~Lloro\orcid{0000-0001-5966-1434}\inst{\ref{aff106}}
\and G.~Mainetti\orcid{0000-0003-2384-2377}\inst{\ref{aff107}}
\and D.~Maino\inst{\ref{aff93},\ref{aff8},\ref{aff108}}
\and E.~Maiorano\orcid{0000-0003-2593-4355}\inst{\ref{aff40}}
\and O.~Mansutti\orcid{0000-0001-5758-4658}\inst{\ref{aff28}}
\and O.~Marggraf\orcid{0000-0001-7242-3852}\inst{\ref{aff109}}
\and N.~Martinet\orcid{0000-0003-2786-7790}\inst{\ref{aff71}}
\and F.~Marulli\orcid{0000-0002-8850-0303}\inst{\ref{aff47},\ref{aff40},\ref{aff41}}
\and R.~Massey\orcid{0000-0002-6085-3780}\inst{\ref{aff110}}
\and S.~Maurogordato\inst{\ref{aff111}}
\and H.~J.~McCracken\orcid{0000-0002-9489-7765}\inst{\ref{aff100}}
\and E.~Medinaceli\orcid{0000-0002-4040-7783}\inst{\ref{aff40}}
\and S.~Mei\orcid{0000-0002-2849-559X}\inst{\ref{aff42},\ref{aff112}}
\and Y.~Mellier\inst{\ref{aff113},\ref{aff100}}
\and M.~Meneghetti\orcid{0000-0003-1225-7084}\inst{\ref{aff40},\ref{aff41}}
\and E.~Merlin\orcid{0000-0001-6870-8900}\inst{\ref{aff14}}
\and G.~Meylan\inst{\ref{aff79}}
\and A.~Mora\orcid{0000-0002-1922-8529}\inst{\ref{aff114}}
\and M.~Moresco\orcid{0000-0002-7616-7136}\inst{\ref{aff47},\ref{aff40}}
\and L.~Moscardini\orcid{0000-0002-3473-6716}\inst{\ref{aff47},\ref{aff40},\ref{aff41}}
\and S.~Mourre\orcid{0009-0005-9047-0691}\inst{\ref{aff111},\ref{aff115}}
\and E.~Munari\orcid{0000-0002-1751-5946}\inst{\ref{aff28},\ref{aff29}}
\and R.~Nakajima\orcid{0009-0009-1213-7040}\inst{\ref{aff109}}
\and C.~Neissner\orcid{0000-0001-8524-4968}\inst{\ref{aff116},\ref{aff72}}
\and S.-M.~Niemi\inst{\ref{aff5}}
\and J.~W.~Nightingale\orcid{0000-0002-8987-7401}\inst{\ref{aff117}}
\and C.~Padilla\orcid{0000-0001-7951-0166}\inst{\ref{aff116}}
\and S.~Paltani\orcid{0000-0002-8108-9179}\inst{\ref{aff85}}
\and F.~Pasian\orcid{0000-0002-4869-3227}\inst{\ref{aff28}}
\and K.~Pedersen\inst{\ref{aff118}}
\and W.~J.~Percival\orcid{0000-0002-0644-5727}\inst{\ref{aff3},\ref{aff4},\ref{aff119}}
\and S.~Pires\orcid{0000-0002-0249-2104}\inst{\ref{aff20}}
\and G.~Polenta\orcid{0000-0003-4067-9196}\inst{\ref{aff89}}
\and M.~Poncet\inst{\ref{aff120}}
\and L.~A.~Popa\inst{\ref{aff121}}
\and L.~Pozzetti\orcid{0000-0001-7085-0412}\inst{\ref{aff40}}
\and F.~Raison\orcid{0000-0002-7819-6918}\inst{\ref{aff31}}
\and R.~Rebolo\inst{\ref{aff76},\ref{aff122},\ref{aff123}}
\and A.~Renzi\orcid{0000-0001-9856-1970}\inst{\ref{aff124},\ref{aff87}}
\and J.~Rhodes\orcid{0000-0002-4485-8549}\inst{\ref{aff95}}
\and G.~Riccio\inst{\ref{aff68}}
\and E.~Romelli\orcid{0000-0003-3069-9222}\inst{\ref{aff28}}
\and M.~Roncarelli\orcid{0000-0001-9587-7822}\inst{\ref{aff40}}
\and R.~Saglia\orcid{0000-0003-0378-7032}\inst{\ref{aff59},\ref{aff31}}
\and J.~A.~Schewtschenko\inst{\ref{aff16}}
\and M.~Schirmer\orcid{0000-0003-2568-9994}\inst{\ref{aff101}}
\and P.~Schneider\orcid{0000-0001-8561-2679}\inst{\ref{aff109}}
\and T.~Schrabback\orcid{0000-0002-6987-7834}\inst{\ref{aff125}}
\and A.~Secroun\orcid{0000-0003-0505-3710}\inst{\ref{aff88}}
\and E.~Sefusatti\orcid{0000-0003-0473-1567}\inst{\ref{aff28},\ref{aff29},\ref{aff30}}
\and G.~Seidel\orcid{0000-0003-2907-353X}\inst{\ref{aff101}}
\and M.~Seiffert\orcid{0000-0002-7536-9393}\inst{\ref{aff95}}
\and S.~Serrano\orcid{0000-0002-0211-2861}\inst{\ref{aff21},\ref{aff126},\ref{aff22}}
\and P.~Simon\inst{\ref{aff109}}
\and C.~Sirignano\orcid{0000-0002-0995-7146}\inst{\ref{aff124},\ref{aff87}}
\and G.~Sirri\orcid{0000-0003-2626-2853}\inst{\ref{aff41}}
\and A.~Spurio~Mancini\orcid{0000-0001-5698-0990}\inst{\ref{aff127}}
\and L.~Stanco\orcid{0000-0002-9706-5104}\inst{\ref{aff87}}
\and J.-L.~Starck\orcid{0000-0003-2177-7794}\inst{\ref{aff20}}
\and J.~Steinwagner\orcid{0000-0001-7443-1047}\inst{\ref{aff31}}
\and P.~Tallada-Cresp\'{i}\orcid{0000-0002-1336-8328}\inst{\ref{aff1},\ref{aff72}}
\and A.~N.~Taylor\inst{\ref{aff16}}
\and I.~Tereno\inst{\ref{aff83},\ref{aff128}}
\and S.~Toft\orcid{0000-0003-3631-7176}\inst{\ref{aff129},\ref{aff130}}
\and R.~Toledo-Moreo\orcid{0000-0002-2997-4859}\inst{\ref{aff131}}
\and F.~Torradeflot\orcid{0000-0003-1160-1517}\inst{\ref{aff72},\ref{aff1}}
\and L.~Valenziano\orcid{0000-0002-1170-0104}\inst{\ref{aff40},\ref{aff90}}
\and J.~Valiviita\orcid{0000-0001-6225-3693}\inst{\ref{aff104},\ref{aff105}}
\and T.~Vassallo\orcid{0000-0001-6512-6358}\inst{\ref{aff59},\ref{aff28}}
\and G.~Verdoes~Kleijn\orcid{0000-0001-5803-2580}\inst{\ref{aff132}}
\and A.~Veropalumbo\orcid{0000-0003-2387-1194}\inst{\ref{aff65},\ref{aff18},\ref{aff19}}
\and Y.~Wang\orcid{0000-0002-4749-2984}\inst{\ref{aff133}}
\and J.~Weller\orcid{0000-0002-8282-2010}\inst{\ref{aff59},\ref{aff31}}
\and A.~Zacchei\orcid{0000-0003-0396-1192}\inst{\ref{aff28},\ref{aff29}}
\and G.~Zamorani\orcid{0000-0002-2318-301X}\inst{\ref{aff40}}
\and F.~M.~Zerbi\inst{\ref{aff65}}
\and E.~Zucca\orcid{0000-0002-5845-8132}\inst{\ref{aff40}}
\and V.~Allevato\orcid{0000-0001-7232-5152}\inst{\ref{aff68}}
\and M.~Bolzonella\orcid{0000-0003-3278-4607}\inst{\ref{aff40}}
\and E.~Bozzo\orcid{0000-0002-8201-1525}\inst{\ref{aff85}}
\and C.~Burigana\orcid{0000-0002-3005-5796}\inst{\ref{aff134},\ref{aff90}}
\and M.~Calabrese\orcid{0000-0002-2637-2422}\inst{\ref{aff135},\ref{aff8}}
\and D.~Di~Ferdinando\inst{\ref{aff41}}
\and J.~A.~Escartin~Vigo\inst{\ref{aff31}}
\and S.~Matthew\orcid{0000-0001-8448-1697}\inst{\ref{aff16}}
\and N.~Mauri\orcid{0000-0001-8196-1548}\inst{\ref{aff75},\ref{aff41}}
\and R.~B.~Metcalf\orcid{0000-0003-3167-2574}\inst{\ref{aff47},\ref{aff40}}
\and A.~A.~Nucita\inst{\ref{aff136},\ref{aff137},\ref{aff138}}
\and M.~P\"ontinen\orcid{0000-0001-5442-2530}\inst{\ref{aff104}}
\and C.~Porciani\orcid{0000-0002-7797-2508}\inst{\ref{aff109}}
\and V.~Scottez\inst{\ref{aff113},\ref{aff139}}
\and M.~Tenti\orcid{0000-0002-4254-5901}\inst{\ref{aff41}}
\and M.~Viel\orcid{0000-0002-2642-5707}\inst{\ref{aff29},\ref{aff28},\ref{aff26},\ref{aff30},\ref{aff27}}
\and M.~Wiesmann\orcid{0009-0000-8199-5860}\inst{\ref{aff94}}
\and Y.~Akrami\orcid{0000-0002-2407-7956}\inst{\ref{aff140},\ref{aff141}}
\and I.~T.~Andika\orcid{0000-0001-6102-9526}\inst{\ref{aff142},\ref{aff143}}
\and R.~E.~Angulo\orcid{0000-0003-2953-3970}\inst{\ref{aff144},\ref{aff145}}
\and S.~Anselmi\orcid{0000-0002-3579-9583}\inst{\ref{aff87},\ref{aff124},\ref{aff7}}
\and M.~Archidiacono\orcid{0000-0003-4952-9012}\inst{\ref{aff93},\ref{aff108}}
\and F.~Atrio-Barandela\orcid{0000-0002-2130-2513}\inst{\ref{aff146}}
\and A.~Balaguera-Antolinez\orcid{0000-0001-5028-3035}\inst{\ref{aff76}}
\and M.~Bethermin\orcid{0000-0002-3915-2015}\inst{\ref{aff147}}
\and A.~Blanchard\orcid{0000-0001-8555-9003}\inst{\ref{aff24}}
\and H.~B\"ohringer\orcid{0000-0001-8241-4204}\inst{\ref{aff31},\ref{aff39},\ref{aff148}}
\and S.~Borgani\orcid{0000-0001-6151-6439}\inst{\ref{aff52},\ref{aff29},\ref{aff28},\ref{aff30},\ref{aff27}}
\and M.~L.~Brown\orcid{0000-0002-0370-8077}\inst{\ref{aff53}}
\and S.~Bruton\orcid{0000-0002-6503-5218}\inst{\ref{aff149}}
\and A.~Calabro\orcid{0000-0003-2536-1614}\inst{\ref{aff14}}
\and B.~Camacho~Quevedo\orcid{0000-0002-8789-4232}\inst{\ref{aff21},\ref{aff22}}
\and A.~Cappi\inst{\ref{aff40},\ref{aff111}}
\and F.~Caro\inst{\ref{aff14}}
\and C.~S.~Carvalho\inst{\ref{aff128}}
\and T.~Castro\orcid{0000-0002-6292-3228}\inst{\ref{aff28},\ref{aff30},\ref{aff29},\ref{aff27}}
\and F.~Cogato\orcid{0000-0003-4632-6113}\inst{\ref{aff47},\ref{aff40}}
\and S.~Conseil\orcid{0000-0002-3657-4191}\inst{\ref{aff78}}
\and S.~Contarini\orcid{0000-0002-9843-723X}\inst{\ref{aff31}}
\and A.~R.~Cooray\orcid{0000-0002-3892-0190}\inst{\ref{aff150}}
\and O.~Cucciati\orcid{0000-0002-9336-7551}\inst{\ref{aff40}}
\and F.~De~Paolis\orcid{0000-0001-6460-7563}\inst{\ref{aff136},\ref{aff137},\ref{aff138}}
\and G.~Desprez\orcid{0000-0001-8325-1742}\inst{\ref{aff132}}
\and A.~D\'iaz-S\'anchez\orcid{0000-0003-0748-4768}\inst{\ref{aff151}}
\and J.~M.~Diego\orcid{0000-0001-9065-3926}\inst{\ref{aff49}}
\and P.~Dimauro\orcid{0000-0001-7399-2854}\inst{\ref{aff14},\ref{aff152}}
\and A.~Enia\orcid{0000-0002-0200-2857}\inst{\ref{aff66},\ref{aff40}}
\and Y.~Fang\inst{\ref{aff59}}
\and P.~G.~Ferreira\orcid{0000-0002-3021-2851}\inst{\ref{aff34}}
\and A.~Finoguenov\orcid{0000-0002-4606-5403}\inst{\ref{aff104}}
\and A.~Franco\orcid{0000-0002-4761-366X}\inst{\ref{aff137},\ref{aff136},\ref{aff138}}
\and K.~Ganga\orcid{0000-0001-8159-8208}\inst{\ref{aff42}}
\and J.~Garc\'ia-Bellido\orcid{0000-0002-9370-8360}\inst{\ref{aff140}}
\and T.~Gasparetto\orcid{0000-0002-7913-4866}\inst{\ref{aff28}}
\and V.~Gautard\inst{\ref{aff153}}
\and R.~Gavazzi\orcid{0000-0002-5540-6935}\inst{\ref{aff71},\ref{aff100}}
\and E.~Gaztanaga\orcid{0000-0001-9632-0815}\inst{\ref{aff22},\ref{aff21},\ref{aff2}}
\and F.~Giacomini\orcid{0000-0002-3129-2814}\inst{\ref{aff41}}
\and F.~Gianotti\orcid{0000-0003-4666-119X}\inst{\ref{aff40}}
\and G.~Gozaliasl\orcid{0000-0002-0236-919X}\inst{\ref{aff154},\ref{aff104}}
\and M.~Guidi\orcid{0000-0001-9408-1101}\inst{\ref{aff66},\ref{aff40}}
\and C.~M.~Gutierrez\orcid{0000-0001-7854-783X}\inst{\ref{aff155}}
\and A.~Hall\orcid{0000-0002-3139-8651}\inst{\ref{aff16}}
\and S.~Hemmati\orcid{0000-0003-2226-5395}\inst{\ref{aff156}}
\and C.~Hern\'andez-Monteagudo\orcid{0000-0001-5471-9166}\inst{\ref{aff123},\ref{aff76}}
\and H.~Hildebrandt\orcid{0000-0002-9814-3338}\inst{\ref{aff157}}
\and J.~Hjorth\orcid{0000-0002-4571-2306}\inst{\ref{aff118}}
\and J.~J.~E.~Kajava\orcid{0000-0002-3010-8333}\inst{\ref{aff158},\ref{aff159}}
\and Y.~Kang\orcid{0009-0000-8588-7250}\inst{\ref{aff85}}
\and V.~Kansal\orcid{0000-0002-4008-6078}\inst{\ref{aff160},\ref{aff161}}
\and D.~Karagiannis\orcid{0000-0002-4927-0816}\inst{\ref{aff37},\ref{aff162}}
\and K.~Kiiveri\inst{\ref{aff103}}
\and C.~C.~Kirkpatrick\inst{\ref{aff103}}
\and S.~Kruk\orcid{0000-0001-8010-8879}\inst{\ref{aff63}}
\and J.~Le~Graet\orcid{0000-0001-6523-7971}\inst{\ref{aff88}}
\and F.~Lepori\orcid{0009-0000-5061-7138}\inst{\ref{aff163}}
\and G.~Leroy\orcid{0009-0004-2523-4425}\inst{\ref{aff164},\ref{aff110}}
\and J.~Lesgourgues\orcid{0000-0001-7627-353X}\inst{\ref{aff17}}
\and L.~Leuzzi\orcid{0009-0006-4479-7017}\inst{\ref{aff47},\ref{aff40}}
\and T.~I.~Liaudat\orcid{0000-0002-9104-314X}\inst{\ref{aff165}}
\and S.~J.~Liu\orcid{0000-0001-7680-2139}\inst{\ref{aff86}}
\and A.~Loureiro\orcid{0000-0002-4371-0876}\inst{\ref{aff166},\ref{aff167}}
\and J.~Macias-Perez\orcid{0000-0002-5385-2763}\inst{\ref{aff168}}
\and G.~Maggio\orcid{0000-0003-4020-4836}\inst{\ref{aff28}}
\and M.~Magliocchetti\orcid{0000-0001-9158-4838}\inst{\ref{aff86}}
\and F.~Mannucci\orcid{0000-0002-4803-2381}\inst{\ref{aff169}}
\and R.~Maoli\orcid{0000-0002-6065-3025}\inst{\ref{aff170},\ref{aff14}}
\and J.~Mart\'{i}n-Fleitas\orcid{0000-0002-8594-569X}\inst{\ref{aff114}}
\and C.~J.~A.~P.~Martins\orcid{0000-0002-4886-9261}\inst{\ref{aff171},\ref{aff69}}
\and L.~Maurin\orcid{0000-0002-8406-0857}\inst{\ref{aff57}}
\and M.~Migliaccio\inst{\ref{aff172},\ref{aff173}}
\and M.~Miluzio\inst{\ref{aff63},\ref{aff174}}
\and P.~Monaco\orcid{0000-0003-2083-7564}\inst{\ref{aff52},\ref{aff28},\ref{aff30},\ref{aff29}}
\and G.~Morgante\inst{\ref{aff40}}
\and C.~Murray\inst{\ref{aff42}}
\and S.~Nadathur\orcid{0000-0001-9070-3102}\inst{\ref{aff2}}
\and K.~Naidoo\orcid{0000-0002-9182-1802}\inst{\ref{aff2}}
\and A.~Navarro-Alsina\orcid{0000-0002-3173-2592}\inst{\ref{aff109}}
\and S.~Nesseris\orcid{0000-0002-0567-0324}\inst{\ref{aff140}}
\and F.~Passalacqua\orcid{0000-0002-8606-4093}\inst{\ref{aff124},\ref{aff87}}
\and K.~Paterson\orcid{0000-0001-8340-3486}\inst{\ref{aff101}}
\and L.~Patrizii\inst{\ref{aff41}}
\and A.~Pisani\orcid{0000-0002-6146-4437}\inst{\ref{aff88},\ref{aff175}}
\and D.~Potter\orcid{0000-0002-0757-5195}\inst{\ref{aff163}}
\and S.~Quai\orcid{0000-0002-0449-8163}\inst{\ref{aff47},\ref{aff40}}
\and M.~Radovich\orcid{0000-0002-3585-866X}\inst{\ref{aff92}}
\and P.~Reimberg\orcid{0000-0003-3410-0280}\inst{\ref{aff113}}
\and I.~Risso\orcid{0000-0003-2525-7761}\inst{\ref{aff176}}
\and P.-F.~Rocci\inst{\ref{aff57}}
\and S.~Sacquegna\orcid{0000-0002-8433-6630}\inst{\ref{aff136},\ref{aff137},\ref{aff138}}
\and M.~Sahl\'en\orcid{0000-0003-0973-4804}\inst{\ref{aff177}}
\and E.~Sarpa\orcid{0000-0002-1256-655X}\inst{\ref{aff26},\ref{aff27},\ref{aff30}}
\and J.~Schaye\orcid{0000-0002-0668-5560}\inst{\ref{aff13}}
\and A.~Schneider\orcid{0000-0001-7055-8104}\inst{\ref{aff163}}
\and M.~Sereno\orcid{0000-0003-0302-0325}\inst{\ref{aff40},\ref{aff41}}
\and A.~Silvestri\orcid{0000-0001-6904-5061}\inst{\ref{aff12}}
\and L.~C.~Smith\orcid{0000-0002-3259-2771}\inst{\ref{aff55}}
\and J.~Stadel\orcid{0000-0001-7565-8622}\inst{\ref{aff163}}
\and C.~Tao\orcid{0000-0001-7961-8177}\inst{\ref{aff88}}
\and G.~Testera\inst{\ref{aff18}}
\and R.~Teyssier\orcid{0000-0001-7689-0933}\inst{\ref{aff175}}
\and S.~Tosi\orcid{0000-0002-7275-9193}\inst{\ref{aff19},\ref{aff176}}
\and A.~Troja\orcid{0000-0003-0239-4595}\inst{\ref{aff124},\ref{aff87}}
\and M.~Tucci\inst{\ref{aff85}}
\and C.~Valieri\inst{\ref{aff41}}
\and A.~Venhola\orcid{0000-0001-6071-4564}\inst{\ref{aff178}}
\and D.~Vergani\orcid{0000-0003-0898-2216}\inst{\ref{aff40}}
\and F.~Vernizzi\orcid{0000-0003-3426-2802}\inst{\ref{aff179}}
\and G.~Verza\orcid{0000-0002-1886-8348}\inst{\ref{aff180}}
\and N.~A.~Walton\orcid{0000-0003-3983-8778}\inst{\ref{aff55}}}
										   
\institute{Centro de Investigaciones Energ\'eticas, Medioambientales y Tecnol\'ogicas (CIEMAT), Avenida Complutense 40, 28040 Madrid, Spain\label{aff1}
\and
Institute of Cosmology and Gravitation, University of Portsmouth, Portsmouth PO1 3FX, UK\label{aff2}
\and
Waterloo Centre for Astrophysics, University of Waterloo, Waterloo, Ontario N2L 3G1, Canada\label{aff3}
\and
Department of Physics and Astronomy, University of Waterloo, Waterloo, Ontario N2L 3G1, Canada\label{aff4}
\and
European Space Agency/ESTEC, Keplerlaan 1, 2201 AZ Noordwijk, The Netherlands\label{aff5}
\and
Center for Data-Driven Discovery, Kavli IPMU (WPI), UTIAS, The University of Tokyo, Kashiwa, Chiba 277-8583, Japan\label{aff6}
\and
Laboratoire Univers et Th\'eorie, Observatoire de Paris, Universit\'e PSL, Universit\'e Paris Cit\'e, CNRS, 92190 Meudon, France\label{aff7}
\and
INAF-IASF Milano, Via Alfonso Corti 12, 20133 Milano, Italy\label{aff8}
\and
Dipartimento di Fisica, Universit\`a degli Studi di Torino, Via P. Giuria 1, 10125 Torino, Italy\label{aff9}
\and
INFN-Sezione di Torino, Via P. Giuria 1, 10125 Torino, Italy\label{aff10}
\and
INAF-Osservatorio Astrofisico di Torino, Via Osservatorio 20, 10025 Pino Torinese (TO), Italy\label{aff11}
\and
Institute Lorentz, Leiden University, Niels Bohrweg 2, 2333 CA Leiden, The Netherlands\label{aff12}
\and
Leiden Observatory, Leiden University, Einsteinweg 55, 2333 CC Leiden, The Netherlands\label{aff13}
\and
INAF-Osservatorio Astronomico di Roma, Via Frascati 33, 00078 Monteporzio Catone, Italy\label{aff14}
\and
INFN-Sezione di Roma, Piazzale Aldo Moro, 2 - c/o Dipartimento di Fisica, Edificio G. Marconi, 00185 Roma, Italy\label{aff15}
\and
Institute for Astronomy, University of Edinburgh, Royal Observatory, Blackford Hill, Edinburgh EH9 3HJ, UK\label{aff16}
\and
Institute for Theoretical Particle Physics and Cosmology (TTK), RWTH Aachen University, 52056 Aachen, Germany\label{aff17}
\and
INFN-Sezione di Genova, Via Dodecaneso 33, 16146, Genova, Italy\label{aff18}
\and
Dipartimento di Fisica, Universit\`a di Genova, Via Dodecaneso 33, 16146, Genova, Italy\label{aff19}
\and
Universit\'e Paris-Saclay, Universit\'e Paris Cit\'e, CEA, CNRS, AIM, 91191, Gif-sur-Yvette, France\label{aff20}
\and
Institut d'Estudis Espacials de Catalunya (IEEC),  Edifici RDIT, Campus UPC, 08860 Castelldefels, Barcelona, Spain\label{aff21}
\and
Institute of Space Sciences (ICE, CSIC), Campus UAB, Carrer de Can Magrans, s/n, 08193 Barcelona, Spain\label{aff22}
\and
Universit\'e Paris-Saclay, CNRS/IN2P3, IJCLab, 91405 Orsay, France\label{aff23}
\and
Institut de Recherche en Astrophysique et Plan\'etologie (IRAP), Universit\'e de Toulouse, CNRS, UPS, CNES, 14 Av. Edouard Belin, 31400 Toulouse, France\label{aff24}
\and
Laboratoire d'etude de l'Univers et des phenomenes eXtremes, Observatoire de Paris, Universit\'e PSL, Sorbonne Universit\'e, CNRS, 92190 Meudon, France\label{aff25}
\and
SISSA, International School for Advanced Studies, Via Bonomea 265, 34136 Trieste TS, Italy\label{aff26}
\and
ICSC - Centro Nazionale di Ricerca in High Performance Computing, Big Data e Quantum Computing, Via Magnanelli 2, Bologna, Italy\label{aff27}
\and
INAF-Osservatorio Astronomico di Trieste, Via G. B. Tiepolo 11, 34143 Trieste, Italy\label{aff28}
\and
IFPU, Institute for Fundamental Physics of the Universe, via Beirut 2, 34151 Trieste, Italy\label{aff29}
\and
INFN, Sezione di Trieste, Via Valerio 2, 34127 Trieste TS, Italy\label{aff30}
\and
Max Planck Institute for Extraterrestrial Physics, Giessenbachstr. 1, 85748 Garching, Germany\label{aff31}
\and
Institut f\"ur Theoretische Physik, University of Heidelberg, Philosophenweg 16, 69120 Heidelberg, Germany\label{aff32}
\and
Universit\'e St Joseph; Faculty of Sciences, Beirut, Lebanon\label{aff33}
\and
Department of Physics, Oxford University, Keble Road, Oxford OX1 3RH, UK\label{aff34}
\and
Institute for Particle Physics and Astrophysics, Dept. of Physics, ETH Zurich, Wolfgang-Pauli-Strasse 27, 8093 Zurich, Switzerland\label{aff35}
\and
LINKS Foundation, Via Pier Carlo Boggio, 61 10138 Torino, Italy\label{aff36}
\and
Dipartimento di Fisica e Scienze della Terra, Universit\`a degli Studi di Ferrara, Via Giuseppe Saragat 1, 44122 Ferrara, Italy\label{aff37}
\and
Mullard Space Science Laboratory, University College London, Holmbury St Mary, Dorking, Surrey RH5 6NT, UK\label{aff38}
\and
Ludwig-Maximilians-University, Schellingstrasse 4, 80799 Munich, Germany\label{aff39}
\and
INAF-Osservatorio di Astrofisica e Scienza dello Spazio di Bologna, Via Piero Gobetti 93/3, 40129 Bologna, Italy\label{aff40}
\and
INFN-Sezione di Bologna, Viale Berti Pichat 6/2, 40127 Bologna, Italy\label{aff41}
\and
Universit\'e Paris Cit\'e, CNRS, Astroparticule et Cosmologie, 75013 Paris, France\label{aff42}
\and
Department of Physics \& Astronomy, University of Sussex, Brighton BN1 9QH, UK\label{aff43}
\and
DAMTP, Centre for Mathematical Sciences, Wilberforce Road, Cambridge CB3 0WA, UK\label{aff44}
\and
Kavli Institute for Cosmology Cambridge, Madingley Road, Cambridge, CB3 0HA, UK\label{aff45}
\and
Istituto Nazionale di Fisica Nucleare, Sezione di Ferrara, Via Giuseppe Saragat 1, 44122 Ferrara, Italy\label{aff46}
\and
Dipartimento di Fisica e Astronomia "Augusto Righi" - Alma Mater Studiorum Universit\`a di Bologna, via Piero Gobetti 93/2, 40129 Bologna, Italy\label{aff47}
\and
Universit\'e de Gen\`eve, D\'epartement de Physique Th\'eorique and Centre for Astroparticle Physics, 24 quai Ernest-Ansermet, CH-1211 Gen\`eve 4, Switzerland\label{aff48}
\and
Instituto de F\'isica de Cantabria, Edificio Juan Jord\'a, Avenida de los Castros, 39005 Santander, Spain\label{aff49}
\and
Higgs Centre for Theoretical Physics, School of Physics and Astronomy, The University of Edinburgh, Edinburgh EH9 3FD, UK\label{aff50}
\and
Aix-Marseille Universit\'e, Universit\'e de Toulon, CNRS, CPT, Marseille, France\label{aff51}
\and
Dipartimento di Fisica - Sezione di Astronomia, Universit\`a di Trieste, Via Tiepolo 11, 34131 Trieste, Italy\label{aff52}
\and
Jodrell Bank Centre for Astrophysics, Department of Physics and Astronomy, University of Manchester, Oxford Road, Manchester M13 9PL, UK\label{aff53}
\and
School of Physics and Astronomy, Cardiff University, The Parade, Cardiff, CF24 3AA, UK\label{aff54}
\and
Institute of Astronomy, University of Cambridge, Madingley Road, Cambridge CB3 0HA, UK\label{aff55}
\and
Universit\'e Libre de Bruxelles (ULB), Service de Physique Th\'eorique CP225, Boulevard du Triophe, 1050 Bruxelles, Belgium\label{aff56}
\and
Universit\'e Paris-Saclay, CNRS, Institut d'astrophysique spatiale, 91405, Orsay, France\label{aff57}
\and
Departamento de F\'isica, FCFM, Universidad de Chile, Blanco Encalada 2008, Santiago, Chile\label{aff58}
\and
Universit\"ats-Sternwarte M\"unchen, Fakult\"at f\"ur Physik, Ludwig-Maximilians-Universit\"at M\"unchen, Scheinerstrasse 1, 81679 M\"unchen, Germany\label{aff59}
\and
Mathematical Institute, University of Leiden, Einsteinweg 55, 2333 CA Leiden, The Netherlands\label{aff60}
\and
Center for Cosmology and AstroParticle Physics, The Ohio State University, 191 West Woodruff Avenue, Columbus, OH 43210, USA\label{aff61}
\and
Department of Physics, The Ohio State University, Columbus, OH 43210, USA\label{aff62}
\and
ESAC/ESA, Camino Bajo del Castillo, s/n., Urb. Villafranca del Castillo, 28692 Villanueva de la Ca\~nada, Madrid, Spain\label{aff63}
\and
School of Mathematics and Physics, University of Surrey, Guildford, Surrey, GU2 7XH, UK\label{aff64}
\and
INAF-Osservatorio Astronomico di Brera, Via Brera 28, 20122 Milano, Italy\label{aff65}
\and
Dipartimento di Fisica e Astronomia, Universit\`a di Bologna, Via Gobetti 93/2, 40129 Bologna, Italy\label{aff66}
\and
Department of Physics "E. Pancini", University Federico II, Via Cinthia 6, 80126, Napoli, Italy\label{aff67}
\and
INAF-Osservatorio Astronomico di Capodimonte, Via Moiariello 16, 80131 Napoli, Italy\label{aff68}
\and
Instituto de Astrof\'isica e Ci\^encias do Espa\c{c}o, Universidade do Porto, CAUP, Rua das Estrelas, PT4150-762 Porto, Portugal\label{aff69}
\and
Faculdade de Ci\^encias da Universidade do Porto, Rua do Campo de Alegre, 4150-007 Porto, Portugal\label{aff70}
\and
Aix-Marseille Universit\'e, CNRS, CNES, LAM, Marseille, France\label{aff71}
\and
Port d'Informaci\'{o} Cient\'{i}fica, Campus UAB, C. Albareda s/n, 08193 Bellaterra (Barcelona), Spain\label{aff72}
\and
INFN section of Naples, Via Cinthia 6, 80126, Napoli, Italy\label{aff73}
\and
Institute for Astronomy, University of Hawaii, 2680 Woodlawn Drive, Honolulu, HI 96822, USA\label{aff74}
\and
Dipartimento di Fisica e Astronomia "Augusto Righi" - Alma Mater Studiorum Universit\`a di Bologna, Viale Berti Pichat 6/2, 40127 Bologna, Italy\label{aff75}
\and
Instituto de Astrof\'{\i}sica de Canarias, V\'{\i}a L\'actea, 38205 La Laguna, Tenerife, Spain\label{aff76}
\and
European Space Agency/ESRIN, Largo Galileo Galilei 1, 00044 Frascati, Roma, Italy\label{aff77}
\and
Universit\'e Claude Bernard Lyon 1, CNRS/IN2P3, IP2I Lyon, UMR 5822, Villeurbanne, F-69100, France\label{aff78}
\and
Institute of Physics, Laboratory of Astrophysics, Ecole Polytechnique F\'ed\'erale de Lausanne (EPFL), Observatoire de Sauverny, 1290 Versoix, Switzerland\label{aff79}
\and
Institut de Ci\`{e}ncies del Cosmos (ICCUB), Universitat de Barcelona (IEEC-UB), Mart\'{i} i Franqu\`{e}s 1, 08028 Barcelona, Spain\label{aff80}
\and
Instituci\'o Catalana de Recerca i Estudis Avan\c{c}ats (ICREA), Passeig de Llu\'{\i}s Companys 23, 08010 Barcelona, Spain\label{aff81}
\and
UCB Lyon 1, CNRS/IN2P3, IUF, IP2I Lyon, 4 rue Enrico Fermi, 69622 Villeurbanne, France\label{aff82}
\and
Departamento de F\'isica, Faculdade de Ci\^encias, Universidade de Lisboa, Edif\'icio C8, Campo Grande, PT1749-016 Lisboa, Portugal\label{aff83}
\and
Instituto de Astrof\'isica e Ci\^encias do Espa\c{c}o, Faculdade de Ci\^encias, Universidade de Lisboa, Campo Grande, 1749-016 Lisboa, Portugal\label{aff84}
\and
Department of Astronomy, University of Geneva, ch. d'Ecogia 16, 1290 Versoix, Switzerland\label{aff85}
\and
INAF-Istituto di Astrofisica e Planetologia Spaziali, via del Fosso del Cavaliere, 100, 00100 Roma, Italy\label{aff86}
\and
INFN-Padova, Via Marzolo 8, 35131 Padova, Italy\label{aff87}
\and
Aix-Marseille Universit\'e, CNRS/IN2P3, CPPM, Marseille, France\label{aff88}
\and
Space Science Data Center, Italian Space Agency, via del Politecnico snc, 00133 Roma, Italy\label{aff89}
\and
INFN-Bologna, Via Irnerio 46, 40126 Bologna, Italy\label{aff90}
\and
School of Physics, HH Wills Physics Laboratory, University of Bristol, Tyndall Avenue, Bristol, BS8 1TL, UK\label{aff91}
\and
INAF-Osservatorio Astronomico di Padova, Via dell'Osservatorio 5, 35122 Padova, Italy\label{aff92}
\and
Dipartimento di Fisica "Aldo Pontremoli", Universit\`a degli Studi di Milano, Via Celoria 16, 20133 Milano, Italy\label{aff93}
\and
Institute of Theoretical Astrophysics, University of Oslo, P.O. Box 1029 Blindern, 0315 Oslo, Norway\label{aff94}
\and
Jet Propulsion Laboratory, California Institute of Technology, 4800 Oak Grove Drive, Pasadena, CA, 91109, USA\label{aff95}
\and
Department of Physics, Lancaster University, Lancaster, LA1 4YB, UK\label{aff96}
\and
Felix Hormuth Engineering, Goethestr. 17, 69181 Leimen, Germany\label{aff97}
\and
Technical University of Denmark, Elektrovej 327, 2800 Kgs. Lyngby, Denmark\label{aff98}
\and
Cosmic Dawn Center (DAWN), Denmark\label{aff99}
\and
Institut d'Astrophysique de Paris, UMR 7095, CNRS, and Sorbonne Universit\'e, 98 bis boulevard Arago, 75014 Paris, France\label{aff100}
\and
Max-Planck-Institut f\"ur Astronomie, K\"onigstuhl 17, 69117 Heidelberg, Germany\label{aff101}
\and
NASA Goddard Space Flight Center, Greenbelt, MD 20771, USA\label{aff102}
\and
Department of Physics and Helsinki Institute of Physics, Gustaf H\"allstr\"omin katu 2, University of Helsinki, 00014 Helsinki, Finland\label{aff103}
\and
Department of Physics, P.O. Box 64, University of Helsinki, 00014 Helsinki, Finland\label{aff104}
\and
Helsinki Institute of Physics, Gustaf H{\"a}llstr{\"o}min katu 2, University of Helsinki, 00014 Helsinki, Finland\label{aff105}
\and
SKA Observatory, Jodrell Bank, Lower Withington, Macclesfield, Cheshire SK11 9FT, UK\label{aff106}
\and
Centre de Calcul de l'IN2P3/CNRS, 21 avenue Pierre de Coubertin 69627 Villeurbanne Cedex, France\label{aff107}
\and
INFN-Sezione di Milano, Via Celoria 16, 20133 Milano, Italy\label{aff108}
\and
Universit\"at Bonn, Argelander-Institut f\"ur Astronomie, Auf dem H\"ugel 71, 53121 Bonn, Germany\label{aff109}
\and
Department of Physics, Institute for Computational Cosmology, Durham University, South Road, Durham, DH1 3LE, UK\label{aff110}
\and
Universit\'e C\^{o}te d'Azur, Observatoire de la C\^{o}te d'Azur, CNRS, Laboratoire Lagrange, Bd de l'Observatoire, CS 34229, 06304 Nice cedex 4, France\label{aff111}
\and
CNRS-UCB International Research Laboratory, Centre Pierre Bin\'etruy, IRL2007, CPB-IN2P3, Berkeley, USA\label{aff112}
\and
Institut d'Astrophysique de Paris, 98bis Boulevard Arago, 75014, Paris, France\label{aff113}
\and
Aurora Technology for European Space Agency (ESA), Camino bajo del Castillo, s/n, Urbanizacion Villafranca del Castillo, Villanueva de la Ca\~nada, 28692 Madrid, Spain\label{aff114}
\and
OCA, P.H.C Boulevard de l'Observatoire CS 34229, 06304 Nice Cedex 4, France\label{aff115}
\and
Institut de F\'{i}sica d'Altes Energies (IFAE), The Barcelona Institute of Science and Technology, Campus UAB, 08193 Bellaterra (Barcelona), Spain\label{aff116}
\and
School of Mathematics, Statistics and Physics, Newcastle University, Herschel Building, Newcastle-upon-Tyne, NE1 7RU, UK\label{aff117}
\and
DARK, Niels Bohr Institute, University of Copenhagen, Jagtvej 155, 2200 Copenhagen, Denmark\label{aff118}
\and
Perimeter Institute for Theoretical Physics, Waterloo, Ontario N2L 2Y5, Canada\label{aff119}
\and
Centre National d'Etudes Spatiales -- Centre spatial de Toulouse, 18 avenue Edouard Belin, 31401 Toulouse Cedex 9, France\label{aff120}
\and
Institute of Space Science, Str. Atomistilor, nr. 409 M\u{a}gurele, Ilfov, 077125, Romania\label{aff121}
\and
Consejo Superior de Investigaciones Cientificas, Calle Serrano 117, 28006 Madrid, Spain\label{aff122}
\and
Universidad de La Laguna, Departamento de Astrof\'{\i}sica, 38206 La Laguna, Tenerife, Spain\label{aff123}
\and
Dipartimento di Fisica e Astronomia "G. Galilei", Universit\`a di Padova, Via Marzolo 8, 35131 Padova, Italy\label{aff124}
\and
Universit\"at Innsbruck, Institut f\"ur Astro- und Teilchenphysik, Technikerstr. 25/8, 6020 Innsbruck, Austria\label{aff125}
\and
Satlantis, University Science Park, Sede Bld 48940, Leioa-Bilbao, Spain\label{aff126}
\and
Department of Physics, Royal Holloway, University of London, Surrey TW20 0EX, UK\label{aff127}
\and
Instituto de Astrof\'isica e Ci\^encias do Espa\c{c}o, Faculdade de Ci\^encias, Universidade de Lisboa, Tapada da Ajuda, 1349-018 Lisboa, Portugal\label{aff128}
\and
Cosmic Dawn Center (DAWN)\label{aff129}
\and
Niels Bohr Institute, University of Copenhagen, Jagtvej 128, 2200 Copenhagen, Denmark\label{aff130}
\and
Universidad Polit\'ecnica de Cartagena, Departamento de Electr\'onica y Tecnolog\'ia de Computadoras,  Plaza del Hospital 1, 30202 Cartagena, Spain\label{aff131}
\and
Kapteyn Astronomical Institute, University of Groningen, PO Box 800, 9700 AV Groningen, The Netherlands\label{aff132}
\and
Infrared Processing and Analysis Center, California Institute of Technology, Pasadena, CA 91125, USA\label{aff133}
\and
INAF, Istituto di Radioastronomia, Via Piero Gobetti 101, 40129 Bologna, Italy\label{aff134}
\and
Astronomical Observatory of the Autonomous Region of the Aosta Valley (OAVdA), Loc. Lignan 39, I-11020, Nus (Aosta Valley), Italy\label{aff135}
\and
Department of Mathematics and Physics E. De Giorgi, University of Salento, Via per Arnesano, CP-I93, 73100, Lecce, Italy\label{aff136}
\and
INFN, Sezione di Lecce, Via per Arnesano, CP-193, 73100, Lecce, Italy\label{aff137}
\and
INAF-Sezione di Lecce, c/o Dipartimento Matematica e Fisica, Via per Arnesano, 73100, Lecce, Italy\label{aff138}
\and
ICL, Junia, Universit\'e Catholique de Lille, LITL, 59000 Lille, France\label{aff139}
\and
Instituto de F\'isica Te\'orica UAM-CSIC, Campus de Cantoblanco, 28049 Madrid, Spain\label{aff140}
\and
CERCA/ISO, Department of Physics, Case Western Reserve University, 10900 Euclid Avenue, Cleveland, OH 44106, USA\label{aff141}
\and
Technical University of Munich, TUM School of Natural Sciences, Physics Department, James-Franck-Str.~1, 85748 Garching, Germany\label{aff142}
\and
Max-Planck-Institut f\"ur Astrophysik, Karl-Schwarzschild-Str.~1, 85748 Garching, Germany\label{aff143}
\and
Donostia International Physics Center (DIPC), Paseo Manuel de Lardizabal, 4, 20018, Donostia-San Sebasti\'an, Guipuzkoa, Spain\label{aff144}
\and
IKERBASQUE, Basque Foundation for Science, 48013, Bilbao, Spain\label{aff145}
\and
Departamento de F{\'\i}sica Fundamental. Universidad de Salamanca. Plaza de la Merced s/n. 37008 Salamanca, Spain\label{aff146}
\and
Universit\'e de Strasbourg, CNRS, Observatoire astronomique de Strasbourg, UMR 7550, 67000 Strasbourg, France\label{aff147}
\and
Max-Planck-Institut f\"ur Physik, Boltzmannstr. 8, 85748 Garching, Germany\label{aff148}
\and
California Institute of Technology, 1200 E California Blvd, Pasadena, CA 91125, USA\label{aff149}
\and
Department of Physics \& Astronomy, University of California Irvine, Irvine CA 92697, USA\label{aff150}
\and
Departamento F\'isica Aplicada, Universidad Polit\'ecnica de Cartagena, Campus Muralla del Mar, 30202 Cartagena, Murcia, Spain\label{aff151}
\and
Observatorio Nacional, Rua General Jose Cristino, 77-Bairro Imperial de Sao Cristovao, Rio de Janeiro, 20921-400, Brazil\label{aff152}
\and
CEA Saclay, DFR/IRFU, Service d'Astrophysique, Bat. 709, 91191 Gif-sur-Yvette, France\label{aff153}
\and
Department of Computer Science, Aalto University, PO Box 15400, Espoo, FI-00 076, Finland\label{aff154}
\and
Instituto de Astrof\'\i sica de Canarias, c/ Via Lactea s/n, La Laguna 38200, Spain. Departamento de Astrof\'\i sica de la Universidad de La Laguna, Avda. Francisco Sanchez, La Laguna, 38200, Spain\label{aff155}
\and
Caltech/IPAC, 1200 E. California Blvd., Pasadena, CA 91125, USA\label{aff156}
\and
Ruhr University Bochum, Faculty of Physics and Astronomy, Astronomical Institute (AIRUB), German Centre for Cosmological Lensing (GCCL), 44780 Bochum, Germany\label{aff157}
\and
Department of Physics and Astronomy, Vesilinnantie 5, University of Turku, 20014 Turku, Finland\label{aff158}
\and
Serco for European Space Agency (ESA), Camino bajo del Castillo, s/n, Urbanizacion Villafranca del Castillo, Villanueva de la Ca\~nada, 28692 Madrid, Spain\label{aff159}
\and
ARC Centre of Excellence for Dark Matter Particle Physics, Melbourne, Australia\label{aff160}
\and
Centre for Astrophysics \& Supercomputing, Swinburne University of Technology,  Hawthorn, Victoria 3122, Australia\label{aff161}
\and
Department of Physics and Astronomy, University of the Western Cape, Bellville, Cape Town, 7535, South Africa\label{aff162}
\and
Department of Astrophysics, University of Zurich, Winterthurerstrasse 190, 8057 Zurich, Switzerland\label{aff163}
\and
Department of Physics, Centre for Extragalactic Astronomy, Durham University, South Road, Durham, DH1 3LE, UK\label{aff164}
\and
IRFU, CEA, Universit\'e Paris-Saclay 91191 Gif-sur-Yvette Cedex, France\label{aff165}
\and
Oskar Klein Centre for Cosmoparticle Physics, Department of Physics, Stockholm University, Stockholm, SE-106 91, Sweden\label{aff166}
\and
Astrophysics Group, Blackett Laboratory, Imperial College London, London SW7 2AZ, UK\label{aff167}
\and
Univ. Grenoble Alpes, CNRS, Grenoble INP, LPSC-IN2P3, 53, Avenue des Martyrs, 38000, Grenoble, France\label{aff168}
\and
INAF-Osservatorio Astrofisico di Arcetri, Largo E. Fermi 5, 50125, Firenze, Italy\label{aff169}
\and
Dipartimento di Fisica, Sapienza Universit\`a di Roma, Piazzale Aldo Moro 2, 00185 Roma, Italy\label{aff170}
\and
Centro de Astrof\'{\i}sica da Universidade do Porto, Rua das Estrelas, 4150-762 Porto, Portugal\label{aff171}
\and
Dipartimento di Fisica, Universit\`a di Roma Tor Vergata, Via della Ricerca Scientifica 1, Roma, Italy\label{aff172}
\and
INFN, Sezione di Roma 2, Via della Ricerca Scientifica 1, Roma, Italy\label{aff173}
\and
HE Space for European Space Agency (ESA), Camino bajo del Castillo, s/n, Urbanizacion Villafranca del Castillo, Villanueva de la Ca\~nada, 28692 Madrid, Spain\label{aff174}
\and
Department of Astrophysical Sciences, Peyton Hall, Princeton University, Princeton, NJ 08544, USA\label{aff175}
\and
INAF-Osservatorio Astronomico di Brera, Via Brera 28, 20122 Milano, Italy, and INFN-Sezione di Genova, Via Dodecaneso 33, 16146, Genova, Italy\label{aff176}
\and
Theoretical astrophysics, Department of Physics and Astronomy, Uppsala University, Box 516, 751 37 Uppsala, Sweden\label{aff177}
\and
Space physics and astronomy research unit, University of Oulu, Pentti Kaiteran katu 1, FI-90014 Oulu, Finland\label{aff178}
\and
Institut de Physique Th\'eorique, CEA, CNRS, Universit\'e Paris-Saclay 91191 Gif-sur-Yvette Cedex, France\label{aff179}
\and
Center for Computational Astrophysics, Flatiron Institute, 162 5th Avenue, 10010, New York, NY, USA\label{aff180}}

\date{\today}

\authorrunning{Euclid Collaboration: S. Joudaki et al.}

\titlerunning{\cloe code implementation}
 
\abstract{We provide a description of the code implementation and structure of Cosmology Likelihood for Observables in Euclid (\cloe), developed by members of the Euclid Consortium. \cloe is a modular \python code for computing the theoretical predictions of cosmological observables and evaluating them against state-of-the-art data from galaxy surveys such as \Euclid in a unified likelihood. This primarily includes the core observables of weak gravitational lensing, photometric galaxy clustering, galaxy-galaxy lensing, and spectroscopic galaxy clustering, but also extended probes such as the clusters of galaxies and cross-correlations of galaxy positions and shapes with the cosmic microwave background. While \cloe has been developed to serve as the unified framework for the parameter inferences in \Euclid, it has general capabilities that can serve the broader cosmological community. It is different from other comparable cosmological tools in that it is written entirely in \python, performs the full likelihood calculation, and includes both photometric and spectroscopic observables. We focus on the primary probes of \Euclid and describe the overall code structure, rigorous code development practices, extensive documentation, unique features, speed optimization, and future development plans. \cloe is publicly available at \url{https://github.com/cloe-org/cloe}.
}


\maketitle

\hypersetup{linkcolor=reddy}

\tableofcontents
   
\section{Introduction}

The next generation of cosmological galaxy surveys, such as the \Euclid space mission \citep{Mellier24}, the Rubin Observatory Legacy Survey of Space and Time (LSST;~\citealt{Ivezic19}), and the {\it Roman} Space Telescope \citep{Spergel15} will observe the positions and shapes of billions of galaxies to an unprecedented precision during the coming decade. By measuring the spatial correlations of these galaxy positions and shapes with one another, they will have the statistical capability to provide order-of-magnitude level improvements in the constraints on the underlying cosmology. In order to achieve this, the measurements are typically compared against the theoretical predictions across the parameter space in a Monte Carlo analysis. Hence, to reach the scientific objectives of the cosmological galaxy surveys, we need a correspondingly accurate inference framework that computes the theoretical predictions and evaluates the probability density in a way that allows for unbiased parameter constraints.

This entails an improvement in the precision with which the theoretical predictions are computed, such as the integrations, interpolations, and approximations involved. It further entails an increase in the complexity of the modeling of the systematic uncertainties in the data and theory, such as the intrinsic galaxy alignments, photometric redshift uncertainties, shear calibration uncertainties, magnification bias, sample impurities, interloper contamination, galaxy bias, and baryonic feedback. The improved numerical precision, achieved for example through a higher density of redshifts and scales, naturally comes at the expense of lower computational speed. Moreover, the increase in the complexity of the modeling of the systematic uncertainties is often associated with an expansion of the parameter space, which naturally leads to a slower convergence of the Monte Carlo chains. Throughout this effort, it is therefore important to maintain or consistently work to improve the computational speed despite the inherent challenges.

\begin{figure*}[h]
\includegraphics[width=0.9\linewidth, trim = {0.5cm 7.5cm 7cm 0.8cm}]{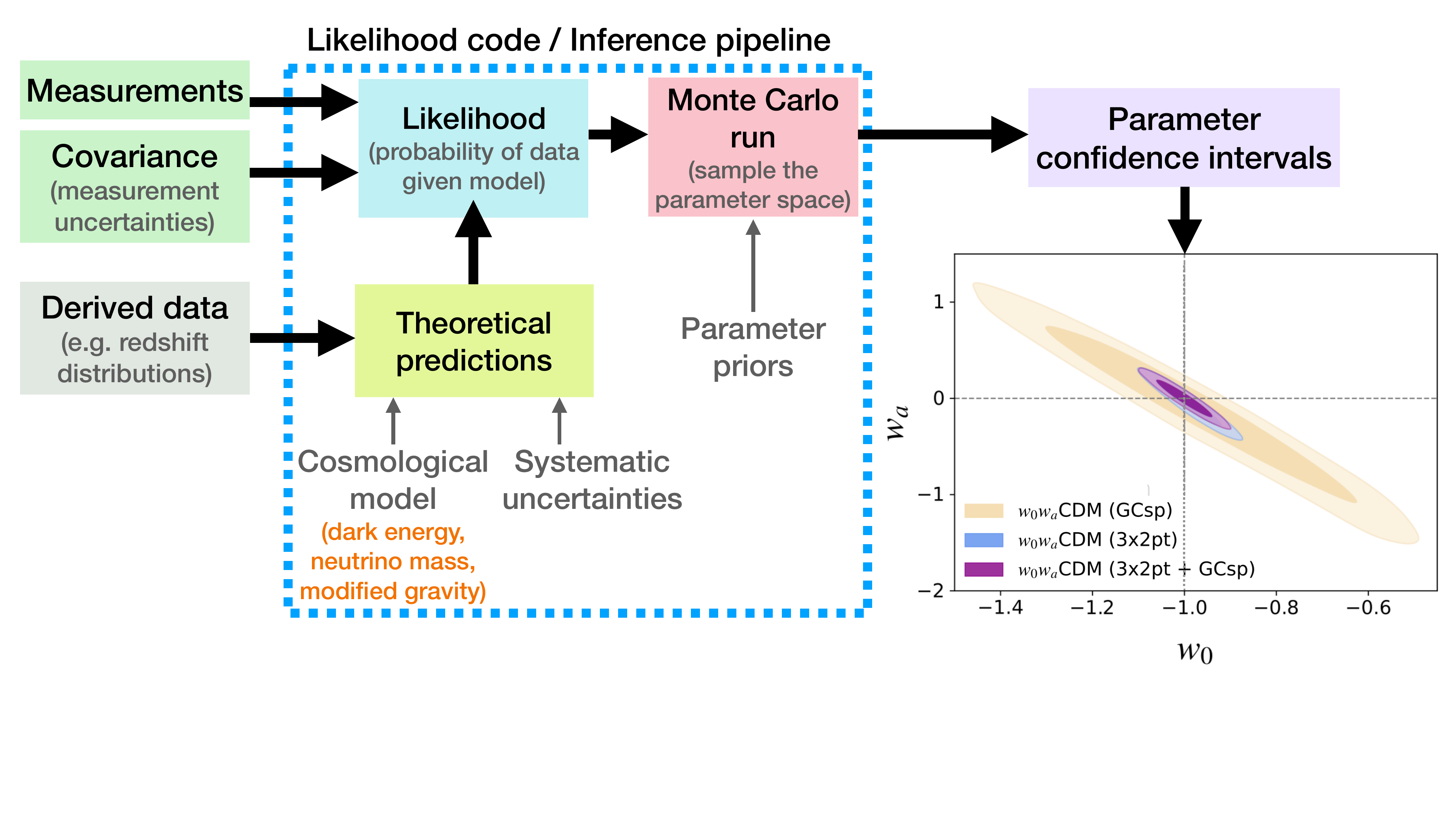}
\caption{Illustration of how the cosmological parameter constraints are obtained. The bottom-right inset is adapted from the forecasted \Euclid results in \citet{Mellier24}, produced with the cosmological inference pipeline \cloe.}
\label{pipelinefig}
\end{figure*}

In the creation of the \Euclid inference pipeline, named the Cosmology Likelihood for Observables in Euclid (\cloe;~interchangeably referred to as the likelihood code, which includes the computation of the theoretical predictions), it has been crucial to enforce a rigorous development workflow. To this end, we have followed development practices inspired by Agile management \citep{Abrahamsson02}, with a task force structure that includes dedicated roles for each of the members. \cloe has a high percentage (>80\%) of unit test coverage, in accordance with modern software development guidelines.\footnote{The unit test coverage is an indicator of the code robustness. However, beyond merely a high coverage, it is the quality of the unit tests that help maintain a bug-free code. Our coverage strikes a balance between an extensive test suite that covers critical parts of the code and the diminishing returns of aiming for an even higher percentage.} It further enforces continuous integration, which triggers an automated build with testing to immediately detect possible errors. It is also important to have the inference pipeline benchmarked at a higher precision than in current survey analyses, given the increased constraining power of the next-generation surveys. The benchmarking entails a comparison of \cloe against other independent codes, and we have enforced a requirement on their agreement to be less than 10\% of the error bars for \Euclid across a wide range of cosmologies and full range of scales \citep{Martinelli24}. This is similar to the precision required for comparable theory codes such as the Core Cosmology Library\footnote{\ccladdress}\href{https://github.com/LSSTDESC/CCL}{\color{black}\faGithub} (\ccl), which had a requirement of 10\% of the error bars for LSST \citep{Chisari19}.

A further consideration includes the development of a single unified likelihood across the different cosmological probes, such as the 3$\times$2pt observables of cosmic shear, galaxy-galaxy lensing, and galaxy clustering. In the case of the 3$\times$2pt observables, this was done for the first time in the analyses of the Kilo Degree Survey and the Dark Energy Survey \citep{joudaki18a, Uitert18, Desy1} and it has become the standard approach in subsequent analyses (e.g.,~\citealt{Heymans21, Desy3, HSCy3s, Hscy3m}). Given their cross-covariance, it would be  incorrect to have separate likelihoods for each of these probes. The cross-correlations between the \Euclid photometric and spectroscopic datasets are currently not included due to their partial redshift overlap (see e.g. \citealt{paganin24}). However, we preserve the ability to include these additional cross-correlations in a straightforward manner by creating a single unified likelihood. This further extends to combined analyses of the photometric and spectroscopic probes with the weak lensing of the cosmic microwave background (CMB;~\citealt{Desy1cmb, Desy3cmb}).

Additional considerations are the user-friendliness and flexibility of the code. This includes the ability for a \cloe user to choose specific probe combinations, summary statistics, and scales in the analysis. It further includes the ability to choose between distinct sampling platforms such as \cobaya\footnote{\cobayaaddress}\href{https://github.com/CobayaSampler/cobaya}{\color{black}\faGithub} \citep{Torrado21} and \cosmosis\footnote{\cosmosisaddress}\href{https://github.com/joezuntz/cosmosis}{\color{black}\faGithub} \citep{Zuntz15}, Boltzmann codes such as \camb\footnote{\cambaddress}\href{https://github.com/cmbant/CAMB}{\color{black}\faGithub} \citep{lewis2000} and \class\footnote{\classaddress}\href{https://github.com/lesgourg/class_public}{\color{black}\faGithub} \citep{Lesgourgues11a, Lesgourgues11b, Blas11}, and different nonlinear matter power spectrum recipes such as \hmcode\footnote{\hmcodeaddress}\href{https://github.com/alexander-mead/HMcode}{\color{black}\faGithub} \citep{Mead15, Mead16, Mead21}, \halofit\href{https://github.com/cmbant/CAMB/blob/master/fortran/halofit.f90}{\color{black}\faGithub} 
\citep{Smith03, Takahashi12}, \bacco\footnote{\baccoaddress} \citep{Angulo21}, and \eecode\footnote{\eecodeaddress}\href{https://github.com/miknab/EuclidEmulator2}{\color{black}\faGithub} \citep{Knabenhans19, Knabenhans21}. The user-friendliness of \cloe also stems from the decision to write it entirely in \python, albeit with dependencies on external non-\python codes such as \camb and \class. This can be contrasted with \ccl, which is written in \ccode with a \python frontend. Despite the intrinsic speed limitation of \python, we find that \cloe is able to produce the likelihood for a \Euclid survey setup in approximately one second, which is a performance that is superior to other comparable codes (as detailed in Sect.~\ref{speedsec}).

A more subtle consideration is the cultivation of expertise within the Euclid Consortium (EC). The development and road-testing of the new parameter inference pipeline has allowed the EC to expand on the necessary expertise ahead of its first set of data analyses. This is highlighted by a series of doctoral theses that include the contributions of early-career researchers toward this effort \citep{Ajanithesis, GCHthesis, Sciottithesis, Gohthesis, Keilthesis}. By developing the inference pipeline in-house, it will also be simpler to tailor it for \Euclid's specific needs moving forward. For instance, \ccl does not have the capability to compute the theory predictions for redshift-space galaxy clustering, as LSST is a photometric survey and the collaboration therefore does not have the direct need for it. On the other hand, \Euclid is both a photometric and spectroscopic survey, and needs to have a single inference pipeline for both. To lower the threshold for entry into parameter inferences by the rest of the EC and beyond, \cloe has an extensive \sphinx documentation and \jupyternotebooks that illustrate how to use and modify the code.

At its core, the operational objective of \cloe is exceptionally simple and can be boiled down to a single equation:
\begin{equation}
    \ln ({\rm Likelihood}) = -\frac{1}{2} ({\rm Data} -  {\rm Theory}) \, {\rm Cov}^{-1} \, ({\rm Data} -  {\rm Theory})^{\sf T} \, + \, A ,
\end{equation}
where the data and theory are expressed in terms of row vectors, ${\rm Cov}$ refers to the covariance that encodes the uncertainty on the measurements, $\sf T$ denotes the transpose operation, $A$ is a normalization constant, and there is an implicit assumption that the data has a Gaussian sampling distribution.\footnote{Note that when the covariance matrix is estimated from  simulations, the Gaussian form of the likelihood is replaced by a $t$-distribution (still assuming that the data intrinsically have a Gaussian sampling distribution; \citealt{SH16, Percival22}; also see \citealt{Upham21, Krywonos24}), as further described in Sect.~\ref{sec3h}.} Note that $A$ is immaterial in a parameter inference, as any procedure is only sensitive to the changes in the likelihood between different points in the parameter space. Here, the likelihood can be expressed as $P(D|\theta,\mathcal{M})$ and encapsulates the probability of the data, $D$, given the model, $\mathcal{M}$, and a set of parameter values, $\theta$. It is the quantity that needs to be computed at each point in parameter space as we perform a Monte Carlo run to obtain the posterior, $P(\theta|D,\mathcal{M})$, which encapsulates the probability of the parameters given the data and model. In detail, Bayes' theorem allows us to relate the two probabilities in the form
\begin{equation} 
P(\theta|D,\mathcal{M}) = {P(D|\theta,\mathcal{M}) \, P(\theta|\mathcal{M})}/{P(D|\mathcal{M})} \; ,
\label{bayeseqn}
\end{equation}
where $P(\theta|\mathcal{M})$ is the prior probability of the parameters given the model, or put differently, the initial probability before taking into account the data. Meanwhile, $P(D|\mathcal{M})$ is known as the Bayesian evidence and corresponds to the probability of the data given the model. In other words, the Bayesian evidence has the parameter dependence of the likelihood integrated out, such that 
\begin{equation}
P(D|\mathcal{M}) = \int P(D|\theta,\mathcal{M})\,P(\theta|\mathcal{M})\;{\rm d}\theta \; .
\label{evidenceeqn}
\end{equation}
Note that for parameter inference in the context of a single model, the evidence is a constant as we sample across the parameter space during a Monte Carlo computation. It therefore simply normalizes the posterior and does not affect the parameter constraints. The Bayesian evidence value, however, plays an important role in model comparison when two or more models are considered (see~\citealt{Trotta08} for a review).\footnote{We note that the Bayesian evidence is computed in nested sampling runs with \cloe. Another relevant model selection statistics 
that can be obtained from MCMC chains using \cloe is the deviance information criterion \citep{spiegelhalter02}. There are moreover a range of data concordance estimators that can be straightforwardly computed from the outputs of MCMC runs, such as the $\log \mathcal{I}$ statistic \citep{joudaki17a, joudaki22}, $Q_{\rm DMAP}$ statistic \citep{raveri19}, and suspiciousness statistic \citep{handley19, lemos20}.}

As illustrated in Fig.~\ref{pipelinefig}, \cloe thereby computes the likelihood, multiplies it with the prior probability based on the choices for the parameter priors by a user, and passes this forward to a sampler. In computing the likelihood, \cloe reads in the data and covariance and computes the theory predictions given a model of the underlying cosmology and systematic uncertainties. This is an important distinction from \ccl, which  produces the theory predictions and then interfaces with \firecrown\footnote{\firecrownaddress}\href{https://github.com/LSSTDESC/firecrown}{\color{black}\faGithub} for the likelihood calculation and sampling. In the present version of \cloe, it is interfaced with both \cobaya and \cosmosis, and the possible samplers are therefore those included in these platforms. This includes the Metropolis--Hastings fast-slow dragging algorithm~\href{https://github.com/CobayaSampler/cobaya/blob/master/cobaya/samplers/mcmc/mcmc.py}{\color{black}\faGithub}, which is a type of Markov chain Monte Carlo (MCMC) method \citep{metropolis53, hastings70, Neal05, lewis13}, and the nested sampling algorithm of \polychord\footnote{\polychordaddress}\href{https://github.com/PolyChord}{\color{black}\faGithub} \citep{handley15a, handley15b}. 

In the forecasted parameter inferences of \citet{GCH24}, \citet{Goh24}, and \citet{Blot24}, \cloe is moreover interfaced with the external sampler
\nautilus\footnote{\nautilusaddress}\href{https://github.com/johannesulf/nautilus}{\color{black}\faGithub} \citep{lange23} via the \cobaya backend. For simpler access to outside samplers, in particular gradient-based samplers such as Hamiltonian Monte Carlo \citep{duane87, Neal96} and extensions such as the No-U-Turn Sampler \citep{hoffman11}, it would be advantageous to disentangle the inference framework from any given sampling platform.

This publication is the second in a series of papers on \cloe and is focused on the code implementation. The first paper contains a description of the theoretical recipe \citep{Cardone24}, the third paper contains the final cosmological parameter forecasts for \Euclid \citep{GCH24}, the fourth paper contains the validation of the code \citep{Martinelli24}, the fifth paper contains forecasted constraints on extended cosmological models \citep{Goh24}, and the sixth paper contains an investigation of the impact of systematic uncertainties on the cosmological constraints of \Euclid \citep{Blot24}. In Sect.~\ref{sec2}, we provide a description of the code development workflow, and in Sect.~\ref{sec3}, we describe the code structure. In Sect.~\ref{sec4}, we focus on some of the unique features of \cloe. In Sect.~\ref{speedsec}, we report on the computational speed of \cloe, and in Sect.~\ref{sec7}, we discuss its future development plans. Lastly, we conclude in Sect.~\ref{sec9}.

\begin{table*}
\caption{Summary of the current \cloe features, associated prescriptions, and references.}
\begin{tabular}{l|l}
\toprule
\cloe feature & Prescription or relevant quantity \\
\midrule
Cosmological probe (1) & Photometric 3$\times$2pt: angular power spectra (\S\ref{sec3e}) \\
Cosmological probe (2) & Photometric 3$\times$2pt: pseudo-$C_\ell$ (\S\ref{sec3e}) \\
Cosmological probe (3) & Photometric 3$\times$2pt: correlation functions (\S\ref{sec3e}) \\
Cosmological probe (4) & Spectroscopic galaxy clustering: multipole power spectra (\S\ref{sec3f}) \\
Cosmological probe (5) & Spectroscopic galaxy clustering: window-convolved multipole power spectra (\S\ref{sec3f}) \\
Cosmological probe (6) & Spectroscopic galaxy clustering: multipole correlation functions (\S\ref{sec3f}) \\
Cosmological probe (7) & CMB lensing and temperature correlations (${\rm G}\phi$, ${\rm L}\phi$, $\phi\phi$, ${\rm GT}$; \S\ref{cmbsec})
\\
Cosmological probe (8) & Clusters of galaxies (see Euclid Collaboration:~Sakr et al., in prep.) \\
User selection of probes, scales, redshifts & Masking vector formalism (\S\ref{sec3i}) \\
Einstein-Boltzmann solver (1) & \camb (\S\ref{sec3d}) \\
Einstein-Boltzmann solver (2) & \class (\S\ref{sec3d}) \\
Nonlinear matter power spectrum (1) & \halofit (\S\ref{sec3d}) \\
Nonlinear matter power spectrum (2) & \hmcode (2016 and 2020 versions; \S\ref{sec3d}) \\
Nonlinear matter power spectrum (3) & \bacco (\S\ref{sec3g}) \\
Nonlinear matter power spectrum (4) & \euclidemu (\S\ref{sec3g}) \\
Baryonic feedback (1) & \hmcode (2016 and 2020 versions; \S\ref{sec3d}) \\
Baryonic feedback (2) & \bacco (\S\ref{sec3g}) \\
Baryonic feedback (3) & \bcemu (\S\ref{sec3g}) \\
Reweighting of lensing kernels & BNT transformation (\S\ref{bntsec}, Appendix \ref{eq:BNT_appendix}) \\
Large-angle corrections & Extended Limber and curved-sky (\S\ref{sec3e1}) \\
Redshift space distortions (1) & Photometric galaxy clustering and galaxy-galaxy lensing (\S\ref{sec3e1}) \\
Redshift space distortions (2) & Spectroscopic galaxy clustering (\S\ref{sec3f1}) \\
Intrinsic alignments (1) & Extended nonlinear linear alignment model (\S\ref{sec3b}, \S\ref{sec3e1}) \\
Intrinsic alignments (2) & Tidal alignment and tidal torquing model (\S\ref{sec3g}) \\
Photometric source redshift uncertainties & Shift ($\delta z_{\rm s}$) in mean redshift of source galaxy distribution (\S\ref{sec3b}, \S\ref{sec3e1}) \\
Photometric lens redshift uncertainties & Shift ($\delta z_{\rm l}$) in mean redshift of lens galaxy distribution (\S\ref{sec3b}, \S\ref{sec3e1}) \\
Spectroscopic redshift uncertainties (1) & Constant across redshift bins (\S\ref{sec3b}, \S\ref{sec3f1}) \\ 
Spectroscopic redshift uncertainties (2) & Linear in redshift (\S\ref{sec3b}, \S\ref{sec3f1}) \\ 
Photometric galaxy bias (1) & Linear interpolation of input values (\S\ref{sec3b}, \S\ref{sec3e1}) \\
Photometric galaxy bias (2) & Constant in each tomographic bin (\S\ref{sec3b}, \S\ref{sec3e1}) \\
Photometric galaxy bias (3) & Cubic polynomial in redshift (\S\ref{sec3b}, \S\ref{sec3e1}) \\
Photometric galaxy bias (4) & 1-loop perturbation theory (\S\ref{sec3g}) \\
Galaxy power spectrum (1) & Linear theory (\S\ref{sec3f1})\\
Galaxy power spectrum (2) & Nonlinear prescription: \eftoflss (\S\ref{sec3g}) \\
Shear calibration uncertainties & Multiplicative bias parameter for each tomographic bin \\
Weak lensing generalization & Weyl power spectrum (see \citealt{Goh24}) \\
Photometric magnification bias (1) & Linear interpolation of input values (\S\ref{sec3b}, \S\ref{sec3e1}) \\
Photometric magnification bias (2) & Constant nuisance parameter for each bin (\S\ref{sec3b}, \S\ref{sec3e1}) \\
Photometric magnification bias (3) & Cubic polynomial in redshift (\S\ref{sec3b}, \S\ref{sec3e1}) \\
Spectroscopic magnification bias & Standard formalism (see \citealt{Goh24})  \\ 
Spectroscopic sample impurities (1) & Redshift-independent outlier fraction (\S\ref{sec3b})  \\ 
Spectroscopic sample impurities (2) & Outlier fraction for each redshift bin (\S\ref{sec3b}) \\ 
Data reader & Both generic and \Euclid-specific data formats (\S\ref{sec3j}) \\
Code robustness & 
Unit tests, continuous integration, \docker images (\S\ref{sec3l}, \S\ref{sec3q}, \S\ref{sec3r}) \\
Code benchmarking & 
Primary \Euclid observables (see \citealt{Martinelli24}) \\
Efficient integration & \fftlog (\S\ref{fftlogsec}) \\
Plotting routines & Cosmological observables and chains (\S\ref{sec3a}, \S\ref{sec3k}) \\
Likelihood shape (1) & Gaussian (analytic covariance; \S\ref{sec3c}, \S\ref{sec3h}) \\
Likelihood shape (2) & Non-Gaussian (simulated covariance; \S\ref{sec3c}, \S\ref{sec3h}) \\
User interface (1) & Executable 
(\S\ref{sec3a}) \\
User interface (2) & \jupyter demonstration and validation notebooks (Appendix \ref{appnote}) \\
User interface (3) & Graphical user interface for creating configuration files 
(\S\ref{sec3n}) \\
Code documentation & Docstrings (\sphinx \numpydoc; \S\ref{sec3s}) \\
Sampling platform (1) & \cobaya (\S\ref{sec3ccobaya}) \\
Sampling platform (2) & \cosmosis (\S\ref{sec3ccosmosis}) \\
Extended cosmology (1) & Evolving dark energy ($w_0$--$w_a$; \S\ref{sec3m}) \\
Extended cosmology (2) & Modified gravity (via modified growth index $\gamma_{\rm MG}; \S\ref{sec3m}$) \\
Extended cosmology (3) & Nonzero curvature (\S\ref{sec3m}) \\
Extended cosmology (4) & Sum of neutrino masses (\S\ref{sec3m}) \\
\bottomrule
\end{tabular}
\label{tabfeat}
\end{table*}

\section{Code development and design}
\label{sec2}

The core functionalities of \cloe have been developed within the EC. At the outset, the development focused on three primary goals. This consisted of: (1) creating the likelihood pipeline for the main \Euclid probes; (2) enforcing a modular code structure; and (3) ensuring that the code is well documented, easy to use and modify, and available to all EC members and eventually the broader community.

The development began by brainstorming the architecture of the inference pipeline within the EC, in particular its key functionalities and use cases. These included interfaces to Boltzmann codes such as \camb and \class, which output the linear matter power spectrum and background quantities such as comoving distances and growth rates. The planned architecture also comprised a theory subpackage to compute the core photometric and spectroscopic observables, a data subpackage to read in the measurements, redshift distributions, and covariances, and a masking subpackage to modify the desired range of the theory and data vectors in redshift, $z$, and wavenumber, $k$. Additional  functionalities encompassed nonlinear corrections for the matter power spectrum and bias models, propagation of systematic uncertainties (e.g.,~in shear calibration and photometric redshifts), calculation of the likelihood for any probe combination, sampling over the likelihood to obtain parameter constraints, and the use of likelihoods of other datasets such as the \Planck CMB \citep{planck18cosmo, planck18like}.

We defined possible use cases for how the planned pipeline could achieve the intended objectives. An example included the execution of a combined analysis of the \Euclid photometric and spectroscopic primary probes to test the standard cosmological model. Project acceptance criteria referring to the performance requirements and necessary code conditions were further agreed on, such as accounting for all major systematic uncertainties and core extended cosmologies (e.g.,~involving massive neutrinos and evolving dark energy).

\subsection{Agile development}
\label{sec2a}

To foster a rigorous development of the inference pipeline, we followed development practices inspired by Agile management \citep{Beck99a, Beck99b, becketal, Abrahamsson02}, with a task force structure comprising two distinct teams with their own dedicated scrum masters, developers, reviewers, and experts. The teams were also able to interact with dedicated consultants and contact points of distinct groups in the EC. This structure ensured that each member continuously improved their ability to, for example, develop or review a piece of code once a proposed implementation had been discussed and approved.

In these scrum-inspired teams, the scrum masters coordinated the coding efforts of their respective teams. The developers created new branches and wrote new code once a proposed implementation for a task had been discussed and approved. The reviewers, 
preferably with more coding experience, were responsible for assessing implementations and were the only ones authorized to merge new code into the repository. The role of the experts was to ensure that the implemented code is sound from the perspective of the underlying physics, that the documentation is up-to-date, and that there are no other inconsistencies in the code to be merged. The final layer of code review and approval was then performed by the task force leads, who had oversight of the entire development.

In ensuring a consistent form and standard of the code entering \cloe, we created contributing guidelines, which includes a set of style guidelines for increased readability and simplified maintenance of the code. In addition to maintaining adherence to these guidelines, the code reviews verified that all continuous integration tests have passed (for further details, see Sect.~\ref{sec3q}), identified bugs or issues not covered by the existing tests, and ensured that all new code is well documented. 
The reviews also encouraged the use of established libraries (e.g.,~\numpy, \scipy) rather than reinventing functionality, 
and highlighted areas for improvement, in particular ways in which the code could be made faster, cleaner, less repetitive, and more object-oriented.

\subsection{Development training}
\label{trainsec}

In order to promote high-quality code implementation in \cloe, each new developer and reviewer underwent a dedicated training procedure, focused on two exercises. The objective of these exercises was to train potential contributors on good coding practices, but also on how to interact in a collaborative development project. Following the two exercises, we decided on suitable roles for new members, which could change over time depending on their own preferences.

In the first exercise, participants were asked to open a merge request in a dedicated training repository, where they implemented basic cosmological functions and corresponding unit tests. The trainer then reviewed the merge request to the same standards as in the main \cloe repository, and evaluated the participants' performance based on the number of mistakes made and the quality of their responses to feedback.

In the second exercise, the roles were reversed: the trainer acted as a developer and submitted a dummy merge request implementing similar cosmological functions with various deliberate errors. The participants had to review it, identify items that required fixing, and request changes. The exercise was assessed according to how many mistakes the participants detected and how effectively they managed the interaction.

\subsection{\git development platform}
\label{gitsec}

To maintain consistency with other projects in the EC, we adopted \gitlab as our development platform using the EC’s self-hosted instance. Each development task was tracked through an issue, with an associated merge request containing only the minimal changes needed to resolve it. This workflow reduced the risk of concurrent edits causing merge conflicts. Issues were also grouped into milestones with well-defined content limits, corresponding to earlier internal-to-EC releases. For broader accessibility, however, we publicly release \cloe on \github.

\subsection{Sampling platform}
\label{sec2b}

Ahead of embarking on the code development, we decided that the sampling platforms of \cobaya and \cosmosis were preferred as compared to other prominent platforms such as \cosmomc\footnote{\cosmomcaddress}\href{https://github.com/cmbant/CosmoMC}{\color{black}\faGithub} \citep{lewis02} and \montepython\footnote{\montepythonaddress}\href{https://github.com/brinckmann/montepython_public}{\color{black}\faGithub} \citep{audren13, brinckmann19}. We considered factors such as the programming language, support for Boltzmann solvers, modularity, extensibility, maintenance, range of samplers and analysis tools, and the availability of expertise for a given framework within the EC.

We placed particularly high value on the programming language being \python and support for both \camb and \class. The former predisposition was driven by the preference for increased accessibility and compatibility with other \Euclid codes. Meanwhile, the latter predisposition was driven by the preference for increased flexibility in the modeling, as for instance extensions to the standard model are typically implemented in either \camb or \class, but not both. This particularly includes modified gravity extensions such as \mgcamb\footnote{\mgcambaddress}\href{https://github.com/sfu-cosmo/MGCAMB}{\color{black}\faGithub} \citep{mgcamb09, wang23}, \mgclass\footnote{\mgclassaddress}\href{https://gitlab.com/zizgitlab/mgclass--ii}{\color{black}\faGitlab} \citep{mgclass1, mgclass2}, \hiclass\footnote{\hiclassaddress}\href{https://miguelzuma.github.io/hi_class_public}{\color{black}\faGithub} \citep{hiclass16, bellini2020},  \eftcamb\footnote{\eftcambaddress}\href{http://eftcamb.org/}{\color{black}\faGithub} \citep{hu14,raveri14}, 
and \isitgr\footnote{\isitgraddress}\href{https://github.com/mishakb/ISiTGR}{\color{black}\faGithub} \citep{isitgr11, isitgr19}.
As \cosmomc is written in \fortran and only compatible with \camb, while \montepython is only compatible with \class, these considerations favored \cobaya and \cosmosis as the backends for \cloe.

\section{Structure of \cloe}
\label{sec3}

As part of the cosmological parameter inference (see Fig.~\ref{pipelinefig}), \cloe reads in probe-specific data products such as photometric redshift distributions and the survey window function. It then computes the theoretical predictions of the considered probes for a given cosmological model and treatment of the systematic uncertainties. These theoretical predictions are used together with the measurements and covariance to obtain the likelihood. Lastly, the likelihood is sampled
across the parameter space to obtain the posterior probability. For the \Euclid mission, the primary probes are defined by the following set:
\begin{itemize}
    \item Cosmic shear  tomography
    \item Photometric galaxy clustering tomography
    \item Photometric galaxy-galaxy lensing tomography
    \item Spectroscopic / Redshift-space galaxy clustering.
\end{itemize}
As shown in Table~\ref{tabfeat}, \cloe allows the user to consider these probes in either Fourier or configuration space. For the photometric probes (first three above), this implies either angular power spectra, also commonly expressed as harmonic space power spectra, or angular correlation functions. Likewise, for the spectroscopic probe (fourth above), it implies either multipole power spectra or multipole correlation functions, following a Legendre expansion of the redshift-space galaxy power spectrum.
These probes can be considered either separately or in a self-consistent combined analysis. In the case of the photometric probes, the combined analysis of cosmic shear, galaxy-galaxy lensing, and galaxy clustering is referred to as a 3$\times$2pt analysis.

\Euclid, and \cloe by extension, currently does not include any cross-correlations between the photometric and spectroscopic galaxies (given their limited redshift overlap in the case of \Euclid, which moreover is expected to give rise to a negligible cross-covariance between the photometric and spectroscopic probes, as illustrated in \citealt{TM22}). 
However, \cloe already extends to a subset of the \Euclid non-primary probes. This includes the cross-correlations of the shear and positions of galaxies with the lensing and temperature anisotropies of the CMB, along with the cluster number counts (and eventually cluster lensing profile and correlation function; Euclid Collaboration:~Sakr et al., in prep.). It is further possible to analyze the \Euclid data alongside other external datasets such as the \Planck CMB temperature and polarization (\citealt{planck18cosmo, planck18like}; see Appendix~\ref{sec3p}). 

The repository is structured in terms of a variety of directories and files described in this section, and further outlined in Appendix~\ref{apprepo}. This includes the directory \texttt{cloe}, which contains the main source code in the form of numerous subpackages and modules for the calculation of the theoretical predictions and likelihood. The directory \texttt{configs} contains configuration \yaml files that include the user specifications for the Monte Carlo runs, such as which probes to consider, treatment of systematic uncertainties, scale cuts, and parameter priors. The directory \texttt{data} contains the data files, such as the measurements, covariances, mixing matrices, and redshift distributions. The directory \texttt{cosmosis} contains the interface to \cosmosis. The directory \texttt{gui} further contains the source code of the graphical user interface, and \texttt{docs} contains the automatically generated \sphinx documentation. 

In addition, the directory \texttt{notebooks} contains \jupyter demonstration and validation notebooks, \texttt{mcmc\_scripts} contains example \python scripts to run MCMC chains, \texttt{scripts} contains example \python scripts to create synthetic data, and \texttt{chains} is automatically created when performing a Monte Carlo run whereby it stores the corresponding chains. The file \texttt{run\_cloe.py} is the main script for running the \cloe user interface, \texttt{setup.py} is the installation and testing configuration script, \texttt{environment.yml} is the \conda environment file, and \texttt{LICENCE} contains the GNU Lesser General Public License (LGPL), which allows \cloe's redistribution and use.

We note that \cloe does not compute the theory predictions entirely on its own, as it has dependencies on external codes such as the \numpy and \scipy open-source \python libraries, \camb and \class for the linear matter power spectrum computation, nonlinear prescriptions such as \hmcode and \bacco for the baryonic feedback modeling, and \cobaya and \cosmosis along with their samplers for the posterior estimation, as further described in this section and Appendix~\ref{sec3p}.

\subsection{The \cloe overlayer}
\label{sec3a}

\begin{figure}[h]
\includegraphics[width=2.95\linewidth]{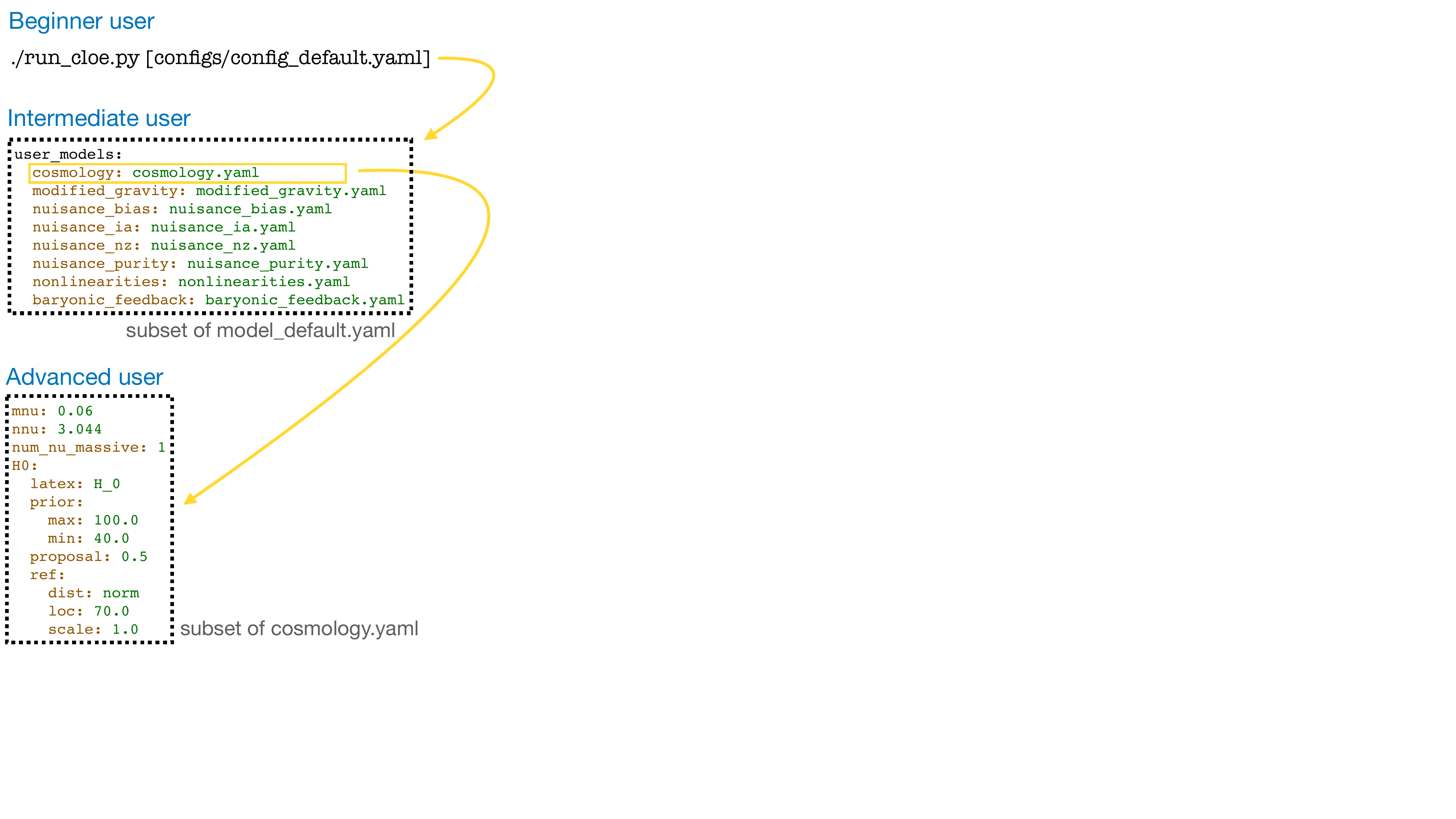}
\vspace{-9.6em}
\caption{Illustration of the different types of \cloe users and their approaches to the code. The square brackets indicate that the argument is optional.}
\label{usersfig}
\end{figure}

The overlayer is \cloe's top level user interface, defined via  \texttt{cloe/user\_interface/likelihood\_ui.py} and \texttt{run\_cloe.py}.
The overlayer allows the user to run \cloe with the Boltzmann solvers and parameter samplers of \cobaya and \cosmosis. It further allows for a wide range of user options, such as the selection of datasets, observables, tomographic bins, scales, summary statistics, nuisance modeling, and parameter priors. These user options are encapsulated within \yaml files that are located in the directory \texttt{cloe/configs} (see Sect.~\ref{sec3b}). In creating the overlayer, our intention has been to provide \cloe with an interface that caters to all potential users, as illustrated in Fig.~\ref{usersfig}. This includes:
\begin{itemize}
    \item The beginner user, who wishes to treat \cloe as a ``black box.'' This user has the option to utilize one of the available configuration files. An example configuration file is \texttt{config\_default.yaml}, which we will henceforth consider as the default configuration file in \cloe. Additional configuration files for non-standard cosmologies are available in the \cloe repository as well.
    \item The intermediate user, who wishes to have more control over the modeling. This user has the option to modify the modeling of the underlying cosmology and systematic uncertainties from the available model files. An example model file is \texttt{model\_default.yaml}, referred to within the default configuration file. Other examples include the set of files this refers to itself.
    \item The advanced user, who wishes to have full access to the modeling of the underlying cosmology and systematic uncertainties.
    This user has the option to directly  modify and even create new model files. An example model file is \texttt{cosmology.yaml}, referred to within the default model file. 
\end{itemize}

The structure of the default configuration file is such that it contains the option to choose between \cobaya and \cosmosis through the \texttt{backend} key and  between \camb and \class via the \texttt{solver} key. The default configuration moreover contains a \texttt{likelihood} block for \Euclid, a \texttt{params} block for the parameter priors, and \texttt{output} block that governs the location of the chains. The \texttt{likelihood} block further defines the arrays of wavenumber and redshift, allows for different modeling choices, and contains information on which probes to consider in the analysis.

In its simplest form, \cloe can be run by executing the command \texttt{./run\_cloe.py}. This utilizes the overlayer's optional arguments, which include the user specification of the configuration file. It is therefore equivalent to executing the following more explicit command:
\begin{lstlisting}
python run_cloe.py configs/config_default.yaml
\end{lstlisting}
In the command above, \texttt{config\_default.yaml} can be replaced with the user's own bespoke configuration file, either of which is then passed to the \texttt{LikelihoodUI} class of the \texttt{likelihood\_ui} module as this class is instantiated in \texttt{run\_cloe.py}. Another optional argument is the specification of the action to be performed (via \texttt{-a} followed by the name of the action in the command to run \cloe). The available actions are named \texttt{run}, \texttt{process}, and \texttt{plot}, where \texttt{run} proceeds with the likelihood inference and is the default action. 

\begin{table*}
\begin{center}
\caption{Summary of the connections between a subset of the equations in \citet{Cardone24} and the corresponding \cloe methods and variable names.
}
\begin{tabular}{l|l|l|l}
\toprule
Quantity & \cloe Method/Variable/Key & Expression & Equation \\
\midrule
Cosmic shear power spectrum & \texttt{Cl\_WL} & $C_{ij}^{\rm LL}(\ell)$ & $(32)$\\
Photometric galaxy power spectrum & \texttt{Cl\_GC\_phot} & $C_{ij}^{\rm GG}(\ell)$ & $(40)$\\
Galaxy-galaxy lensing power spectrum & \texttt{Cl\_cross} & $C_{ij}^{\rm GL}(\ell)$ & $(54)$\\
Modified gravity growth index & \texttt{use\_gamma\_MG} & $\gamma_{\rm MG} \neq 6/11$ & $(27)$\\
Redshift-dependent outlier fraction & \texttt{f\_out\_z\_dep} & $f_{\rm out}$ & $(100)$\\
Spectroscopic redshift error & \texttt{GCsp\_z\_err} & $\sigma_{\rm err}(z) \neq 0$ & $(98)$ \\
Photometric galaxy bias model & \texttt{bias\_model} & $b^{\rm photo}(z)$ & $(52)$ \\
Magnification bias model & \texttt{magbias\_model} & $b^{\rm mag}(z)$ & $(51)$ \\
Photometric redshift space distortions & \texttt{add\_phot\_RSD}  & $W_i^{\rm RSD}(\ell, z) \neq 0$ & $(47)$ \\
BNT transformation & \texttt{BNT\_transform} & $\mathfrak{M}_{ij}$ & $(88)$ \\
Cosmic shear tracer prefactor & \texttt{set\_prefactor} & $F_\gamma(\ell)$ & $(36)$\\
Photometric galaxy tracer prefactor & \texttt{set\_prefactor} & $F_\mu(\ell)$ & $(44)$ \\
Source galaxy redshift distribution & \texttt{nz\_WL} & $n^{\rm S}_i(z)$ & $(37)$ \\
Lens galaxy redshift distribution & \texttt{nz\_GC} & $n^{\rm L}_i(z)$ & $(46)$ \\
Photometric redshift uncertainties & \texttt{evaluates\_n\_i\_z} & $\Delta z_i^{\rm{S}/\rm{L}}$ & $(117)$ \\
Multiplicative shear calibration bias & \texttt{multbias} & $m_i$ & $(123)$ \\
Cosmic shear window function & \texttt{WL\_window} & $W_{i}^{\gamma}(z)$ & $(37)$ \\
Intrinsic alignment window function & \texttt{IA\_window} & $W_{i}^{\rm IA}(z)$ & $(39)$\\
Galaxy clustering window function & \texttt{GC\_window} & $W_{i}^{\delta}(z)$ & $(46)$ \\
Redshift space distortion window function & \texttt{GC\_window\_RSD} & $W_{i}^{\rm RSD}(\ell, z)$ & $(47)$ \\
Magnification bias window function & \texttt{magnification\_window} & $W_{i}^{\mu}(z)$ & $(50)$ \\
Photometric 3$\times$2pt pseudo-$C_{\ell}$ & \texttt{pseudo\_Cl\_3x2pt} & $\tilde{C}_{ij}(\ell)$ &  $(61)$ \\
Cosmic shear correlation function & \texttt{corr\_func\_3x2pt} & $\xi_{ij}^\pm(\theta)$ &  $(71)$ \\
Photometric galaxy correlation function & \texttt{corr\_func\_3x2pt} & $\xi_{ij}^{\rm GG}(\theta)$ &  $(74)$ \\
Galaxy-galaxy lensing correlation function & \texttt{corr\_func\_3x2pt} & $\xi_{ij}^{\rm GL}(\theta)$ &  $(75)$ \\
Multipole power spectrum & \texttt{multipole\_spectra} & $P_{\rm obs, \ell}(k^{\rm fid}, z)$ & $(95)$ \\
Multipole correlation function & \texttt{multipole\_correlation\_function} & $\xi_{{\rm obs},\ell}(s^{\rm fid},z)$ & $(105)$ \\
Convolved multipole power spectrum & \texttt{convolved\_power\_spectrum\_multipoles} & $\tilde{P}_{\ell}(k)$ & $(108)$ \\
\bottomrule
\end{tabular}
\label{tabpri}
\end{center}
\end{table*}

As the default \texttt{run} action is considered, the \texttt{run} method of \texttt{LikelihoodUI} is called within \texttt{run\_cloe.py}. This \texttt{run} method in turn either calls the internal method \texttt{run\_cobaya} or \texttt{run\_cosmosis} depending on the user choice for the \texttt{backend} key in the default configuration file. When either of these two methods is called, \cloe proceeds by either executing \texttt{cobaya.run} or \texttt{cosmosis.run\_cosmosis}, respectively, passing the user configurations to the sampling platforms in the process. The Monte Carlo output is subsequently written to the \texttt{chains} folder that is generated at runtime.

In addition, \texttt{process} refers to the processing of the chains, which at the moment is restricted to the creation of triangle contours using \getdist\footnote{\getdistaddress}\href{https://github.com/cmbant/getdist}{\color{black}\faGithub} \citep{lewis19}. Lastly, the action \texttt{plot} refers to the creation of figures of the theoretical predictions of the observables as a function of redshift and scale. This utilizes \cloe's plotting module \texttt{cloe/auxiliary/plotter.py} and specifically refers to the angular power spectrum of the cosmic shear, galaxy-galaxy lensing, and photometric galaxy clustering, along with the multipole power spectrum of the spectroscopic galaxy positions in redshift space.

Other optional arguments of \texttt{run\_cloe.py} include \texttt{-ps} for plot settings, \texttt{-d} for additional dictionary arguments that take precedence over those in the configuration file (by default \texttt{config\_default.py}), and \texttt{-v} for the verbosity level of the run. While plot settings are supplied directly through the command line, a \yaml-based specification is also compatible with the overall configuration structure. In the case of the verbosity level, via the \python module \texttt{logging} imported within \texttt{cloe/auxiliary/logger.py}, it can be set to either \texttt{debug}, \texttt{info}, \texttt{warning}, \texttt{error}, or \texttt{critical}. These are the standard options, where \texttt{debug} is the lowest logging level that provides details to help diagnose coding bugs, while \texttt{critical} is the highest logging level that highlights errors that can prevent the code from continuing to run. In between these two levels, \texttt{info} provides confirmation that the code runs as expected, \texttt{warning} points out unexpected behavior, and \texttt{error} indicates the inability of the code to perform one of its functions.

We note that \cloe can also be run directly from a platform such as \cobaya. This would allow for the use of \cobaya's own user interface, either from the shell (via the \texttt{cobaya-run} command) or interactively from a \python script or interpreter (via an import of the \texttt{run} method from the \texttt{cobaya.run} module). For additional user flexibility, we refer the reader to Sect.~\ref{sec3m} for this separate, but entirely self-consistent, approach to run \cloe that bypasses the overlayer.

\subsection{Configuration \yaml files}
\label{sec3b}

\subsubsection{Top layer}

\begin{figure*}[h]
\centering
\includegraphics[width=0.98\linewidth]{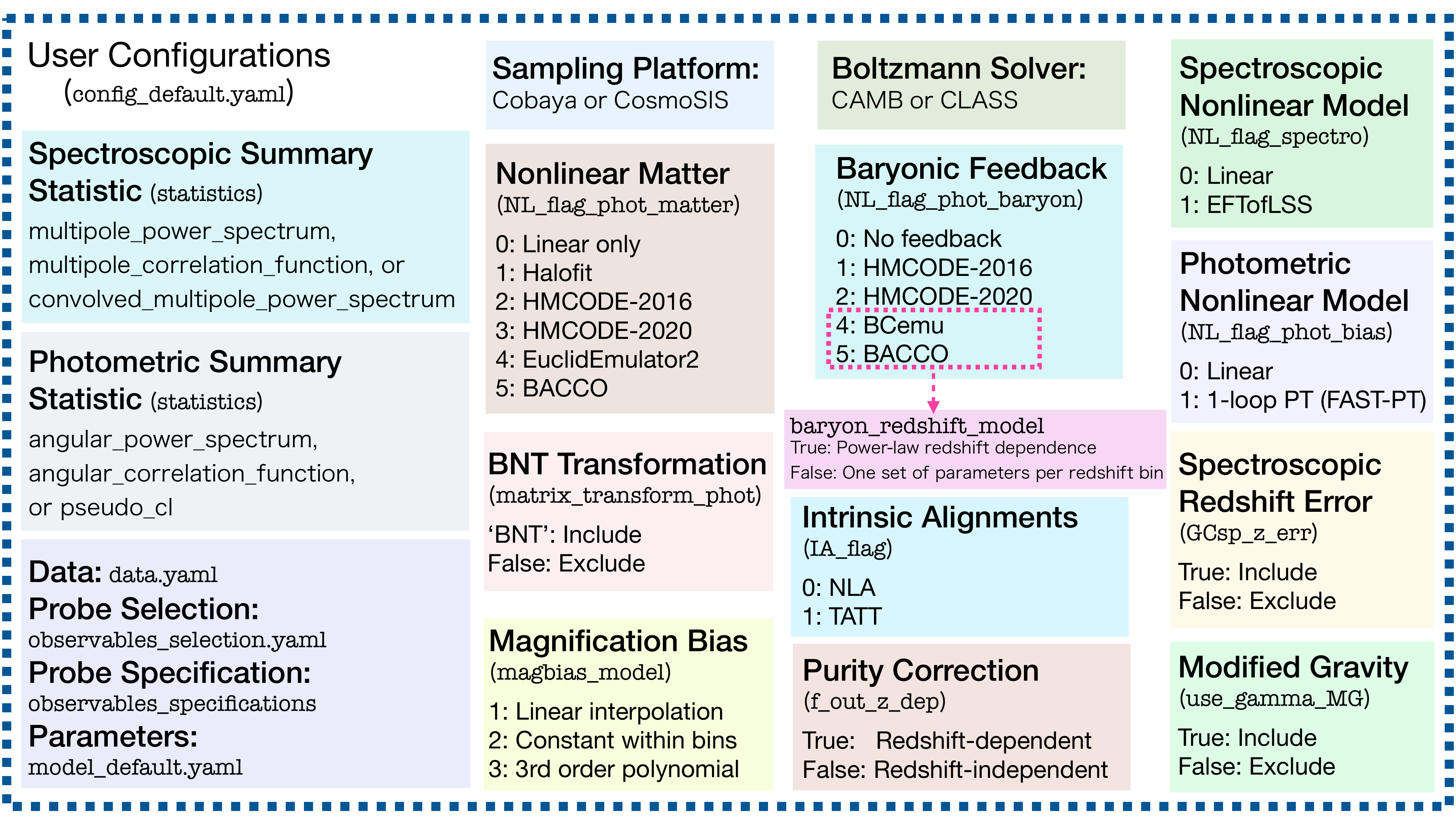}
\caption{Illustration of a subset of the configurations to be selected by the user as part of the theory and likelihood calculation, as described in Sect.~\ref{sec3b}. We note that the entries of a given key will be distinct for a given version of \cloe.}
\label{configsfig}
\end{figure*}

As discussed in Sect.~\ref{sec3a}, \cloe uses the \yaml format for its configuration files, which are stored in the \texttt{configs} directory. As illustrated in Fig.~\ref{configsfig}, the primary configuration file is \texttt{config\_default.yaml}, which contains a range of keys and references to other \yaml files that determine the use of \cloe. This file has expanded in terms of its functionalities as \cloe has evolved and the values of distinct keys are specific to each version of the code. The top two keys are \texttt{backend} and \texttt{solver}, which respectively allow the user to choose between \cobaya and \cosmosis for the sampling platform, along with \camb and \class for the Boltzmann solver. When \cobaya is chosen, there are references to other \yaml files in a nested structure. Meanwhile, for \cosmosis, there are references to other \ini files as this is the file format understood by this platform (Sect.~\ref{sec3ccosmosis}). A natural generalization would be to further unify the configurations for these two platforms.

In \texttt{config\_default.yaml}, the user can modify the keys that govern the setup of the parameter inference. This includes the nonlinear matter power spectrum, where the key \texttt{NL\_flag\_phot\_matter} allows the user to choose between the prescriptions of \halofit, \hmcode, \bacco, and \eecode. Moreover, this includes the baryonic feedback corrections to the matter power spectrum, where the key \texttt{NL\_flag\_phot\_baryon} allows the user to choose between \hmcode, \bacco, and \bcemu\footnote{\bcemuaddress}\href{https://github.com/sambit-giri/BCemu}{\color{black}\faGithub} \citep{gs21}. Here, \bcemu is an emulator of the baryonic correction model, also known as the baryonification model \citep{st15, schneider19, arico20}. In the case of \bacco and \bcemu, the redshift dependence of the baryonic feedback is set via the key \texttt{Baryon\_redshift\_model}. Meanwhile, the keys \texttt{NL\_flag\_phot\_bias} and \texttt{NL\_flag\_spectro} determine the use of either the linear or nonlinear clustering of the photometric and spectroscopic galaxies, respectively. 

The intrinsic alignment (IA) model key, \texttt{IA\_flag}, allows the user to choose between the nonlinear linear alignment model (NLA; \citealt{HS04, BK07, Joachimi2011}) and the tidal alignment and tidal torquing model (TATT; \citealt{Blazek19}). The redshift dependence of the photometric linear galaxy bias is set by \texttt{bias\_model}, the magnification bias model is set by \texttt{magbias\_model}, the spectroscopic redshift error is set by \texttt{GCsp\_z\_err}, and the redshift-dependent sample purity correction is set by \texttt{f\_out\_z\_dep} (corresponding, respectively, to Eqs.~52, 51, 98, and 100 in \citealt{Cardone24}, as summarized in Table~\ref{tabpri}). Other keys include \texttt{IR\_resum} for the infrared resummation scheme of the galaxy power spectrum (Gaussian filter or discrete sine transform; described in e.g. \citealt{pezzotta25}), \texttt{use\_Weyl} for use of the Weyl power spectrum instead of the matter power spectrum (further described in \citealt{Goh24}), and \texttt{print\_theory} for whether the user wishes to have the theory predictions written to file. The modified gravity growth index, $\gamma_{\rm MG}$, in Eq.~(27) of \citet{Cardone24} is currently also set by the user in this top-layer configuration file.

The user can choose the summary statistics for the observables via the \texttt{statistics} key. In the current version of \cloe, this includes either angular power spectra, pseudo-$C_\ell$, or correlation functions for the photometric probes, and it includes either multipole power spectra, window-convolved multipole power spectra, or multipole correlation functions for the spectroscopic probes.\footnote{This entry can also accommodate additional summary statistics such as COSEBIs \citep{Schneider2010, Asgari2012} and band powers \citep{Schneider02, Uitert18}.} In \texttt{config\_default.yaml}, we refer to \texttt{data.yaml} for the specifications of the observational data and to \texttt{observables\_selection.yaml} for the choices on the set of observables to consider in the analysis. We moreover introduce the \texttt{observables\_specifications} key that allows the user to have full control over the redshifts and scales that enter into the parameter inference (as further discussed below and in Sect.~\ref{sec3i}). For the auto-correlations, the specifications for each of the observables are given in files of the form \texttt{X-Z.yaml}, where \texttt{X} $\in$ \{\texttt{WL}, \texttt{GCphot}, \texttt{GCspectro}\} for the cosmic shear, photometric galaxy clustering, and spectroscopic galaxy clustering, while \texttt{Z} $\in$ \{\texttt{FourierSpace}, \texttt{ConfigurationSpace}\} for the Fourier space and configuration space analyses, respectively. Similarly, for the cross-correlations, the observables specifications are given in files of the form \texttt{X-Y-Z.yaml}, where additionally \texttt{Y} $\in$ \{\texttt{WL}, \texttt{GCphot}, \texttt{GCspectro}\}. 

The specifications of the galaxy cluster observables are included in the file \texttt{CG.yaml} (Euclid Collaboration:~Sakr et al., in prep.). For the CMB cross-correlations, the specifications are included in \texttt{CMBlensing.yaml}, \texttt{CMBlensing-WL.yaml}, \texttt{CMBlensing-GCphot.yaml}, and \texttt{iSW-GCphot.yaml}, as further discussed in Sect.~\ref{cmbsec}. The cosmological and systematics models are then specified in \texttt{model\_default.yaml}. We further note that the configuration file allows the user to specify the range and density of redshifts and wavenumbers for which the matter power spectrum will be computed. Moreover, the base file name of the Monte Carlo chains is specified in the key \texttt{output} and the choice of sampler in the key \texttt{sampler}. There is also an \texttt{action} key that allows the user to either execute a Monte Carlo run or create plots as previously detailed in Sect.~\ref{sec3a}. 

\subsubsection{Middle layer}
\label{midlayersec}

Moving one layer down, in \texttt{configs/data.yaml} we specify the names of the measurement and covariance files, along with ancillary information such as the redshift bin edges and whether the covariance is constructed analytically or via numerical simulations. In \texttt{observables\_selection.yaml}, we introduce the following upper triangular structure:
\begin{figure}[h]
\includegraphics[width=0.61\linewidth, trim = {-1cm 1.7cm 9cm 0.95cm}]{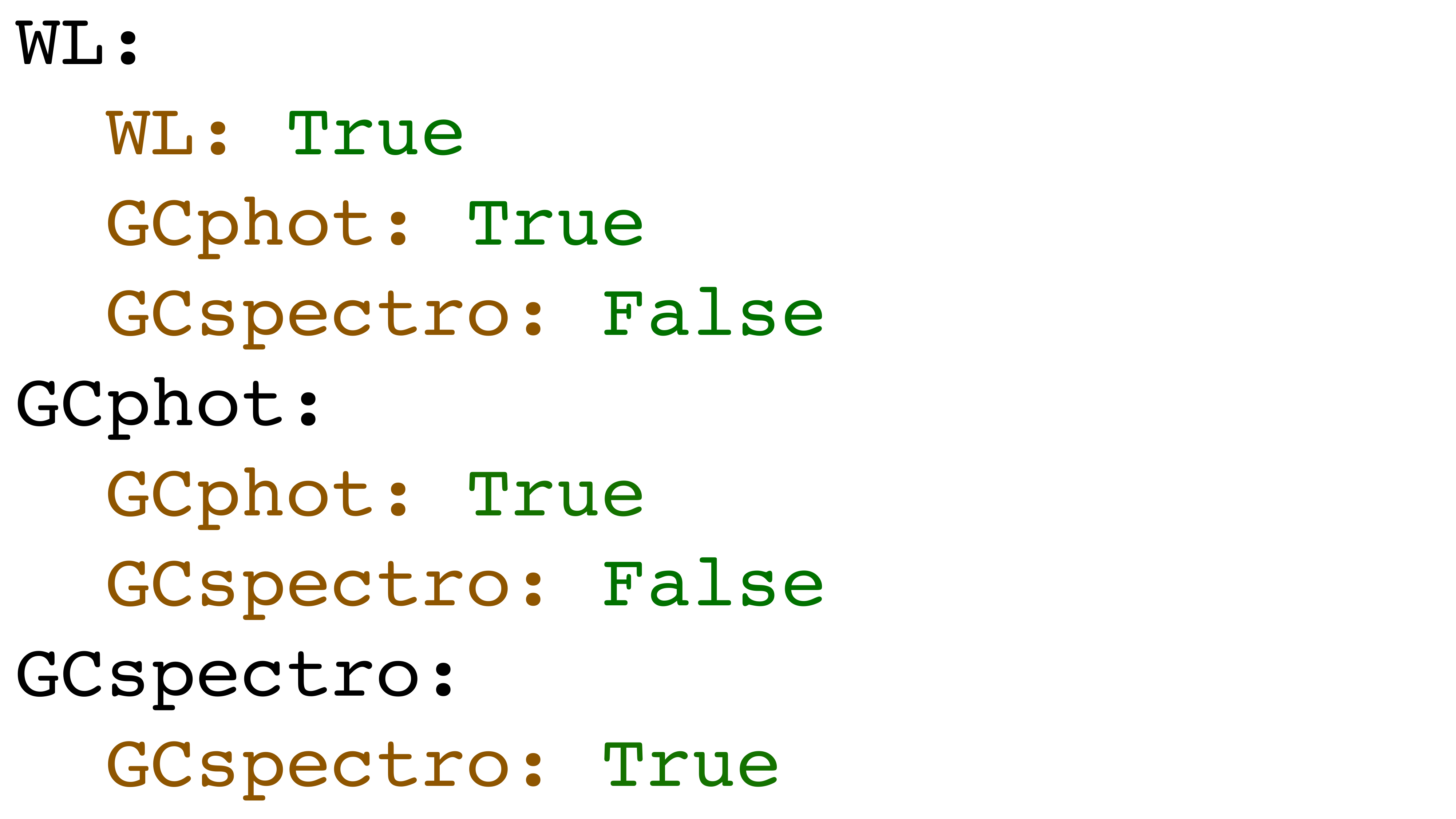}
\vspace{0.2em}
\end{figure}
\\ \indent 
Here, the user specifies \texttt{True} for the selected observables and \texttt{False} otherwise. In this specific example, the selection corresponds to a combined photometric 3$\times$2pt + spectroscopic galaxy clustering analysis. In the same file, the user can moreover specify whether to include redshift space distortions for the photometric galaxy clustering and galaxy-galaxy lensing via the key \texttt{add\_phot\_RSD} and whether to perform a Bernardeau--Nishimichi--Taruya (BNT;~\citealt{bnt14}) transformation of the cosmic shear and galaxy-galaxy lensing via the key \texttt{matrix\_transform\_phot} (as further discussed in Sect.~\ref{bntsec}).

Given the selection of observables and summary statistics, the key \texttt{observables\_specifications} encodes the user specified cuts in scale and redshift via a set of \yaml files for the different probes. As an example, in the case of shear angular power spectra, the file \texttt{WL-FourierSpace.yaml} has the following simple form for a scenario with three tomographic bins: 
\newpage
\begin{figure}[h]
\vspace{-0.23em}
\includegraphics[width=1.06\linewidth, trim = {0cm 2cm 9cm -0.5cm}]{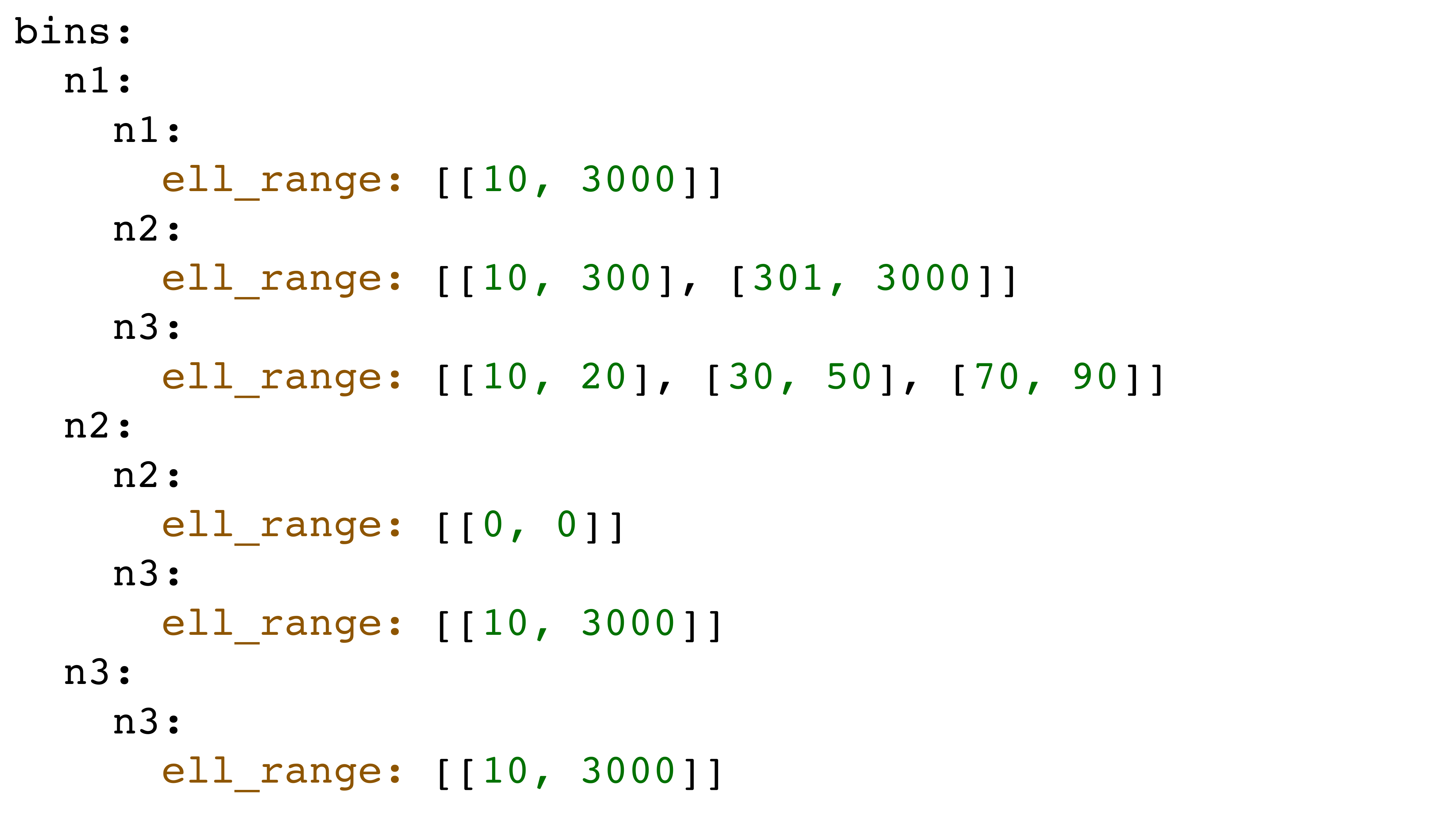}
\end{figure}
\indent 
Here, the user specifies that they would like to consider a multipole range between $10 \leq \ell \leq 3000$ for the tomographic bin combinations 1--1, 1--2, 2--3, and 3--3 (where the bin combinations refer to the indices $i$ and $j$ in the angular power spectrum $C_{ij}(\ell)$, and likewise for the correlation functions). The user moreover specifies the multipole range $\{10 \leq \ell \leq 20\} \, \cup \, \{30 \leq \ell \leq 50\} \, \cup \, \{70 \leq \ell \leq 90\}$ for the 1--3 bin combination, along with the neglect of the 2--2 bin combination entirely. The observables specifications follow a similar format for the other probes.

Lastly, \texttt{model\_default.yaml} contains the paths for the model specifications under the key \texttt{user\_models}:
\begin{figure}[h]
\includegraphics[width=1.32\linewidth, trim = {0cm 11.5cm 4cm 1.5cm}]{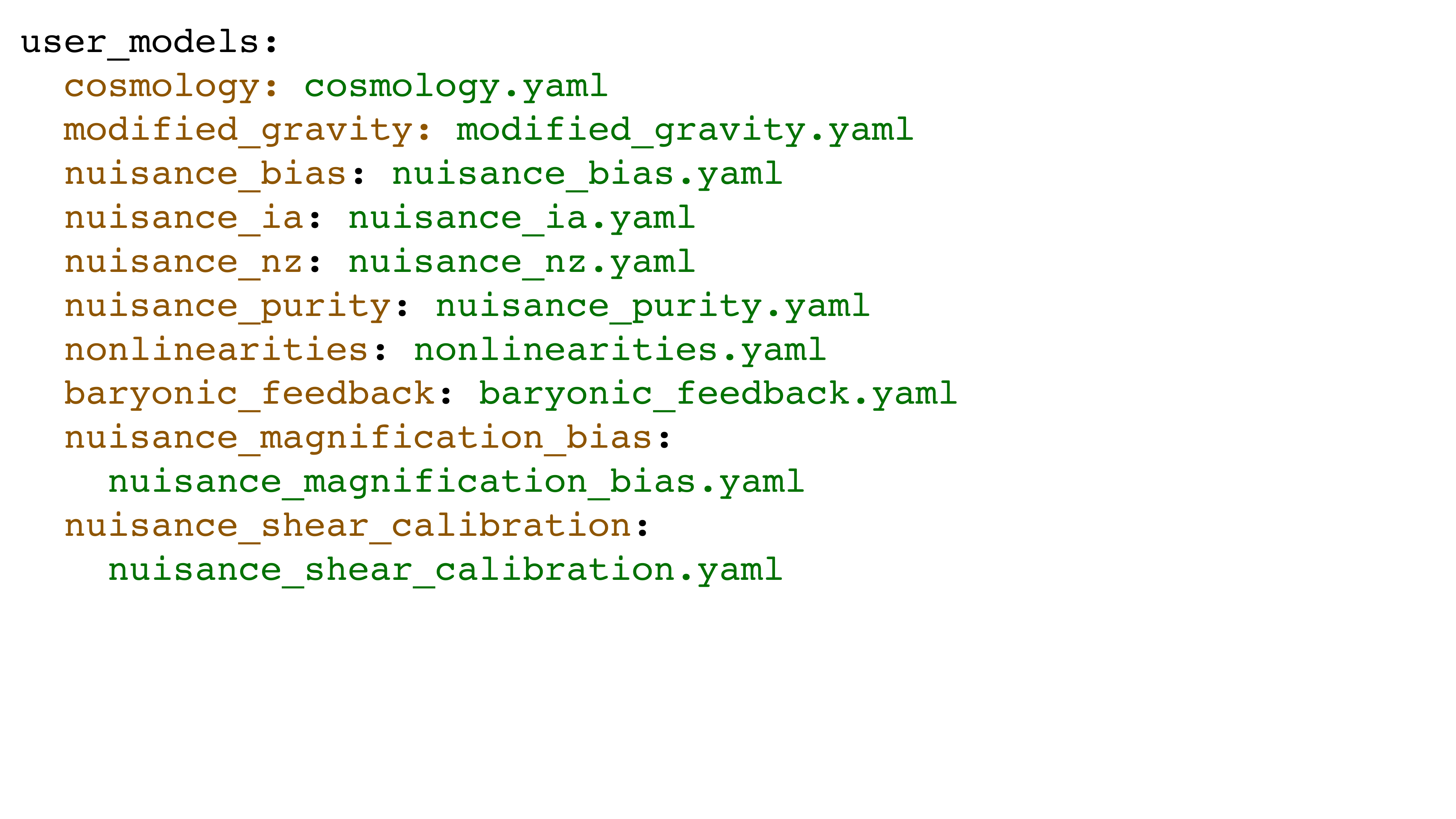}
\end{figure}
\\ \indent 
We next proceed to describe these files, which contain the cosmological and nuisance parameter priors.

\subsubsection{Bottom layer}

In the third configuration layer located within the directory \texttt{configs/models}, the file \texttt{cosmology.yaml} contains the specifications of the cosmological parameters and their priors. The file \texttt{modified\_gravity.yaml} allows the user to modify the prior on the modified gravity growth index, $\gamma_{\rm MG}$, and can be naturally expanded to include the deviations in other parameters such as the effective gravitational couplings to matter and light via the $G_{\rm matter}$ and $G_{\rm light}$ parameters \citep{JZ08, Amendola08}. 

Meanwhile, \texttt{nuisance\_bias.yaml} contains the linear photometric and spectroscopic galaxy bias parameters,  \texttt{nuisance\_ia.yaml} contains the IA parameters, and \texttt{nuisance\_nz.yaml} contains the parameters governing the photometric and spectroscopic redshift uncertainties. In the case of the photometric redshift uncertainties, this corresponds to the $\delta z_{\rm source, i}$ and $\delta z_{\rm lens, j}$ shift parameters of the mean redshift distributions of the source and lens galaxies in tomographic bins $i$ and $j$, respectively.

In addition, \texttt{nuisance\_purity.yaml} contains the parameters $f_{\rm out}$ that quantify the fraction of random outliers in each spectroscopic redshift bin (Eq.~100 in \citealt{Cardone24}). The file \texttt{nonlinearities.yaml} contains the nonlinear photometric and spectroscopic galaxy bias parameters, and \texttt{baryonic\_feedback.yaml} contains the baryonic feedback nuisance parameters associated with the \hmcode, \bacco, and \bcemu prescriptions in \cloe. Lastly, \texttt{nuisance\_magnification\_bias.yaml} and \texttt{nuisance\_shear\_calibration.yaml} contain the magnification bias and multiplicative shear calibration bias parameters for each tomographic bin, respectively. 

\subsubsection{Adding new models and parameters}

As new models of the underlying cosmology and systematic uncertainties are developed, the corresponding parameters can be incorporated in a straightforward way. A new parameter is first defined in the constructor of \texttt{cloe/cosmo/cosmology.py}, and then specified in the configuration files of the chosen backend. 

For the \cobaya backend, this entails adding the parameter to \texttt{configs/params.yaml} and to the relevant file under \texttt{configs/models/<name-of-model>.yaml}, which is then referred to in \texttt{configs/model\_default.yaml}. We note that the default model configuration file also provides the option to overwrite the file \texttt{params.yaml}, which contains all of the \cloe-specific parameters required by \cobaya. For \cosmosis, the corresponding parameters are instead added to \texttt{cosmosis/cosmosis\_values.ini}, as further discussed in Sect.~\ref{sec3ccosmosis}. 

In the case of an entirely new type of model, users may also introduce a control key that determines whether it is considered in the analysis (e.g., analogous to \texttt{IA\_flag} or \texttt{bias\_model}), along with extended unit tests. This architecture enables the seamless integration of new models without modifying the core likelihood or sampling infrastructure.

\subsection{Interfaces to \cobaya and \cosmosis}
\label{sec3c}

This subsection describes the independent interfaces of \cloe with \cobaya and \cosmosis, which enable the user to utilize the full range of samplers and likelihoods in either of these two platforms. This includes samplers such as Metropolis--Hastings, \polychord, and \nautilus, and the likelihoods of surveys such as \Euclid (currently with synthetic data; \citealt{Mellier24}), \Planck \citep{planck18cosmo, planck18like}, the Atacama Cosmology Telescope (ACT; \citealt{actdr6}), the Kilo Degree Survey (KiDS; \citealt{wright25}), the Dark Energy Survey (DES; \citealt{Desy3}), and the Dark Energy Spectroscopic Instrument (DESI; \citealt{desi2024}). The user can choose to run \cloe with either platform by modifying the \texttt{backend} key in the default configuration file  
(\texttt{config\_default.yaml}) to either \texttt{Cobaya} or \texttt{Cosmosis}. The sampling-related options (e.g., maximum number of steps, convergence criterion, and number of live points in the case of nested samplers) are set in this same file for \cobaya, and in \texttt{run\_cosmosis.ini} for \cosmosis. We next describe the \cloe interfaces to these two sampling platforms.

\begin{figure*}[h]
\centering
\includegraphics[width=0.75\linewidth, trim = {14cm 1.6cm 14cm 1.35cm}]{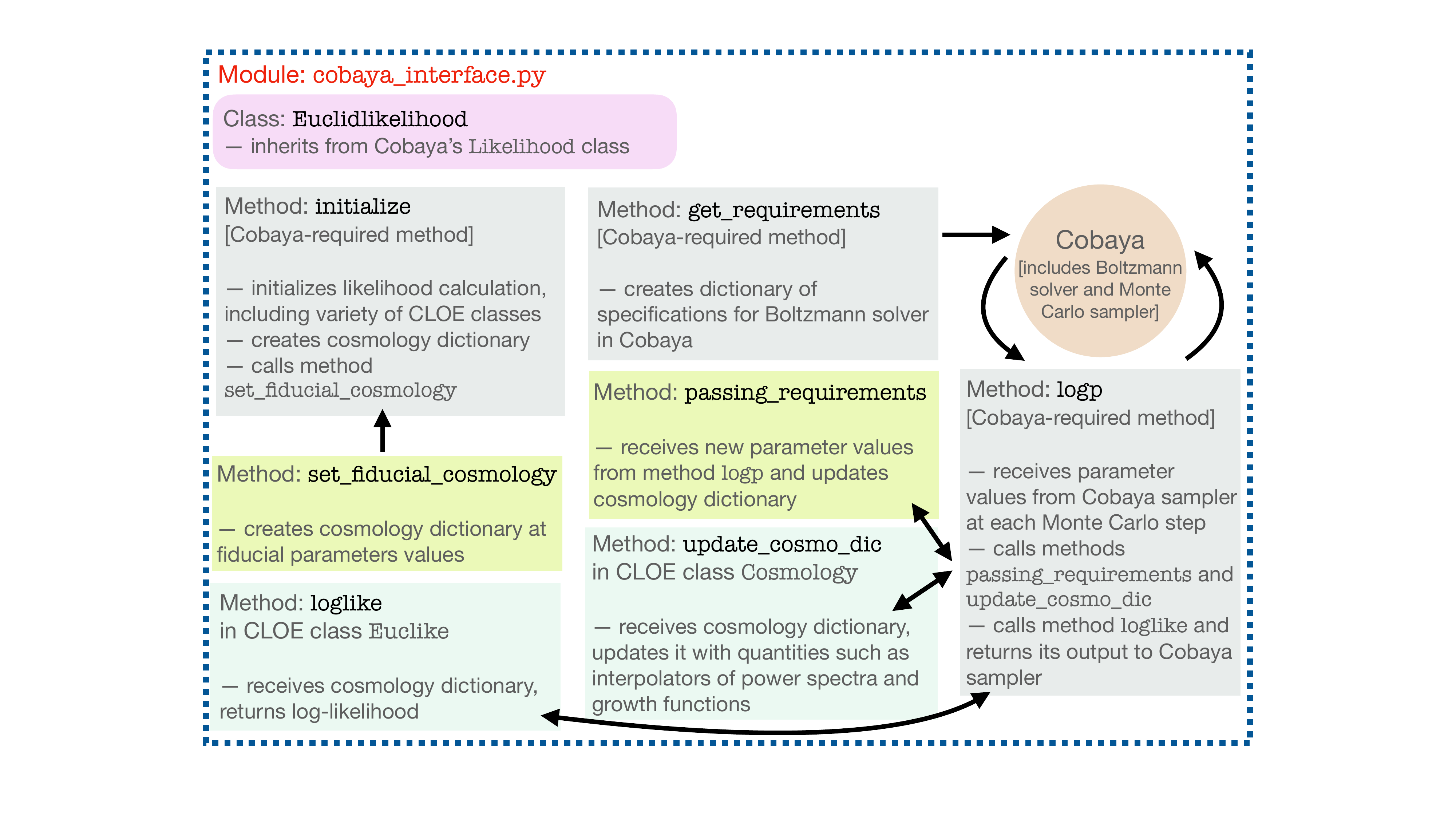}
\vspace{-0.5em}
\caption{Illustration of the structure of the interface between \cloe and \cobaya, encoded in \texttt{cobaya\_interface.py}. The color coding takes on the following form: red denotes the \cloe module, pink denotes the \cloe class, gray denotes the \cobaya-required methods of the class, green denotes additional methods of the class, cyan denotes methods existing in other \cloe classes, and brown denotes \cobaya. The arrows represent the flow of information passed, for instance, from one method to another.}
\label{cloecobayafig}
\end{figure*}

\subsubsection{\cobaya interface}
\label{sec3ccobaya}

The interface between \cloe and \cobaya is defined in \texttt{cloe/cobaya\_interface.py}. As illustrated in Fig.~\ref{cloecobayafig}, this module contains the \texttt{EuclidLikelihood} class, which inherits from the general \texttt{Likelihood} class  of \cobaya. The overall objective of  \texttt{EuclidLikelihood} is to provide the log-likelihood from \cloe to \cobaya as the parameter values are updated in the Monte Carlo sampling. In connection to this interface, we note that \cobaya requires the existence of \texttt{EuclidLikelihood.yaml}, which contains the default user settings and informs \cobaya of the attributes to expect for the \texttt{EuclidLikelihood} class. We note that only the keys and parameters defined here, along with those known to \camb and \class, can be used in a given run. When a given entry is also found in \texttt{config\_default.yaml}, it takes precedence in that it is used in the execution of the code instead.

The \texttt{cobaya\_interface} module imports the \cloe classes of \texttt{Cosmology} and \texttt{Euclike} from the single-class modules \texttt{cosmology} and \texttt{euclike}, respectively. The former stores all of the parameters and user flags in a dictionary (\texttt{cosmo\_dic}) and includes methods for the computation of basic cosmological quantities, such as the growth factor and interpolated distances, while the latter includes the computation of the full multi-probe \Euclid likelihood. 

The \cobaya interface further imports all of the methods in \cloe's \texttt{observables\_dealer} module, which reads in the user selections for the redshift and scale cuts into a single dictionary. It also imports the \texttt{camb\_to\_classy} dictionary in \cloe's \texttt{params\_converter} module, which allows for a translation between the parameter naming conventions of \camb and \class. Lastly, it imports \cobaya's \texttt{get\_model} wrapper to obtain quantities such as distances, growth rates, and linear matter power spectra at the fiducial cosmology. This is for instance needed to compute the Alcock--Paczy\'nski (AP; \citealt{AP79}) distortions for the spectroscopic galaxy clustering.

In the \texttt{initialize} method of the \texttt{EuclidLikelihood} class, the density and range of redshifts and wavenumbers are defined for the calculation of the theoretical predictions. Given the information that the user specifies in the \cloe configuration file (\texttt{config\_default.yaml}), the code takes the user preferences for which probes to consider (for example, the spectroscopic clustering), the specific summary statistics to assign to the probes (for example, Fourier or configuration space), and the specific choices of redshifts and scales for each of the chosen probes. The code feeds these specifications into a dedicated method in the module \texttt{observables\_dealer}, which creates a single dictionary that is returned to the interface module where it is named \texttt{observables}.

At this stage, the \texttt{initialize} method creates an instance of the \texttt{Euclike} class named \texttt{likefinal}. The initialization of this class takes the \texttt{data} and \texttt{observables} dictionaries as arguments, where the former dictionary contains the specifications for the loading and handling of the data in \cloe. Next, it uses the \texttt{camb\_to\_classy} dictionary to set the naming convention for the cosmological parameters that enter the selected Boltzmann solver (\camb or \class). The \texttt{initialize} method then creates an instance of the \texttt{Cosmology} class named \texttt{cosmo}. The initialization of this class includes the creation of \texttt{cosmo\_dic} and further initialization of the \texttt{Nonlinear} class with \texttt{cosmo\_dic} as its argument. While the \texttt{observables} dictionary has already been passed to \texttt{Euclike}, the information on the selection of observables is also passed to \texttt{cosmo\_dic}, as it is needed to optimize the initialization of the \texttt{Photo} class. Lastly, the \texttt{initialize} method sets the fiducial cosmology by calling the internal method \texttt{set\_fiducial\_cosmology}, at which stage it adds the fiducial Hubble function, comoving distance, and angular diameter distance to \texttt{cosmo\_dic}.\footnote{We note that \texttt{initialize} moreover has a dedicated call to the method \texttt{get\_masked\_data} within \texttt{Euclike}, which computes the BNT transformation matrix and masks the measurements and covariance (as further discussed in Sects.~\ref{sec3h}, \ref{sec3i}, and \ref{bntsec}). This solution is subject to further optimization.} As noted earlier, these fiducial quantities are for instance needed to compute the AP effect.

Aside from \texttt{initialize}, the interface contains the methods \texttt{set\_fiducial\_cosmology}, \texttt{get\_requirements}, \texttt{passing\_requirements}, and \texttt{logp}. The first of these methods creates an instance of the \texttt{Cosmology} class named \texttt{fiducial\_cosmology} and then updates the corresponding \texttt{cosmo\_dic} to the fiducial cosmological parameter values. Subsequently, it creates an \texttt{info\_fiducial} dictionary, containing the fiducial parameter values from \texttt{cosmo\_dic} along with other specifications for the chosen Boltzmann solver, which it passes to \texttt{get\_model} to create an instance of \cobaya's \texttt{Model} class. Given a set of scales, $k$, and redshifts, $z$, this method returns quantities such as the fiducial expansion rate, comoving distance, growth rate, and linear power spectrum  of the matter and Weyl potential from \camb/\class. Here, the Weyl potential refers to the sum of the two metric potentials, $\Psi$ and $\Phi$. As a result, the Weyl power spectrum is given by $P_{\rm Weyl}(k, z) = P_\frac{\Psi+\Phi}{2}(k, z)$, to which the weak lensing observables are directly sensitive. 

The second method, \texttt{get\_requirements}, is specific to \cobaya and creates a dictionary of specifications for the chosen Boltzmann solver. Meanwhile, the method \texttt{passing\_requirements} takes the cosmological and nuisance parameter values from \cobaya at each step of a Monte Carlo run, obtains derived quantities such as the growth rate and linear matter power spectrum from \camb/\class, and passes the parameters and derived quantities to \texttt{cosmo\_dic}.\footnote{We note that \texttt{passing\_requirements} first attempts to execute a \python \texttt{try} block of code that utilizes the user configurations it obtains from \texttt{config\_default.yaml} (or similarly the MCMC scripts discussed in Sect.~\ref{sec3m}). It otherwise turns to an \texttt{except} block of code, which it executes by utilizing a received \texttt{info} dictionary of user configurations and \cobaya \texttt{model} wrapper (see Appendix~\ref{appnote}).} Lastly, \texttt{logp} is a \cobaya-specific method that takes the cosmological parameters from \cobaya and passes them to \texttt{passing\_requirements} as it calls this method. The \texttt{logp} method then calls the \texttt{update\_cosmo\_dic} method of \texttt{cosmo}, which takes as input the given set of redshifts and single wavenumber as it updates a wide range of interpolators of quantities such as distances, growth rates, and matter and Weyl power spectra. This method also updates the nonlinear module, by calling the \texttt{update\_dic} method of the nonlinear instance. 

\begin{figure*}[h]
\centering
\includegraphics[width=0.75\linewidth, trim = {9.9cm 1.6cm 15cm 1.35cm}]
{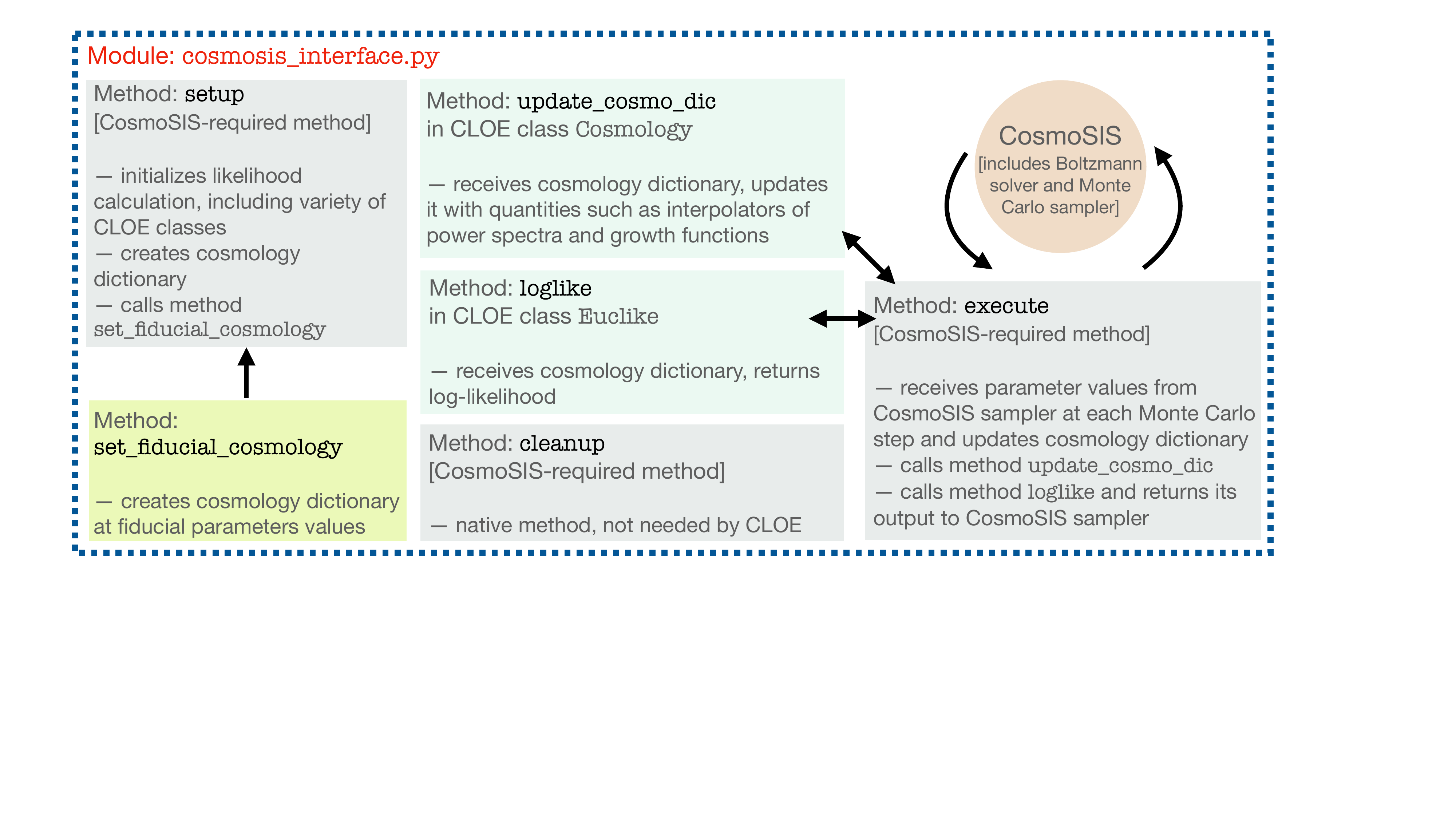}
\vspace{-9em}
\caption{Illustration of the structure of the interface between \cloe and \cosmosis, encoded in \texttt{cosmosis\_interface.py}. The color coding and arrows follow that of Fig.~\ref{cloecobayafig}.}
\label{cloecosmosisfig}
\end{figure*}

The \texttt{logp} method then computes the number of sampled parameters, which is needed when the covariance matrix is estimated from numerical simulations. In this case, the unknown true covariance matrix needs to be marginalized over, leading to a likelihood function that is a multivariate $t$-distribution. Considering a frequency-matching prior for the covariance matrix, the marginalized likelihood function is given by \citet{Percival22}. The marginalized likelihood function for an independence Jeffreys prior for the covariance matrix is further obtained in the limit of a large number of simulations \citep{SH16, Percival22}. Note that these distributions still assume that the data have an underlying Gaussian sampling distribution, and thus do not account for any intrinsic non-Gaussianity. Lastly, the natural logarithm of the likelihood is computed by calling the \texttt{loglike} method of \texttt{likefinal}, which requires \texttt{cosmo\_dic} and the number of sampled parameters as its arguments. This external likelihood is then returned to the sampler within \cobaya.

\subsubsection{\cosmosis interface}
\label{sec3ccosmosis}
An independent interface between \cloe and \cosmosis has also been implemented. As illustrated in Fig.~\ref{cloecosmosisfig}, this gives the user the flexibility to choose \cosmosis instead of \cobaya for the parameter sampling, while still relying on \cloe to produce the theoretical predictions and return the value of the log-likelihood. Using \cosmosis allows the user to employ samplers that are not implemented within \cobaya by default, such as \emcee\footnote{\emceeaddress}\href{https://github.com/dfm/emcee}{\color{black}\faGithub} \citep{fm13}, \multinest\footnote{\multinestaddress}\href{https://github.com/farhanferoz/MultiNest}{\color{black}\faGithub} \citep{feroz08, feroz09, feroz19},
\zeus\footnote{\zeusaddress}\href{https://github.com/minaskar/zeus}{\color{black}\faGithub} \citep{karamanis21}, \pocomc\footnote{\pocomcaddress}\href{https://github.com/minaskar/pocomc}{\color{black}\faGithub} \citep{karamanis22}, and \nautilus. We note that similar to \cobaya, there is also the possibility to use \cosmosis with the \polychord and Metropolis--Hastings samplers.

There are currently two approaches for interfacing \cloe with \cosmosis depending on the user preferences. In the main approach, background quantities in the form of distances and growth rates, along with the matter power spectrum, are obtained from a standalone Boltzmann code that is connected to \cosmosis through its DataBlock structure. These quantities are used by \cloe to compute the theoretical predictions and subsequently the log-likelihood, which is passed to the sampler in \cosmosis. In the second approach, the Boltzmann calculation, theoretical predictions, and  log-likelihood are all executed through the \cloe $\times$ \cobaya interface, while \cosmosis performs the sampling of the parameter space.\footnote{We note that the Boltzmann code in the first approach is currently restricted to \camb, while both \camb and \class are allowed in the second approach given the intrinsic capabilities of \cobaya. In principle, the first approach could also support \class and emulators of the linear matter power spectrum.} We note that the first approach is suitable for parameter inference purposes, and the primary motivation for the second approach are additional consistency tests.\footnote{The second approach to the \cosmosis interface is suitable for purposes of added consistency tests, but not parameter inference because it requires the reinitialization of the code at each point in the parameter space during the sampling process.}

In both approaches, the user interacts with the \cosmosis interface, which is specified in the \texttt{cosmosis} directory of \cloe. This includes specifying a parameter file, a values file, and a priors file in the \texttt{ini} format adopted by \cosmosis (encapsulating the parameters to be sampled and their priors). The interfacing code is \texttt{cosmosis\_interface.py} in the first approach and \texttt{cosmosis\_with\_cobaya\_interface.py} in the second approach. In both of these cases, we follow the standard \cosmosis structure in defining \texttt{setup}, \texttt{execute}, and \texttt{cleanup} methods, which respectively initialize the likelihood, compute the likelihood at each point in parameter space, and deallocate memory at the end of the parameter inference. We next highlight the structure of both approaches and the relevant files to consider in each case.
 
In the first approach, the default parameter file is given by \texttt{run\_cosmosis.ini}. It requires two modules: the Boltzmann module, given by \texttt{[camb]} in the current implementation, and the \texttt{[euclid]} module that prompts \cloe to calculate the log-likelihood. The \texttt{[camb]} module is encoded in \texttt{camb\_interface.py}, which calculates the matter power spectrum and background quantities, such as the angular diameter distance, $D_{\rm A}(z)$, linear growth rate, $f(z)$, and root-mean-square of the linear matter overdensity field on $8 \ h^{-1} \ {\rm Mpc}$ scales, $\sigma_8(z)$, according to the user specifications in \texttt{config\_for\_cosmosis.yaml}.\footnote{We note that 
\texttt{camb\_interface.py} has been adapted from the original file in the \cosmosis Standard Library.} These quantities are stored in the \cosmosis DataBlock. In the \texttt{[euclid]} module, encoded in \texttt{cosmosis\_interface.py},\footnote{We note that \texttt{cosmosis\_interface.py} is our \cosmosis equivalent to \texttt{cobaya\_interface.py} described in Sect.~\ref{sec3ccobaya}.} \cloe retrieves these and other quantities from the DataBlock, along with further user specifications for \cloe that are given in  \texttt{config\_for\_cosmosis.yaml}. It inputs all of the quantities and user specifications into an instantiated cosmology dictionary (\texttt{cosmo\_dic}). In calculating the log-likelihood, it further creates an instance of the \texttt{Euclike} class and calls the \texttt{loglike} method with \texttt{cosmo\_dic} as the input. The resulting value of the log-likelihood is  then stored in the \cosmosis DataBlock, which is a process that repeats for each step of the parameter sampling. 

In this first approach, we can broadly view the \texttt{setup} method in \texttt{cosmosis\_interface.py} as  the equivalent to the \texttt{initialize} method in \texttt{cobaya\_interface.py} (described in Sect.~\ref{sec3ccobaya}). Both of the \cosmosis and \cobaya interfaces contain the method \texttt{set\_fiducial\_cosmology}, while the method \texttt{execute} in the \cosmosis interface can be considered as the equivalent to the union of the \texttt{logp} and \texttt{passing\_requirements} methods in the \cobaya interface. In contrast to the \cobaya interface, which provides this natively, the interpolation of the matter power spectrum is added to this method as well.

In the second approach, the \cloe $\times$ \cobaya pipeline performs every aspect of the likelihood calculation as previously described. Hence, the \cosmosis parameter \texttt{ini} file only requires the \texttt{[euclid]} module as an input. The file specified within this module, \texttt{cosmosis\_with\_cobaya\_interface.py}, establishes the interface with \cosmosis by first reading the \texttt{cobaya\_config\_for\_cosmosis.yaml} config file in the form of a \cobaya \texttt{info} dictionary. It retrieves the parameters currently being sampled from the \cosmosis DataBlock and updates the \texttt{info} dictionary with their values. Given this dictionary as input, a \cobaya \texttt{model} instance is created with the method \texttt{get\_model}, and the log-likelihood value is then retrieved with the method \texttt{model.logposterior}. The log-likelihood is subsequently stored in the \cosmosis DataBlock and the process is repeated, as with the first approach. Here, there is no explicit call to the Boltzmann code, as the calculations of the cosmological quantities are performed under the hood when the \texttt{model} instance is created. This approach effectively overlays a \cosmosis wrapper over the \cloe $\times$ \cobaya pipeline. 

\subsection{Matter power spectrum and background quantities: \camb, \class, and beyond}
\label{sec3d}

\cloe obtains the linear matter and Weyl power spectra, along with cosmological background quantities such as comoving distances and growth rates, from either of the two Boltzmann codes: \camb and \class. In contrast to \ccl, which has a direct interface to these codes, \cloe is connected to them through either the \cobaya or \cosmosis platforms.\footnote{However, note that \firecrown + \ccl by default utilizes the Boltzmann solver of the sampling platform as well.} As a result, depending on whether \cobaya or \cosmosis is used, \cloe's interactions with the chosen Boltzmann code occurs in either \texttt{cobaya\_interface.py} or \texttt{cosmosis\_interface.py} (as discussed in Sects.~\ref{sec3ccobaya} and \ref{sec3ccosmosis}). 
In either of these interfaces, it stores the outputs of the selected Boltzmann code in its cosmology dictionary (\texttt{cosmo\_dic}).

The user choice between \camb and \class is set through the \texttt{solver} key in \texttt{configs/config\_default.yaml}. Their accuracy settings can be modified via the nested key \texttt{extra\_args} that takes as optional input the respective accuracy parameter names and values of each code. By default, the standard accuracy settings of \camb and \class are used. We find that the theoretical predictions from \cloe when using either Boltzmann code are in strong agreement, with the 3$\times$2pt angular power spectra and redshift-space multipole power spectra differing by less than $0.05\%$ and $0.4\%$ on average, respectively. These differences correspond to $0.01$--$0.08\sigma$ in terms of the covariance of the full \Euclid survey, rising to $0.2\sigma$ for the most discrepant synthetic data point. The results are robust to variations in the accuracy settings of \camb and \class, indicating that the residual differences are intrinsic to the respective Boltzmann codes.

\begin{figure*}[h]
\centering
\includegraphics[width=0.60\linewidth, trim = {13.3cm 1.6cm 14cm 1.35cm}]{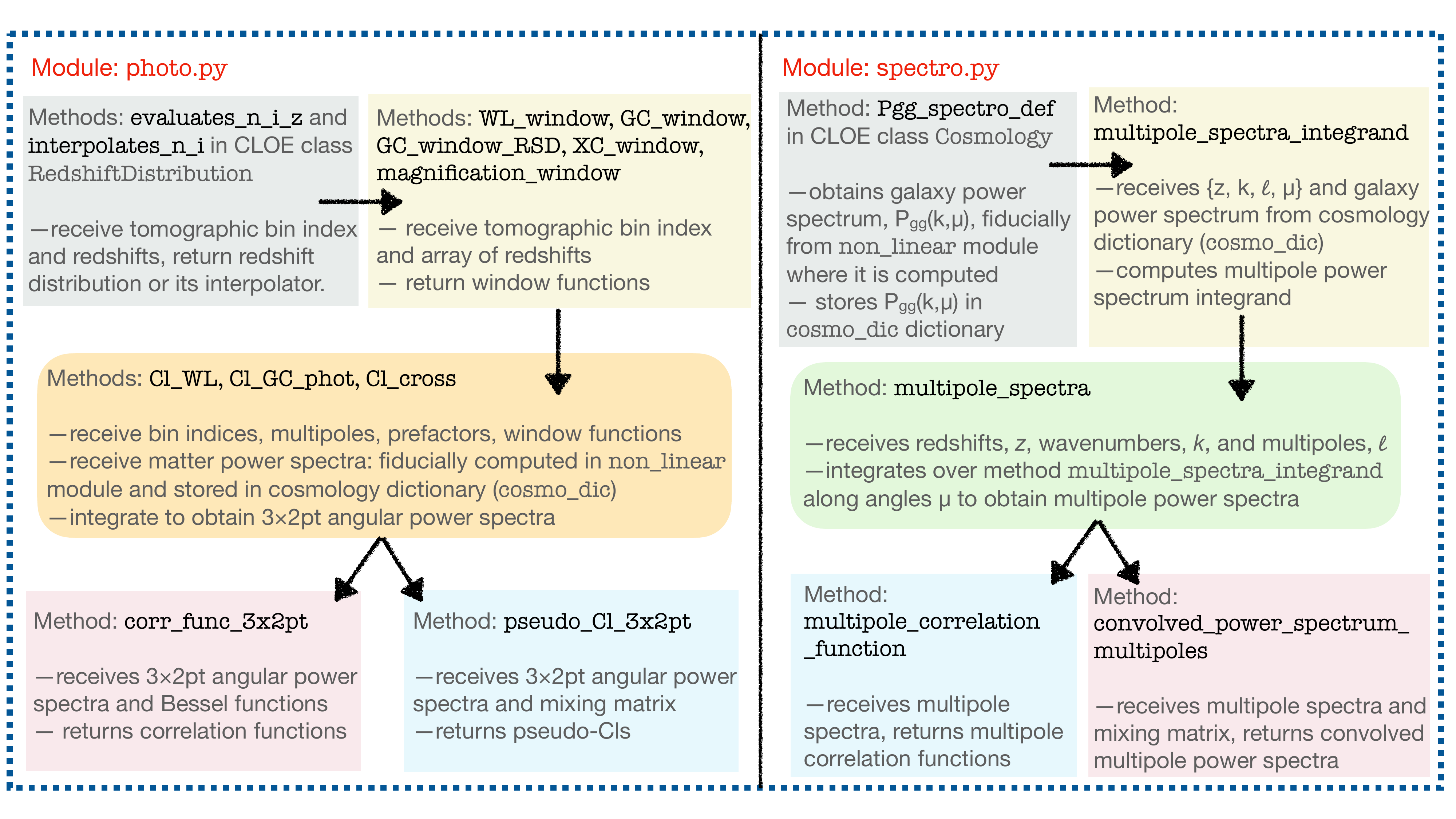}
\vspace{0.4em}
\caption{Illustration of the structure of the photometric and spectroscopic modules in \cloe.}
\label{photspecfig}
\end{figure*}

In the default configuration file, \cloe provides a variety of options for extending the linear matter power spectrum to nonlinear scales through the key \texttt{NL\_flag\_phot\_matter}. This includes \hmcode and \halofit along with the \eecode and \bacco emulators. We note that these nonlinear corrections to the matter power spectrum exclude baryonic feedback. Instead, \cloe includes a separate key, \texttt{NL\_flag\_phot\_baryon}, which allows for baryonic feedback corrections to the matter power spectrum via \hmcode, \bacco, and \bcemu. In the case of \hmcode, we note that this includes both the original \citep{Mead15} and \hmcodetwenty \citep{Mead21} versions.\footnote{We note the caveat that \hmcodetwenty is currently only available with \camb as the backend.} For further details on the implementation of nonlinear corrections to the matter power spectrum in \cloe, we refer to Euclid Collaboration:~Carrilho et al. (in prep.).

\cloe self-consistently considers the Weyl power spectrum in the computation of the theory predictions when modified gravity, which modifies the relation between the metric potentials, is explored.\footnote{We emphasize that weak lensing is sensitive to the sum of the metric potentials (i.e.,~to $\Xi^{\rm Weyl} \equiv {(\Psi + \Phi)}/{2}$), which are identical in GR in the absence of anisotropic stress. In GR, the Poisson equation moreover relates the metric potentials to the matter density contrast, $\delta$. This in turn implies a direct relationship between the Weyl and matter power spectra (i.e.,~$P_{\rm Weyl} \propto P_{\delta}$) that allows the weak lensing theory to be expressed in terms of the matter. However, this feature no longer holds in modified gravity models, which motivates the use of the Weyl power spectrum in constructing the weak lensing theory predictions (as further discussed in e.g.~\citealt{ps16} and \citealt{joudaki22}).} To this end, we assume that the nonlinear corrections to the Weyl power spectrum are identical to those for the matter power spectrum (i.e.,~$P_{\rm Weyl}^{\rm nonlinear}/P_{\rm Weyl}^{\rm linear} = P_{\rm matter}^{\rm nonlinear}/P_{\rm matter}^{\rm linear})$. For further details on the implementation of the Weyl power spectrum feature in \cloe, we refer the reader to \citet{Goh24}.

We note that modified versions of \camb and \class that come with a \python wrapper (e.g., \hiclass, \isitgr, \eftcamb, \mgcamb) are straightforwardly included in \cloe. This particularly holds when considering \cobaya as the backend, as the additional parameters of a given extended cosmology are automatically inherited, such that the user only needs to specify their priors in \texttt{configs/models/cosmology.yaml}. This comes with the caveat that modified Boltzmann solvers that require additional input arguments need to have this specified when called in \cloe, and that the modeling of the nonlinear corrections in particular needs to be updated for the extended cosmologies as well (for a self-consistent example of the latter, see e.g.~\citealt{joudaki22}).\footnote{We moreover advise \cloe users to ensure that the extended model is adequately accounted for in the theoretical prescription of the observables, which can include expressing the equations in terms of the Weyl power spectrum by enabling the \texttt{use\_Weyl} key in 
the default configuration file.
}

\subsection{Photometric module}
\label{sec3e}

As shown in Fig.~\ref{photspecfig}, the primary probes of \cloe are divided into distinct photometric and spectroscopic modules (the latter described in Sect.~\ref{sec3f}). This split is motivated by the negligible cross-covariance and cross-correlations between the photometric and spectroscopic probes in \Euclid (e.g.,~\citealt{Mellier24, TM22, paganin24}). The photometric module file \texttt{photo.py} is located in \texttt{cloe/photometric\_survey} and contains the class \texttt{Photo}. It performs the theory computation of the  photometric 3$\times$2pt observables, which include the angular power spectra and correlation functions for cosmic shear, galaxy-galaxy lensing, and galaxy clustering. The \texttt{photometric\_survey} package further includes the module file \texttt{redshift\_distribution.py}. This module includes the class \texttt{RedshiftDistribution} that describes and allows for the mitigation of uncertainties in the photometric redshift distributions.

\subsubsection{Overall structure}
\label{sec3e1}

The \texttt{Photo} class performs the calculation of the theory predictions for the photometric 3$\times$2pt observables, both in Fourier and configuration space. The constructor of the class takes as input three dictionaries: the cosmology dictionary (\texttt{cosmo\_dic}) and dictionaries containing the redshift distributions of the photometric source and lens samples (\texttt{nz\_dic\_WL} and \texttt{nz\_dic\_GC}, respectively). It also takes as input a key, \texttt{add\_RSD}, which specifies whether to include the impact of redshift space distortions in the photometric galaxy clustering and galaxy-galaxy lensing.\footnote{We refer the reader to \citet{Tanidis24} for an exploration of their impact on the \Euclid constraints.} 

The core operations of the class are performed by the methods \texttt{Cl\_WL}, \texttt{Cl\_GC\_phot}, and \texttt{Cl\_cross}, which respectively compute the angular power spectra of the cosmic shear, galaxy clustering, and galaxy-galaxy lensing corresponding to Eqs.~(32), (40), and (54) in \citet{Cardone24}.\footnote{We note that a natural generalization would consist in merging these methods into a single method.} These methods have certain features in common, in that they all adhere to the following six-component structure:
\begin{enumerate}
    \item{Take as input the tomographic bin indices $i$ and $j$ (for either the source or lens galaxy distributions depending on the probe) along with the array of multipoles, $\ell$, for which to compute the theoretical predictions.}
    \item{Compute the multipole-dependent prefactors due to the extended Limber approximation and spherical-sky generalization \citep{Limber54, LA2008, Hu2000, Kilbinger17}. These prefactors are given by Eqs.~(36) and (44) in \citet{Cardone24} for tracers of cosmic shear and galaxy clustering, respectively. This operation is performed once via the \texttt{set\_prefactor} method and the output is then stored during a Monte Carlo run. 
    }
    \item{Create instances of the \texttt{RedshiftDistribution} class, passing the dictionary of nuisance parameters, such as the $\delta z_i$ shift parameters, and the dictionary of either the lens or source galaxy redshift distributions in the call to the class constructor. The most important methods of this class are \texttt{interpolates\_n\_i} that returns an interpolator of the systematics-shifted redshift distributions and \texttt{evaluates\_n\_i\_z} that evaluates the shifted distributions at a given redshift. A natural extension would be the mitigation of uncertainties in the width of the redshift distributions and nonzero rate of catastrophic outliers \citep{Bernstein09, RM22, Mill25}. Additional strategies to mitigate the photometric redshift uncertainties include the Gaussian mixture model \citep{Stolzner21}, \textsc{Hyperrank} method \citep{Cordero22}, and resampling methods \citep{Hildebrandt17, Zhang23}.
    }
    \item{Interpolate the precomputed window functions for the shear, IA, galaxy density contrast, redshift space distortions, and magnification bias in Eqs.~(37), (39), (46), (47), and (50) of \citet{Cardone24}, respectively, at the grid of lens redshifts used in the integration to obtain the angular power spectra, $C_{ij}(\ell)$. The window functions are precomputed at each point in the Monte Carlo sampling, as a given kernel can be used by multiple angular power spectrum calculations. An example is the IA window function, \texttt{IA\_window}, used to compute the intrinsic-intrinsic and shear-intrinsic components of the cosmic shear, along with the galaxy-intrinsic component of the galaxy-galaxy lensing. The window functions depend on the redshift distributions of either the source or lens galaxies, which are obtained from 
    \texttt{interpolates\_n\_i} and \texttt{evaluates\_n\_i\_z} of the \texttt{RedshiftDistribution} class.
    }
    \item{Evaluate and store the matter, galaxy, and IA auto- and cross-power spectra. This is achieved through the \texttt{evaluate\_power\_WL},~~\texttt{evaluate\_power\_GC\_phot}, and \texttt{evaluate\_power\_XC} methods in the case of cosmic shear, galaxy clustering, and galaxy-galaxy lensing, respectively. These methods obtain the needed subset of the power spectra at a given set of redshifts and wavenumbers (precisely the ones used to integrate over the lens galaxies) from bivariate spline interpolators of the power spectra in the dictionary \texttt{cosmo\_dic}. This dictionary receives the power spectra from \texttt{cloe/cosmo/cosmology.py}, where they are either generated in the linear regime or obtained from the \texttt{non\_linear} subpackage depending on the values of the keys \texttt{NL\_flag\_phot\_matter}, \texttt{NL\_flag\_phot\_baryon}, \texttt{NL\_flag\_phot\_bias}, and \texttt{IA\_flag}. These are keys that the user of \cloe specifies in \texttt{configs/config\_default.yaml}. 
    }
    \item Put the different components composed of the multipole prefactors, window functions, and power spectra from steps 2, 4, and 5 above together according to Eqs.~(32), (40), and (54) in \citet{Cardone24}. Subsequently, perform the final integration over the lens galaxies via the trapezoidal method to obtain the target angular power spectra. At this stage, we multiply the angular power spectra with the tomographic-bin dependent multiplicative shear calibration bias in the case of cosmic shear and galaxy-galaxy lensing (according to Eq.~124 in \citealt{Cardone24}). 
\end{enumerate}
The angular power spectra are used to obtain the pseudo-$C_\ell$, which capture the mode-mixing effect due to masking (see e.g. \citealt{Loureiro22, Upham22}) in the method \texttt{pseudo\_Cl\_3x2pt} following Sect. 3.2 of \citet{Cardone24}. In this regard, pixelization effects arising from finite map resolution remain to be addressed. 

The angular power spectra are also used to obtain the correlation functions in the method \texttt{corr\_func\_3x2pt}. This method follows Eqs.~(71), (74), and (75) of \citet{Cardone24} in transforming $C_{ij}(\ell)$ to $\xi_{ij}(\theta)$. It takes as input the range of angular scales, $\theta$, the tomographic bin indices $i$ and $j$, and a string that specifies the type of correlation function to consider (i.e.,~either $\xi^{\rm GG}$, $\xi^{\rm GL}$, $\xi^+$, or $\xi^-$, representing the galaxy clustering, galaxy-galaxy lensing, and two shear correlations, respectively). For the chosen probe, the method computes the angular power spectrum at a sparse number of multipoles (fiducially chosen to be 128 multipoles) and then performs a cubic spline interpolation to obtain the angular power spectrum at a dense number of multipoles (fiducially chosen to be 100\,000 multipoles). This interpolated angular power spectrum is subsequently used together with the precomputed Bessel functions of the first kind (given by the method \texttt{set\_bessel\_tables} at orders 0, 2, and 4) in a Riemann sum to obtain the correlation functions. 

We note that the integration to obtain the angular power spectrum currently utilizes the extended Limber approximation \citep{Limber54, LA2008}, rather than the full non-Limber calculation (recent examples include \citealt{Campagne2017, Assassi2017, Schoneberg2018, Fang2019, Leonard2023, Feldbrugge2023, Chiarenza2024, Reischke2025, Reymond2025}). This primarily affects the photometric galaxy clustering and galaxy-galaxy lensing on large angular scales (see \citealt{Kilbinger17} for a demonstration of its negligibility for cosmic shear). However, as described in Sect.~\ref{fftlogsec}, \cloe includes an implementation of the \fftlog algorithm that efficiently computes oscillatory integrals involving spherical Bessel functions, including those that arise in non-Limber calculations.

We further note that our integrals over the angular power spectra to obtain the correlation functions currently assume a flat sky and effective scale for each angular bin \citep{Stebbins96, Ng99, Chon04, Kilbinger17, Kitching17, Troxel18, Asgari19, Krause22}. A natural generalization of the integral would therefore be to perform the spherical transform and bin-averaging of angular scales following the prescriptions in earlier work, such as \citet{Kilbinger17} and \cite{Krause22}, as further described in \citet{Cardone24}.

As the Bessel functions only depend on the product of the multipoles and angles, $\ell\theta$, it is sufficient to compute them once in a Monte Carlo run. In \cloe, the precomputation is performed and stored as part of the initialization of the \texttt{Euclike} class in \texttt{cloe/like\_calc/euclike.py}. We note that \cloe also has the ability to generate the angular power spectra and correlation functions outside of the Monte Carlo context, for instance within \jupyternotebooks, where quantities such as the Bessel functions and multipole prefactors are computed only when the methods requiring them are called.

\begin{figure*}[h]
\centering
\sidecaption
\includegraphics[width=0.6012\linewidth, trim = {0cm 3cm 16cm 0cm}]{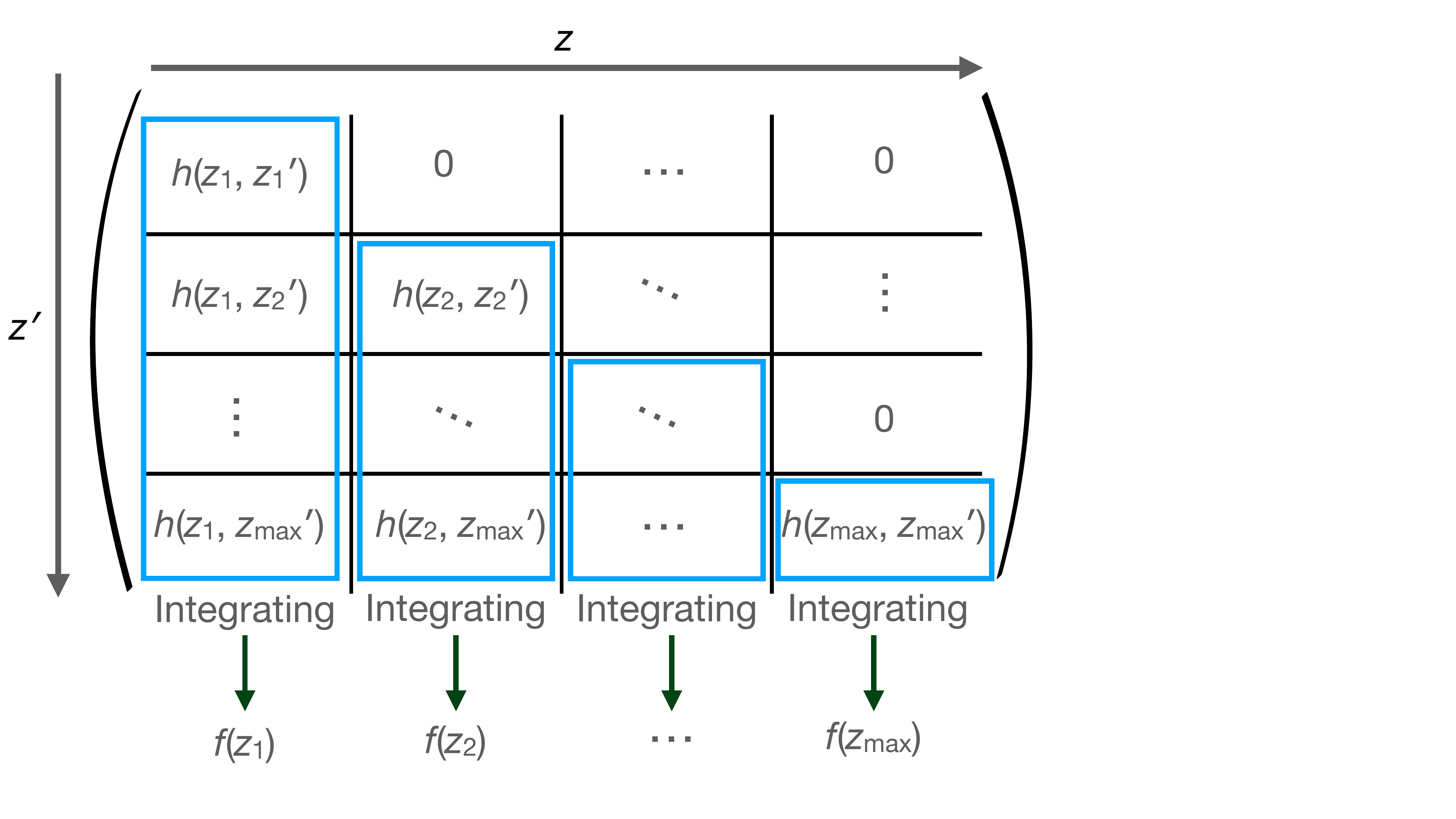}
\caption{Illustration of the integration strategy for the cosmic shear and lensing magnification window functions, generically expressed here as $f(z)$, which enter the corresponding angular power spectra and correlation functions in \cloe. The sizes of the $z$ and $z'$ arrays can be different, as long as $z \in z'$. In practice, we have chosen $z = z'$ for the implementation in \cloe.}
\label{integmat}
\vspace{0.7em}
\end{figure*}

In \citet{Martinelli24}, we benchmark the angular power spectra and correlation functions from the photometric module to a precision of 10\% of the \Euclid error bars for a range of points in the space of cosmological and systematics parameters. This is comparable to the requirement enforced for other Stage IV survey codes such as \ccl \citep{Chisari19}. In achieving a high numerical precision, the values of variables that govern 
it are fixed within the initialization of the \texttt{Photo} class. This particularly includes the minimum redshift used in the integration over the lens galaxies (fiducially set to $z_{\rm min} = 10^{-3}$, as the window functions and thereby lensing signal vanish at $z=0$) and the maximum multipole used in the integration to obtain the correlation functions (fiducially, $\ell_{\rm max} = 10^5$). We note that these variables are not included in the user configuration file (\texttt{configs/config\_default.yaml}), as they should not need to be modified, but can be set directly within \texttt{Photo.py} by advanced users. For further discussion of the impact of accuracy settings, we refer to Sect.~7.4 in \citet{Martinelli24}.

\subsubsection{Parameter space}

In the case of a final \Euclid data release (DR3) scenario, \cloe considers 13 tomographic bins up to a redshift of $z=2.5$ \citep{Pocino21, Mellier24}, which translates into a large number of nuisance parameters. This currently includes one nuisance parameter to capture the shift in the mean of the redshift distribution of each tomographic bin (i.e., a total of 13 parameters for the 13 bins). It further includes one nuisance parameter to capture the multiplicative shear calibration uncertainty of each bin (i.e., 13 parameters in total). For the magnification bias, \cloe either considers a cubic polynomial to capture the redshift dependence (Eq.~120 in \citealt{Cardone24}) or an amplitude shift parameter for each tomographic bin (i.e., either 4 or 13 parameters). For the galaxy bias, \cloe either considers a cubic polynomial (Eq.~121 in \citealt{Cardone24}), an amplitude parameter for each tomographic bin, or a one-loop perturbation theory model that utilizes \fastpt\footnote{\fastptaddress}\href{https://github.com/jablazek/FAST-PT}{\color{black}\faGithub} \citep{McEwen16, Fang17} and introduces four nuisance parameters for each tomographic bin (i.e., either 4, 13, or 52 parameters; in the latter case consistent with the nonlinear approach for spectroscopic clustering, see Euclid Collaboration:~Carrilho et al., in prep. for further details). 

For the intrinsic galaxy alignments, \cloe introduces two nuisance parameters for the NLA model (quantifying the amplitude and redshift dependence; \citealt{HS04, BK07, Joachimi2011}) or it considers a TATT model \citep{Blazek19} that introduces a total six nuisance parameters.
\cloe further requires one parameter to capture the impact of baryonic feedback with \hmcode, but as many as 84 parameters for \bcemu and \bacco. In total, this amounts to between 37 and 183 nuisance parameters to describe the photometric observables for \Euclid in the current version of \cloe. As a result, this is a completely new realm for cosmological parameter inferences, where past and current photometric survey analyses have not required to sample more than $\sim20$ nuisance parameters (e.g.,~\citealt{Desy3}). This puts significant pressure on the computational speed of inference pipelines designed for Stage IV surveys, such as \cloe and \firecrown + \ccl (for further details, see Sect.~\ref{speedsec} and \citealt{GCH24}).

\subsubsection{Integration strategy}
\label{sec3e2}

In the creation of the theoretical predictions for the angular power spectra and correlation functions of the photometric galaxies, we need to pay particular attention to the computational speed. We have therefore devised a particular strategy for the computation of these theoretical quantities, which typically involve integrals of the form
\begin{equation}
    I = \int_0^{z_{\rm max}} \mathrm{d}z \; f(z) \, g(z) \; .
\end{equation}
Here, $z_{\rm max}$ denotes the maximum redshift of the source galaxy sample, which may itself depend on e.g. photometric redshift uncertainties. In some cases, both $f$ and $g$ are direct functions of redshift, $z$, and thus their integrals do not require particular care beyond an appropriate choice of redshift binning and reliance on the vectorization capabilities of \python when computing the integrand. A more complicated case appears, particularly in the case of weak lensing, when either or both $f$ and $g$ are themselves integrals over redshift,
\begin{equation}
    f(z) = \int_{z}^{z_{\rm max}} \mathrm{d}z' \;  h(z, z') \; ,
    \label{eq:integrand}
\end{equation}
where not only the bound of the integral varies with $z$, but it also appears as an argument in the integrand. A naive approach may consist in looping over all possible values of the bound (up to $z_{\rm max}$), each time computing and integrating the integrand over a fixed number of redshift values. However, such an approach would be far from optimal: loops in \python are costly and do not exploit its vectorization capabilities. Moreover, this approach leads to inefficient sampling: an equal amount of computational power would be spent on an increasingly tiny redshift interval, with a too fine sampling when the bound is close to $z_{\rm max}$ and too coarse when it is far from $z_{\rm max}$.

Instead, as illustrated in Fig.~\ref{integmat}, our proposed implementation utilizes a regular square grid in $z$ and $z'$, such that a matrix of the same grid size is filled with the values of the integrand of Eq.~(\ref{eq:integrand}) for each $(z, z')$ combination. The full matrix is computed at once using \python broadcasting and vectorization. The upper triangle part of the matrix is then set to zero, such that the remaining nonzero elements of each column correspond to the integrand of Eq.~(\ref{eq:integrand}) evaluated for a given value of $z$, for a series of $z'$ values between $z$ to $z_{\rm max}$. Finally, we numerically integrate over all columns of the matrix at once, yielding the required array of $f(z)$ integrals. These $f(z)$ are used together with the correspondingly derived $g(z)$ to obtain $I$ via the trapezoidal method in \scipy's \texttt{integrate}\footnote{\url{https://docs.scipy.org/doc/scipy/reference/integrate.html}} subpackage.

The advantages of this approach include a better sampling efficiency (uniform in terms of redshift ``density''), which avoids unnecessary fine sampling for small intervals, and speed improvements due to vectorization. A possible drawback of the approach is an increase in the memory usage which, however, is negligible for the fiducial choice of redshift sampling in \cloe (and reasonable deviations around it).

\subsection{Spectroscopic module}
\label{sec3f}

The spectroscopic module file \texttt{spectro.py} is located in \texttt{cloe/spectroscopic\_survey} and contains the class \texttt{Spectro}. As illustrated in Fig.~\ref{photspecfig}, it performs the theory computation of the spectroscopic observables, which include the  
multipole power spectra and correlation functions, incorporating a range of systematic uncertainties, such as redshift errors, sample impurities, and galaxy biases. The multipoles under consideration are the monopole, quadrupole, and hexadecapole (respectively $\ell=\{0, 2, 4\}$). We next provide an overview of the code structure, size of the parameter space, and integration strategy.

\subsubsection{Overall structure}
\label{sec3f1}

The core operation of the \texttt{Spectro} class is performed by the method \texttt{multipole\_spectra}, which computes the multipole power spectrum, $P_{\ell}(k, z)$, at a given redshift, $z$, scale, $k$, and list of multipoles, $\ell$, following Eq.~(95) in \cite{Cardone24}. It consists of the scaling factors, $(q_\perp^2q_\parallel)^{-1}$, which encapsulate the AP effect \citep{AP79}, multiplied by a Legendre-polynomial weighted integral over the redshift-space galaxy power spectrum, $P_{\rm gg}^{\rm spectro}(k, \mu_k, z)$, with respect to $\mu_k$ that denotes the cosine of the angle between the Fourier mode ${\vec k}$ and the line of sight. This integral is performed using Simpson's rule, which takes as input the method \texttt{multipole\_spectra\_integrand} at a given \{$z$, $k$, $\mu_k$, $\ell$\}. The integrand obtains the galaxy power spectrum from the cosmology dictionary (\texttt{cosmo\_dic}), which in turn acquires it from \texttt{Pgg\_spectro\_def} in \texttt{cloe/cosmo/cosmology.py}.\footnote{We note that \texttt{Pgg\_spectro\_def} either refers to a method in \texttt{cosmology.py} that computes the galaxy power spectrum in the linear regime, or to a method in the \texttt{non\_linear} subpackage that computes it to nonlinear scales using the effective field theory of large-scale structure (\eftoflss; for further details on the \eftoflss and other planned nonlinear implementations, see Euclid Collaboration:~Moretti et al., in prep.). The user can choose between these options via the key \texttt{NL\_flag\_spectro} in \texttt{configs/config\_default.yaml}.}

The method \texttt{multipole\_spectra} is then used to obtain the window-convolved multipole power spectra in the method \texttt{convolved\_power\_spectrum\_multipoles}. This includes reading in the mixing matrix, $W_{\ell\ell'}$, in the \texttt{Reader} module, storing it in the dictionary \texttt{mixing\_matrix\_dict}, and accessing it through the constructor of the \texttt{Spectro} class. \cloe obtains the convolved multipole power spectra by performing a convolution of the mixing matrix with the unconvolved multipole power spectra according to Eq.~(108) in \citet{Cardone24}. These convolved power spectra are the quantities that are then compared against the measurements. The \texttt{multipole\_spectra} method is further used to obtain the multipole correlation functions, $\xi_{{\rm obs},\ell}(s^{\rm fid},z)$, where $s^{\rm fid}$ is the physical separation at the fiducial cosmology, through the method \texttt{multipole\_correlation\_function}. This method follows Eq.~(105) in \citet{Cardone24} in performing a Hankel transform of the unconvolved multipole power spectrum via the \fftlog algoritm for which we created a dedicated subpackage in \texttt{cloe/fftlog} (see Sect.~\ref{fftlogsec}). The \texttt{multipole\_correlation\_function} method moreover includes the additive contributions of spectroscopic magnification bias on the multipole correlation functions, as further described in \citet{Goh24}.

\begin{figure*}[h]
\centering
\includegraphics[width=0.48\linewidth, trim = {14cm 3cm 14cm 0cm}]{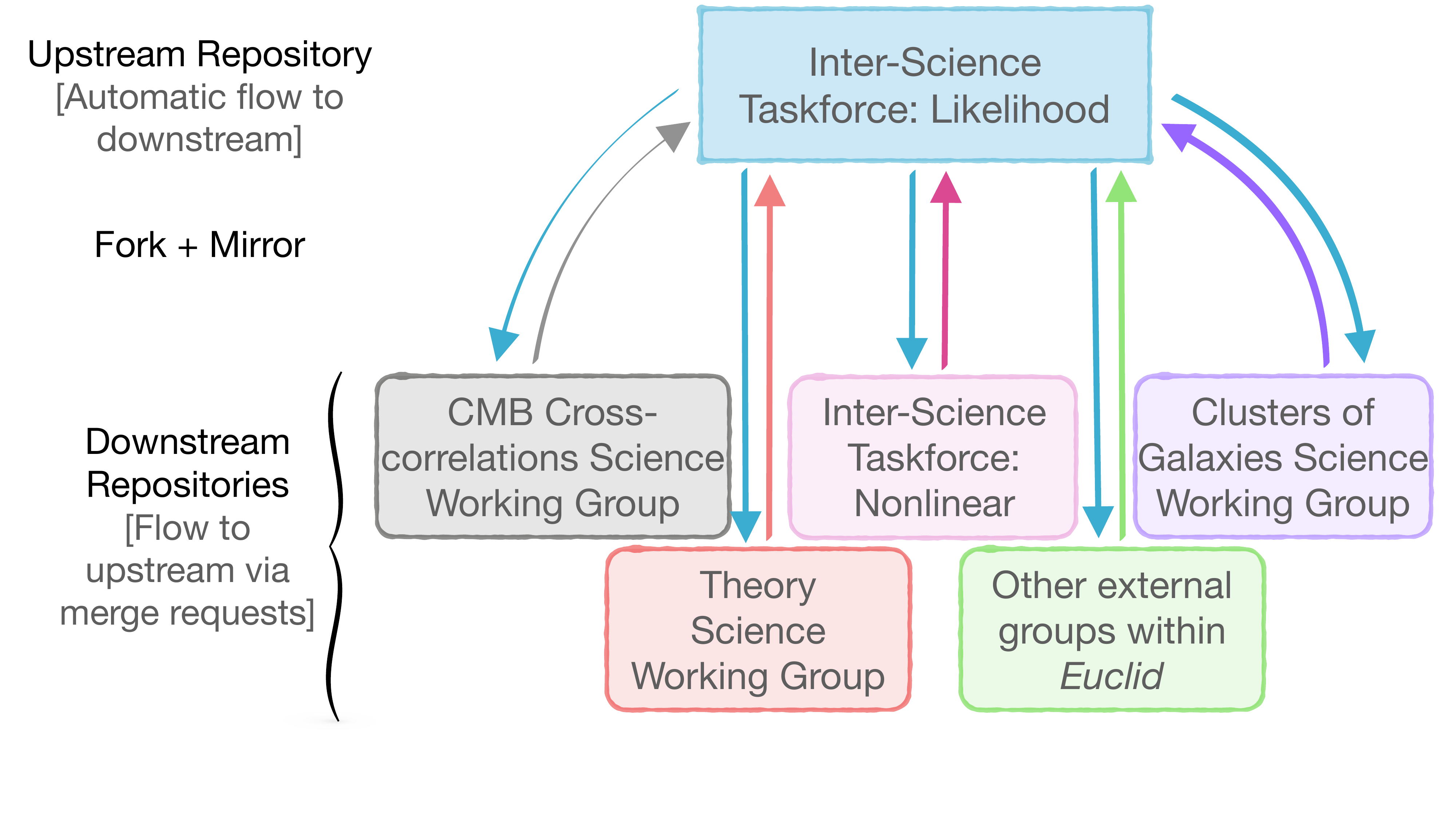}
\caption{Illustration of the flow between the upstream and downstream repositories governing the  development of the current version of \cloe among distinct groups in \Euclid, such as Inter-Science Taskforce:~Likelihood, Inter-Science Taskforce:~Nonlinear, the Theory Science Working Group, the CMB Cross-correlations Science Working Group, and the Clusters of Galaxies Science Working Group.}
\label{repoflow}
\end{figure*}

\subsubsection{Parameter space}

In the method \texttt{multipole\_spectra\_integrand}, we include the impact of systematic uncertainties such as  spectroscopic redshift errors and random catastrophic outliers in the galaxy samples (see section 4.2 in \citealt{Cardone24} for the theoretical prescription of this modeling). For each systematic uncertainty, this results in one nuisance parameter per redshift bin. In the case of \Euclid, with its four spectroscopic redshift bins, this adds up to 8 nuisance parameters. The modeling of the galaxy power spectrum itself, via \texttt{Pgg\_spectro\_def}, also includes a range of nuisance parameters. In the \eftoflss formalism, this includes linear, quadratic, and two nonlocal galaxy biases, along with four counter-terms based on 1-loop perturbation theory (as further described in Euclid Collaboration:~Moretti et al., in prep.; also see
\citealt{Bernardeau02, Baumann12, Carrasco12, Ivanov20, Chudaykin20, Carrilho23}). These eight galaxy power spectrum parameters are distinct for each redshift bin, such that in the case of \Euclid, this adds up to 32 nuisance parameters. 

After the propagation of these systematic uncertainties, the method \texttt{multipole\_spectra\_integrand} further adds the noise contribution to the galaxy power spectrum from the \texttt{non\_linear} module (via \texttt{cosmo\_dic}), which includes a Poissonian shot noise parameter and scale-dependent deviations that are described by two additional nuisance parameters for each redshift bin (\citealt{Carrilho23}; Euclid Collaboration:~Moretti et al., in prep.). As a result, there are up to $8+32+12=52$ nuisance parameters required to describe the spectroscopic observables for \Euclid in the current version of \cloe. However, for purposes of computational speed, it is often prudent to reduce this number in the parameter inference as discussed in Euclid Collaboration:~Moretti et al. (in prep.) and \citet{GCH24}. This includes the analytic marginalization of a subset of the nuisance parameters (as previously considered in the spectroscopic and photometric contexts in for example \citealt{Ivanov20, Ruiz-Zapatero23, Hadzhiyska23}).

\subsubsection{Integration strategy}
\label{sec3f2}

In the creation of the theoretical predictions for the multipole power spectra of the spectroscopic galaxies, \cloe  considers an integration of the galaxy power spectrum over $\mu_k$ using Simpson's rule. The multipole correlation functions are then obtained by integrating the multipole power spectra over $k$ using the \fftlog algorithm (see Sect.~\ref{fftlogsec}).

In \cloe, the \texttt{multipole\_spectra} method obtains the multipole power spectra at a given redshift, $z$, and wavenumber, $k$, by integrating over a linearly spaced grid of (fiducially) 2001 $\mu_k$ values between $\mu_{k; \rm min} = -1$ to $\mu_{k; \rm max} = 1$. The multipole power spectra are multiplied by $i^\ell$ and a volume factor ${k^3}/{2\pi^2}$. The \texttt{multipole\_correlation\_function} method in turn theoretically calculates the multipole correlation function given an input redshift and arrays of physical separations and multipoles. In achieving this, it uses the \fftlog algorithm to integrate the multipole power spectra across wavenumber. To this end, \texttt{multipole\_spectra} fiducially calculates the multipole power spectra for a log-spaced grid of $N=2048$ wavenumbers between $k_{\rm min} = 5\times10^{-5} \, {\rm Mpc}^{-1}$ to $k_{\rm max} = 50 \, {\rm Mpc}^{-1}$.\footnote{We caution the reader that in the current version of \cloe, the nonlinear galaxy power spectrum from the EFTofLSS model is only validated down to $k \simeq 0.2 \, {\rm Mpc}^{-1}$. This is sufficient for the scales considered in the multipole power spectrum analysis, but not for the multipole correlation function analysis, and is an aspect of the nonlinear modeling that remains to be improved.} The obtained array is then passed to the \texttt{fftlog} method for the evaluation of the integral. As the output of \texttt{fftlog} is provided for log-spaced values of separation, an interpolator is further created and used to evaluate the correlation function for the input separation array.

\subsection{Nonlinear subpackage and repository network}
\label{sec3g}

The \texttt{cloe/non\_linear} subpackage contains the code implementation of the nonlinear corrections to the matter, galaxy, and IA power spectra, as described in Euclid Collaboration:~Carrilho et al. (in prep.) and Euclid Collaboration:~Moretti et al. (in prep.). 

In understanding the interface with the \texttt{non\_linear} subpackage, we highlight that it contains a suite of modules, the most central of which is \texttt{nonlinear}. When the user enables nonlinear corrections through the default configuration file (\texttt{config\_default.yaml}), the \texttt{nonlinear} module receives the linear power spectra and other background quantities from the cosmology dictionary (\texttt{cosmo\_dic}), which is passed to it from the \texttt{cosmology} module. These linear power spectra effectively include auto- and cross-terms of the total matter, cold dark matter (CDM), baryonic matter, galaxy, and IA fields, along with the redshift-space galaxy power spectrum $P_{\rm gg}^{\rm spectro}(k, \mu_k, z)$.\footnote{We note that this is the galaxy power spectrum before the application of the damping due to the nonzero outlier fraction and spectroscopic redshift errors in the \texttt{spectro} module, corresponding to Eq.~(103) in \citet{Cardone24}.} 

The \texttt{nonlinear} module computes the nonlinear corrections to the power spectra with the aid of other modules in the subpackage, and then returns these to the \texttt{cosmology} module. These other modules include \texttt{eft}, which computes the redshift-space galaxy power spectrum; \texttt{miscellaneous.py}, which implements the TATT IA model (among other features); \texttt{power\_spectrum}, which provides a shared interface and data structure for the power-spectrum modules; and the modules \texttt{pLL\_phot}, \texttt{pgL\_phot}, \texttt{pgg\_phot}, and \texttt{pLL\_spectro}, which transform the nonlinear boosts into nonlinear matter, IA, and galaxy auto- and cross-power spectra.

As the nonlinear subpackage was developed by a distinct task force within \Euclid, it was important to create a management of the \cloe code repository that prevented any conflicts in the code development, as illustrated in Fig.~\ref{repoflow}. Examples of possible coding conflicts include the simultaneous modification of the same pieces of code, or the modification of code by one group that has become outdated in repository of another group.

In preventing these forms of conflicts, we created a repository network, where the main \cloe repository was forked and automatically mirrored by the various groups in \Euclid. In this context, the forks constitute new repositories that begin as exact copies of the original ``upstream'' repository. Meanwhile, the automatic mirroring refers to the synchronization of the forked, or ``downstream'', repositories with the main upstream repository. This synchronization only applies in one direction, in that any changes to the upstream repository are automatically propagated to the downstream repositories, while any changes in the downstream repositories are only propagated into the upstream repository through merge requests.
 
To this end, as code was developed within the downstream repositories, it first underwent an internal merge request to be accepted into the forked repository. The code was then pushed to the upstream repository through another merge request. The review of this second merge request was focused on possible coding conflicts and the homogenization in the style and documentation of the merged code. 

\begin{figure*}[h]
\centering
\includegraphics[width=0.6\linewidth, trim = {13cm 0cm 13cm 0cm}]{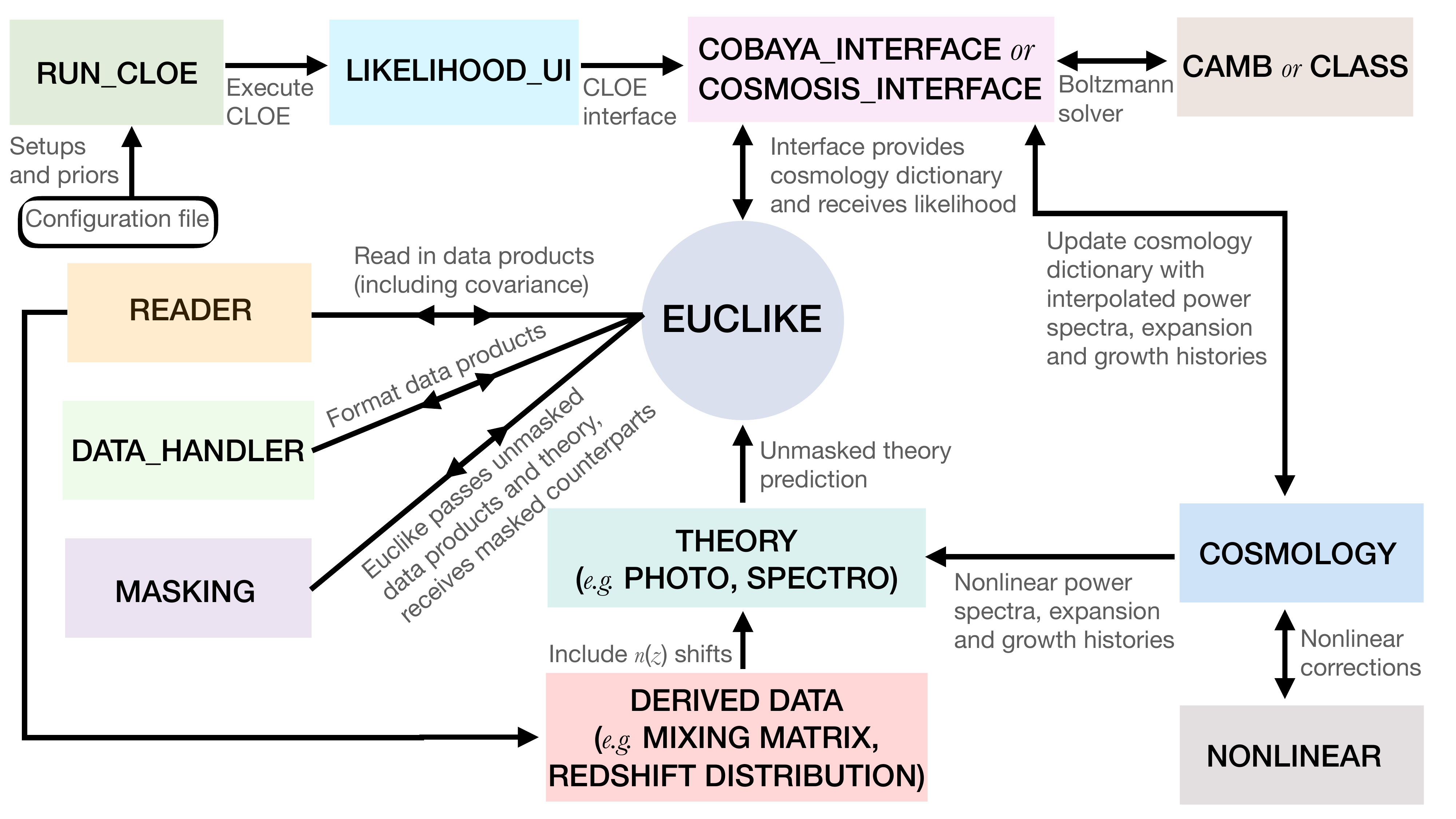}
\vspace{0.4em}
\caption{Illustration of the modules involved in the calculation of the \cloe likelihood that is then passed to the Monte Carlo sampler.}
\label{euclikefig}
\end{figure*}

\subsection{Data reader}
\label{sec3j}

The \texttt{reader} module within \texttt{cloe/data\_reader} contains the \texttt{Reader} class, which is responsible for reading in all of the data products, which includes the measurements, redshift distributions, covariance matrices, mixing matrices, luminosity ratios, and any other files that need to be included in the likelihood calculation (in one of the following file formats: \textsc{ascii}, \textsc{fits}, \textsc{npy}, and \textsc{npz}). As described in Sect.~\ref{sec3h}, this class is instantiated in the constructor of \texttt{Euclike}, which in turn is executed once in a given Monte Carlo run by either the \cobaya or \cosmosis interface. 

The constructor of \texttt{Reader} receives the \texttt{data} dictionary, which contains the user specifications for the files that need to be read in. As described in Sect.~\ref{midlayersec}, this dictionary is obtained from \texttt{configs/data.yaml} in either the \cobaya or \cosmosis interface. It includes all necessary metadata, such as the names and locations of the data products, the redshift-bin definitions, whether the measurements and covariances are provided in Fourier or configuration space, and whether the covariances are constructed analytically or from numerical simulations (including the number of realizations in the latter case). The \texttt{data} dictionary is then used by core methods of the class, such as \texttt{reader\_raw\_nz}, which reads in the photometric source and lens redshift distributions. In the method \texttt{compute\_nz}, the cubic spline interpolators of these distributions are further normalized (such that they integrate to unity) and stored in the dictionaries \texttt{nz\_dict\_WL} and \texttt{nz\_dict\_GC\_Phot}, respectively.

Another illustration of a core method of the \texttt{Reader} class is \texttt{read\_GC\_spectro}, which reads in the spectroscopic galaxy clustering measurements, covariance, and fiducial cosmology to capture the AP effect. Likewise, the method \texttt{read\_phot} reads in the photometric 3$\times$2pt measurements and covariance. This includes the capability to read in data files in the format of the \Euclid Science Ground Segment via \euclidlib\footnote{\euclidlibaddress}\href{https://github.com/euclidlib/euclidlib}{\color{black}\faGithub}. All of the imported measurements and covariances are stored in the dictionary \texttt{data\_dict}, which is then accessed by \texttt{Euclike} and \texttt{Data\_handler} (further described in Sect.~\ref{sec3i}). We note that the measurements can be either real or synthetic, and that the module is sufficiently generic that it can be straightforwardly extended to accommodate additional datasets. The overall design ensures that the numerical data and their associated metadata remain consistent throughout the pipeline, enabling the correct mapping between the measurements, covariances, and theoretical predictions.

\subsection{Auxiliary subpackage}
\label{sec3k}

\cloe's \texttt{auxiliary} subpackage contains a range of modules needed for the cosmological parameter inference. This includes \texttt{observables\_dealer} referred to in Sect.~\ref{sec3c}. Its core method merges the selection and specifications of the observables (i.e.,~the choice of observables, scales, and redshifts) into a single dictionary. It also has methods that examine the soundness of the observables and scales selected for the analysis, along with the visualization of the observables. 

Other modules include \texttt{yaml\_handler}, which contains methods for the reading and writing of \yaml files, along with \texttt{likelihood\_yaml\_handler}, which contains methods to handle the \yaml files and related dictionaries in \cloe. The module \texttt{matrix\_transforms} includes transformations of the cosmic shear and galaxy-galaxy lensing observables, as further described in Sect.~\ref{bntsec}.

The module \texttt{getdist\_routines} contains an interface to \getdist, the module \texttt{matrix\_manipulator} allows for the merging of matrices, and the module \texttt{run\_method} examines if the code is run from a script or interactively. Moreover, the module \texttt{redshift\_bins} contains common methods to operate on redshift bin edges and redshift-dependent nuisance parameters. 

Meanwhile, the \texttt{plotter} module contains the \texttt{Plotter} class that allows for the plotting of the cosmological observables (currently the 3$\times$2pt angular power spectra and the multipole power spectra), and the related \texttt{plotter\_default} module contains a dictionary with the default settings for the plotting routines. Lastly, the \texttt{logger} module contains methods to handle the logging, as described in Sect.~\ref{sec3a}.

\subsection{Likelihood calculation}
\label{sec3h}

As illustrated in Fig.~\ref{euclikefig}, the \Euclid likelihood calculation is performed in \texttt{cloe/like\_calc/euclike.py}. Depending on the user preferences, the \texttt{euclike} module currently computes the likelihood for all or a subset of the 3$\times$2pt observables, the spectroscopic galaxy clustering, cluster observables, and cross-correlations of the photometric galaxy positions and lensing with the CMB temperature and lensing.

The \texttt{euclike} module  imports other \cloe classes for the computation of the theory vectors (\texttt{Photo}, \texttt{Spectro}, \texttt{CG}, \texttt{CMBX}, \texttt{RedshiftDistribution}, and \texttt{BNT\_transform}), along with the reading, handling, and masking of the data and covariance (\texttt{Reader}, \texttt{Data\_handler}, and \texttt{Masking}). The \texttt{euclike} module itself contains the class \texttt{Euclike}, which is instantiated by \cloe's \cobaya or \cosmosis interface, as described in Sect.~\ref{sec3c}. To this end, the class constructor receives the \texttt{data} dictionary, which contains the specifications for data loading and handling, to be passed to the \texttt{reader} module. The class constructor also receives the \texttt{observables} dictionary, which contains the specifications for the chosen observables by the user. 

As part of the initialization of \texttt{Euclike}, the \texttt{Reader} class is instantiated and the \Euclid measurements, redshift distributions, mixing matrices, and covariances are read and stored once. Additional quantities that only need to be computed once, such as the Bessel functions and multipole prefactors for the 3$\times$2pt probes, are computed within the constructor as well. Moreover, depending on the user preferences on the probes to consider in the analysis, theory prediction classes such as \texttt{Photo}, \texttt{Spectro}, and \texttt{CMBX} are instantiated.

\begin{figure*}[h]
\centering
\includegraphics[width=1.0\linewidth, trim = {0cm 23cm 0cm 0cm}]{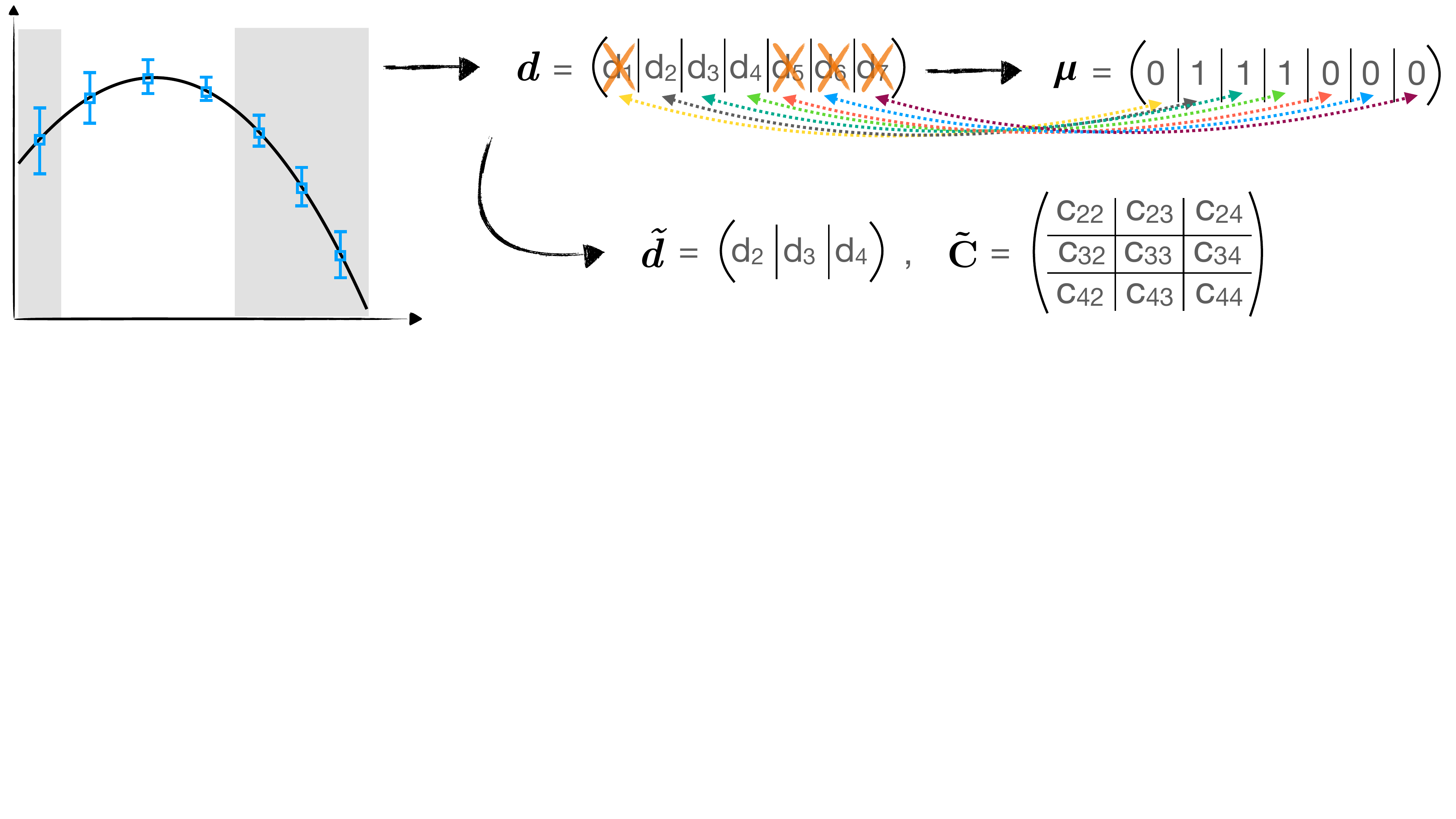}
\vspace{-0.5em}
\caption{Illustration of the masking procedure given a set of measurements in \cloe. In the figure to the left, the gray bands denote the masked out regions. Here, ${\vec{d}}$ is the unmasked data vector, ${\bm{\mu}}$ is the masking vector, $\textbf{\textit{\~d}}$ is the masked data vector, and $\tens{\tilde{C}}$ is the masked covariance matrix. The corresponding masking is enforced for the theory vector. Note that the symmetry of the covariance would in practice enforce $c_{32} = c_{23}$, $c_{42} = c_{24}$, and $c_{43} = c_{34}$.}
\label{maskfig}
\end{figure*}

After \texttt{Euclike} is instantiated, the \cobaya or \cosmosis interface calls the method \texttt{get\_masked\_data} within \texttt{Euclike} once, which arranges and ``masks'' the data products into their final format (by employing the \texttt{Data\_Handler} and \texttt{Masking} classes described in Sect.~\ref{sec3i}). The masking procedure enforces cuts in the measurements and covariances to account for the user preferences on the scales and redshifts to include in the analysis (where the preferences are transmitted via \texttt{config\_default.yaml}).\footnote{We reiterate that ``masking'' here refers to cuts in the data and theory vectors to accommodate the user preferences on the scales and redshifts to include in the analysis. It is distinct from the masking of the observed sky, which necessitates a convolution of the theory predictions with the mixing matrix.} These masked measurements and covariances are the quantities that are then used by the \texttt{loglike} method of \texttt{Euclike} to obtain the log-likelihood. In addition, the method \texttt{get\_masked\_data} computes the BNT matrix once, which is only needed when the user requests the weak lensing observables to be BNT transformed (as further described in Sect.~\ref{bntsec}).\footnote{The implementation of the BNT transformation in \cloe is itself the driver behind the design to call the method \texttt{get\_masked\_data} within the \cobaya or \cosmosis interface, rather than within the constructor of the \texttt{Euclike} class. In either approach, the method is only called once in a given Monte Carlo run.}

As referenced above, the central method of \texttt{Euclike} is \texttt{loglike}, which performs the likelihood calculation and returns its output to one of the samplers in either \cobaya or \cosmosis (via the corresponding interface modules in \cloe). We note that the output of this method is technically the natural logarithm of the likelihood, which for a Gaussian likelihood corresponds to $\ln \mathcal{L} = -\frac{1}{2}\chi^2 + {\rm constant}$. For completeness, we define $\chi^2 = (\textbf{\textit{\~d}} - \textbf{\textit{\~t}}) \, {\tens{\tilde{C}}}^{-1} \, (\textbf{\textit{\~d}} - \textbf{\textit{\~t}})^{{\sf T}}$, where $\textbf{\textit{\~d}}$ refers to the masked data vector, $\textbf{\textit{\~t}}$ to the masked theory vector, $\tens{\tilde{C}}$ to the masked covariance matrix, and ${\sf T}$ to the transpose operation. In the event that the covariance matrix is constructed using numerical simulations, the user can specify this to \cloe via the \texttt{cov\_is\_num} key in \texttt{cloe/data.yaml}. Consequently, the likelihood takes on a non-Gaussian form, such that 
\begin{equation}
\ln \mathcal{L} = -\frac{m}{2} \, \ln \left(1 + \frac{\chi^2}{n_{\rm sim} - 1}\right) + {\rm const.} \, , 
\end{equation}
where $n_{\rm sim}$ corresponds to the number of simulations and $m$ is a fixed quantity that is determined by the number of simulations, data points, and parameters (\citealt{Percival22};~also see \citealt{Hotelling1931, SH16}). As noted earlier, the constant entering $\ln \mathcal{L}$ is immaterial for parameter inference. For instance, in MCMC methods, this is because the sampling depends on differences in the log-posterior, and thereby the log-likelihood, between points in parameter space.

As discussed in Sect.~\ref{sec3c}, \texttt{loglike} is called by either the method \texttt{logp} in the \texttt{cobaya\_interface} module or the method \texttt{execute} in the \texttt{cosmosis\_interface} module. It receives the cosmology dictionary (\texttt{cosmo\_dic}), which it uses to compute
the theory vectors for the chosen probes and summary statistics encoded in the \texttt{observables} dictionary. This is achieved using the methods \texttt{create\_photo\_theory} and \texttt{create\_spectro\_theory} for the 3$\times$2pt observables and spectroscopic galaxy clustering, respectively, and analogously for the CMB cross-correlations and cluster observables.

For a given probe, similar to the masking of the data, the theory vector is subsequently masked in redshift and scale according to the user preferences. This is accomplished by passing the uncut theory vectors to the \texttt{masking} module via the method \texttt{set\_theory\_vector},  
and subsequently executing the method \texttt{get\_masked\_theory\_vector}, 
as further described in Sect.~\ref{sec3i}. The difference between the masked theory and data vectors is then used together with the inverse of the masked covariance to compute the $\chi^2$. As described above, this $\chi^2$ is in turn used to obtain the log-likelihood considering either a Gaussian or non-Gaussian shape of the likelihood depending on whether the covariance is analytic or simulated.

In the current version of \cloe, we assume that there is no covariance between the photometric and spectroscopic probes, and thereby compute their log-likelihoods separately before adding them. The same applies to the cluster probes, which are treated independently as an approximation. However, as described in Sect.~\ref{cmbsec}, \cloe does account for the full covariance between the 3$\times$2pt probes and CMB cross-correlations as part of a 7$\times$2pt analysis. In other words, 
in a scenario where all probes are included, 
\cloe considers
\begin{equation}
\ln {\mathcal{L}} = \ln {\mathcal{L}}_{7\times2{\rm pt}} + \ln {\mathcal{L}}_{\rm spectro} + \ln {\mathcal{L}}_{\rm clusters}.
\end{equation}
Given the creation of the full covariance between the distinct probes, the changes to \cloe to accommodate such a covariance and thereby perform a single likelihood computation are straightforward.

\subsection{Masking of the data, covariance, and theory}
\label{sec3i}

The \cloe masking procedure, which follows the approach in \cosmolss \footnote{\cosmolssaddress}\href{https://github.com/sjoudaki/CosmoLSS}{\color{black}\faGithub} \citep{joudaki18a}, allows the user to have complete control over which probes, scales, and redshifts are included in the likelihood analysis. This amounts to a level of control over the analysis that extends down to individual data points. We note that ``masking'' here refers to cuts to the measurements, covariance matrix, and theory vector and is distinct from the masking of the observed sky. \cloe currently masks the photometric 3$\times$2pt probes, the spectroscopic galaxy clustering, and the CMB cross-correlations, but not the cluster probes which are included in their entirety. 

In Sect.~\ref{sec3b}, we outlined how the user can impose their selection of the probes, scales, and redshifts via the configuration \yaml files in the \texttt{configs} directory. These user selections directly modify a single quantity in \cloe: the masking vector. It denotes an array of the same size as the data vector, where the elements are either 0 or 1 depending on whether the corresponding element in the data vector is included in the analysis.\footnote{We note that there are an uncountable number of other acceptable definitions for the masking vector that would lead to the same intended outcome, and that our current definition is driven by its intuitive nature.} The masking vector is then used as the basis to obtain the masked data vector, masked covariance, and masked theory vector, which together enter the log-likelihood calculation (described in Sect.~\ref{sec3h}).

As illustrated in Fig.~\ref{maskfig}, let us consider a hypothetical scenario in which the cosmological measurements consist of seven data points across different physical scales, such that ${\vec d} = (d_1, d_2, d_3, d_4, d_5, d_6, d_7)$. In this scenario, the user wishes to impose scale cuts that would end up removing the first data point along with the fifth, sixth, and seventh data points out of the likelihood analysis for physical reasons (say to avoid nonlinearities on small scales and residual uncertainties in the additive shear bias calibration on large scales). As a result, the masking vector takes on the form ${\bm \mu} = (0, 1, 1, 1, 0, 0, 0)$. It is then applied to the measurements and theory predictions to obtain the masked data vector, $\textbf{\textit{\~d}} = (d_2, d_3, d_4)$, and masked theory vector, $\textbf{\textit{\~t}} = (t_2, t_3, t_4)$.
Likewise, the covariance matrix
\begin{equation}
{\tens{C}}= 
\begin{pmatrix}
c_{11} & c_{12} & c_{13} & c_{14} & c_{15} & c_{16} & c_{17}\\
c_{21} & c_{22} & c_{23} & c_{24} & c_{25} & c_{26} & c_{27}\\
c_{31} & c_{32} & c_{33} & c_{34} & c_{35} & c_{36} & c_{37}\\
c_{41} & c_{42} & c_{43} & c_{44} & c_{45} & c_{46} & c_{47}\\
c_{51} & c_{52} & c_{53} & c_{54} & c_{55} & c_{56} & c_{57}\\
c_{61} & c_{62} & c_{63} & c_{64} & c_{65} & c_{66} & c_{67}\\
c_{71} & c_{72} & c_{73} & c_{74} & c_{75} & c_{76} & c_{77} 
\end{pmatrix}
\end{equation}
is transformed into the masked covariance that is given by 
\begin{equation}
{\tens{\tilde{C}}}=
\begin{pmatrix}
c_{22} & c_{23} & c_{24}\\
c_{32} & c_{33} & c_{34}\\
c_{42} & c_{43} & c_{44}
\end{pmatrix},
\end{equation}
where $c_{mn}$ for indices $m$ and $n$ refers to the distinct component of $\tens{C}$ and $\tens{\tilde{C}}$ that captures the covariance between the $d_m$ and $d_n$ components of the data vector. Note further that we have expressed the covariances in generality here, as their symmetry property would in practice enforce $c_{nm} = c_{mn}$.

One important subtlety related to this example is that in likelihood analyses where the different probes share a single joint covariance, a concatenated data vector is first constructed. This is relevant for instance in the 3$\times$2pt analysis, where \cloe reads in a single externally produced covariance matrix $\tens{C}_{3\times{\rm 2pt}}$ and thereby first constructs the multi-probe data vector ${\vec d}_{3\times{\rm 2pt}}$ from the single-probe data vectors of 
${\vec d}_{\rm cosmic \, shear}$, ${\vec d}_{\rm galaxy-galaxy \, lensing}$, and ${\vec d}_{\rm galaxy \, clustering}$. The same procedure applies in the construction of the multi-probe theory vector, here given by ${\vec t}_{3\times{\rm 2pt}}$. The multi-probe data vector, covariance matrix, and theory vector are the quantities that are then masked.

We note that the masking of the data vector and covariance matrix is performed only once in a given Monte Carlo run. However, the theory vector is computed and masked at each new point in parameter space. \cloe currently speeds up the likelihood calculation by neglecting the computation of the theory vector for probes that are not selected. This entails setting the relevant components of the theory vector to zero before the construction of the masked theory vector. This approach can be generalized, such that the theory computation is avoided for each of the distinct components of the data vector that are masked out of the analysis (i.e., neglecting the theory computation of distinct scales and redshifts of a given probe that is more broadly included in the analysis).

Let us now provide further details on how the user selections are propagated into a masking of the measurements, covariance, and theory predictions via the modules \texttt{data\_handler} and \texttt{masking} in the directory \texttt{cloe/masking}. In broad terms, the former module creates the masking vector and arranges the measurements into a form that \cloe expects (in the case of the 3$\times$2pt probes, it orders the distinct measurements in the following order: cosmic shear, followed by galaxy-galaxy lensing, and then galaxy clustering). The latter module then applies the masking vector to the data vector, covariance matrix, and theory vector to obtain the corresponding masked quantities.

The \texttt{data\_handler} module holds the \texttt{Data\_Handler} class. Its constructor receives an initialized instance of the class \texttt{Reader}, which specifies the structure and organization of the data. The constructor also receives three distinct dictionaries that respectively contain the measurements, covariances, and user selections of the observables, redshifts, and scales to include in the analysis (which in the case of the third dictionary is decided by the user via the keys \texttt{observables\_specifications} and \texttt{observables\_selection} in \texttt{config\_default.yaml} as discussed in Sect.~\ref{sec3b}). These dictionaries are stored as instance attributes. The constructor subsequently calls the relevant class methods to output the finalized forms of the unmasked data vectors and covariances, which includes the concatenation of covariant data vectors.

In order to perform the masking of these unmasked data vectors and covariances, \cloe first creates the masking vector. It does this within the class constructor as well, which calls the method \texttt{create\_masking\_vector\_phot}  when any of the 3$\times$2pt probes are considered in the likelihood analysis. For each tomographic bin combination of a given 3$\times$2pt probe (for the user-selected summary statistic, say angular power spectra rather than correlation functions), the method obtains the scales for which there are measurements from the data dictionary, along with the ranges of scales that the user has selected for the analysis from the observables dictionary. It feeds these two arrays to the method \texttt{get\_masking}, which creates an array with values of 0 or 1 depending on which elements of the data array are contained in the specified ranges of the observables array. It flattens the resulting two-dimensional array of masking vectors across tomographic bin combinations into a one-dimensional array, and subsequently concatenates the corresponding arrays for the three distinct 3$\times$2pt probes.

This masking procedure is currently taken to be distinct for the photometric and spectroscopic probes, as we are not including possible cross-correlations between them (consistent with the analysis planned for the first data release of \Euclid). There is therefore an analogous masking procedure for the spectroscopic galaxy clustering that creates its own distinct masking vector.\footnote{The method \texttt{create\_masking\_vector\_full} in the data handler module creates a single joint masking vector for the photometric and spectroscopic probes. While not relevant for the first set of cosmological analyses with \Euclid, it has possible future applications for overlapping photometric and spectroscopic survey analyses. A similar method currently exists for the photometric probes and their cross-correlations with the CMB, as further described in Sect.~\ref{cmbsec}.}

The method \texttt{get\_data\_and\_masking\_vector\_phot} returns the finalized unmasked 3$\times$2pt data vector and covariance matrix, along with the corresponding masking vector. Likewise, the  method \texttt{get\_data\_and\_masking\_vector\_spectro} returns the spectroscopic galaxy clustering data, covariance, and masking vector. These two \texttt{Data\_handler} methods are called and their corresponding quantities are stored within the \texttt{get\_masked\_data} method in the \texttt{Euclike} class. 

At this stage of  \texttt{get\_masked\_data}, the \texttt{Masking} class of the \texttt{masking} module is instantiated and receives the data and covariance via its respective methods \texttt{set\_data\_vector} and  \texttt{set\_covariance\_matrix}. It moreover receives the masking vector through its method \texttt{set\_masking\_vector} and stores it as a Boolean data type. The motivation for this conversion is that a \numpy array of the latter can be provided as the index to another \numpy array of the same size, which has the effect of retaining only the array elements where the corresponding mask value is \texttt{True}. Hence, while not necessary, it simplifies the masking operation.\footnote{In principle, the Boolean data type could also be used to define the masking vector at the outset. Here, we have aimed to combine the intuitive nature of the initial definition of integers with the practical nature of the final definition of Booleans.} The method \texttt{get\_masked\_data} then calls the \texttt{Masking} methods \texttt{get\_masked\_data\_vector} and \texttt{get\_masked\_covariance\_matrix}. In these methods, the Boolean nature of the masking vector is utilized to transform the unmasked data vector and covariance matrix into their corresponding masked quantities.

\begin{figure*}[h]
\vspace{-0.29em}
\centering
\sidecaption
\includegraphics[width=0.677\linewidth, trim = {17.4cm 18.1cm 19.4cm 0cm}]{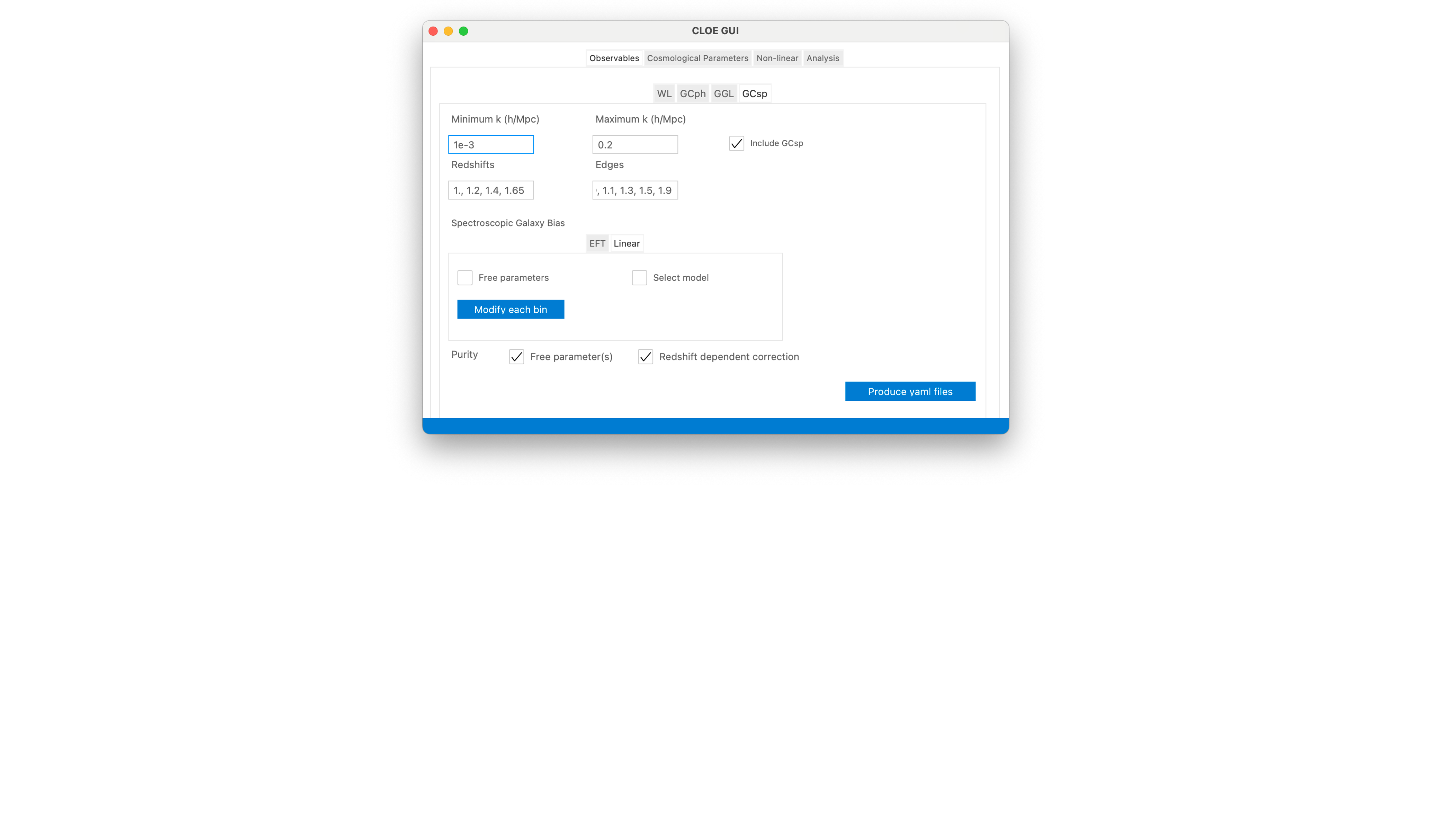}
\vspace{0.69em}
\caption{Illustration of the appearance of the \cloe GUI. Instead of modifying the configuration files through the command line, the GUI allows the user to produce the necessary files in an interactive way.}
\label{guifig}
\end{figure*}

\subsection{Code outputs}

We next describe what \cloe outputs in a given run. Here, it is important to distinguish between what \cloe outputs on its own and what is outputted as a result of its interface with a given sampling platform. In \texttt{config\_default.yaml}, \cloe allows for the key \texttt{print\_theory}, which outputs the theory vector at a given point in parameter space to a \texttt{.dat} file when set to \texttt{True}. This key is most suitably employed when executing a single likelihood evaluation, for instance achieved using the \texttt{evaluate} sampler in \cobaya or the \texttt{test} sampler in \cosmosis. Through the default configuration file, \cloe also allows for the key \texttt{plot\_observables\_selection} that creates a graphical representation of the selected observables when set to \texttt{True}.

As a result of \cloe's interface with sampling platforms such as \cobaya and \cosmosis, when a Monte Carlo run is executed, the files intrinsic to each platform are created within the \texttt{chains} directory. As an example, in the case of an MCMC run with \cobaya, this includes $N$ number of \texttt{.txt} files containing the $N$ chains, a \texttt{checkpoint} file that stores metadata about the sampling state and allows reruns to continue from the latest state rather than from the beginning, and a \texttt{progress} file that stores information about the level of convergence of the chains. The run also produces a \texttt{covmat} file that stores the proposal covariance matrix, an \texttt{input.yaml} file that stores the user-defined configurations, and an \texttt{updated.yaml} file that stores the combined user-defined and default configurations used for the run. In addition, \cobaya creates two auxiliary files: \texttt{updated.dill\_pickle} which is a binary serialization of \texttt{updated.yaml}, and \texttt{input.yaml.locked} which is a temporary lock file for \texttt{input.yaml} that is removed when the run finishes. 

In the corresponding example for \cosmosis, two file types are generated, namely \texttt{.txt} and \texttt{sampler\_status}. The former contains the chains, while the latter records the internal state of the sampler and enables checkpointing. Hence, the exact set of output files depends on factors such as the user-defined options in \texttt{config\_default.yaml} and the choice of sampler within a given platform. Additional diagnostic or auxiliary files may also be produced depending on the specific sampler settings or whether features such as adaptive covariance estimation are enabled.

\subsection{Unit tests}
\label{sec3l}

Unit testing is a standard practice in modern code development. It denotes the process where individual units of the source code are tested to confirm they return the expected outputs. This helps to ensure the quality of the code, as any bugs introduced during development that modify the output of a given unit of code will get flagged immediately.

The more code that is unit tested, the higher the coverage. Here, the coverage is a percentage measure of the degree to which the source code is tested. In the case of \cloe, the unit tests are provided for the different modules in the directory \texttt{cloe/tests} and are executed with \pytest\footnote{\pytestaddress}. \cloe's current unit test coverage is 82\%, which is above our criterion of greater than 80\% coverage. These tests are run automatically for every merge request, as part of the continuous integration. We note that these unit tests are distinct from the validation of \cloe, which tests the accuracy of the likelihood calculation through comparison against independent codes and as a function of changes in the numerical precision, as described in Sect.~\ref{secval} and \citet{Martinelli24}.

\subsection{MCMC scripts}
\label{sec3m}

In the \texttt{mcmc\_scripts} directory, \cloe provides a series of ready-made scripts that allow for the sampling of the parameter space with the \cobaya interface. The scripts contain all of the user settings for different probe combinations, treatment of systematic uncertainties, and cosmological models (including nonzero curvature, evolving dark energy, and modification of gravity). They thereby allow for the bypassing of the \cloe overlayer and constitute an additional approach to execute the cosmological parameter inference.\footnote{To provide further details, there are two related ways in which we can run \cloe with \cobaya. One is through the overlayer with \texttt{cobaya.run} where we pass the dictionary \texttt{cobaya\_dict}, which encapsulates the information in the \cloe configuration \yaml files. The second approach is with the ready-made scripts, where we pass the \texttt{info} dictionary, defined within the script itself, to the \texttt{run} method of \texttt{cobaya.run}. For the same user configurations, these two approaches are identical.}

\cloe further contains an example job submission script, \texttt{example\_mcmc\_script\_for\_cluster.sh}, which permits the user to execute a Monte Carlo run on a high performance computing cluster that uses the SLURM workload manager.

\subsection{Graphical user interface}
\label{sec3n}

We have created a graphical user interface (GUI) to assist users in running \cloe. This GUI targets beginner users in particular, by allowing them to rapidly produce \cloe-compatible configurations. To this end, the GUI is not designed to perform a Monte Carlo run, but rather allows the user to generate the configuration files needed for such a run. The GUI is also not meant to replace the configuration files presented in Sect.\ \ref{sec3a}, but rather provides an interactive way to guide users in the creation of these files.\footnote{We note that sampling platforms such as \cobaya contain their own GUI framework for the generation of the user settings desired for a parameter inference run (which can be executed from the shell via the command \texttt{cobaya-cosmo-generator} following an installation of \pyside). While the \cobaya GUI allows the user to choose between predefined datasets, such as \Planck \citep{planck18cosmo} and the Dark Energy Survey \citep{Desy3}, the \cloe GUI is focused on the creation of the \yaml files needed by its default configuration file, such as the choice of probes, redshifts, scales, and systematic uncertainties (along with the parameter priors in the case of both GUIs).}

Analogously to the \cobaya GUI, the \cloe GUI has been produced using the \python library \pyqt\footnote{\pyqtaddress}. This choice is motivated by its wide multi-platform support, as it can be used on \linux, \mac, and \windows systems, and for its ease in extending and personalizing the GUI.

The GUI is activated by executing \texttt{gui/script\_gui.py}.
As seen in Fig.\ \ref{guifig}, the GUI contains different sections and subsections. These are related to choices in, for instance, the observables to include and their scale cuts, the cosmological parameters to sample and their priors, the nonlinear modeling, and other analysis choices such as the inclusion of the BNT transformation and non-Gaussian terms to the covariance.

The primary limitation of the GUI is that it provides less fine-grained control over \cloe options compared to the direct usage of the configuration files. As a result, more advanced users will likely prefer to either directly modify the example configuration files that we have provided or create entirely new files themselves. Users also have the possibility to modify the GUI-generated configuration files at the command line. Beyond this limitation, maintaining the GUI so that it remains aligned with the underlying configuration structure introduces additional development overhead.

\subsection{Continuous integration}
\label{sec3q}

We set up a series of continuous integration (CI) and continuous delivery (CD) jobs on the \gitlab repository. This was managed through a series of instructions in a dedicated \yaml file (\texttt{.gitlab-ci.yml}).\footnote{\url{https://docs.gitlab.com/ee/ci}}

In this regard, CI refers to the process of regularly testing software to minimise the introduction of bugs and to quickly identify breaking changes to the code. This is good practice for any project but is particularly pertinent for software developed by large groups of people. Meanwhile, CD refers to the process of regularly providing products to users. This is good practice for ensuring that users have quick access to bug fixes and new features and any corresponding documentation. 

We set up a total of four CI jobs. The first job automatically triggers the unit tests described in Sect.~\ref{sec3l} and provides a coverage report every time a commit is pushed to an open merge request. The second job builds the code documentation described in Sect.~\ref{sec3s} on request (i.e., the job is manually triggered) and saves this as an artifact for review.\footnote{\url{https://docs.gitlab.com/ee/ci/jobs/job_artifacts.html}} The third job generates a profiling report on request by running \texttt{cProfile}\footnote{\url{https://docs.python.org/3/library/profile.html}} on the top level script \texttt{run\_cloe.py}, and saves the output as an artifact for review. This report can be visualized with a viewer, such as \snakeviz\footnote{\snakevizaddress}, and is used to assess the overall speed of the code and identify potential bottlenecks. Lastly, the fourth job runs a series of end-to-end verification tests in \texttt{cloe/tests/verification} on request. These tests, namely of the user interface (see Sect.~\ref{sec3a}) and the DEMO notebook (see Appendix~\ref{appnote}), cover a larger scope than individual code units and help determine if the software as a whole is working as expected.

We further set up a single CD job that automatically deploys the code documentation as an \html website. This job is triggered every time the development branch is merged into the main branch. This is only done to release new versions of the software. Hence, every new release of the software automatically includes the latest version of the code documentation.

\subsection{\docker image}
\label{sec3r}

We generate dedicated \docker\footnote{\url{https://www.docker.com}} \citep{merkel2014docker} images for \cloe to facilitate code testing and maximise the reproducibility of the results. The main \texttt{Dockerfile}, which defines the build instructions for the image, uses an \ubuntu v18.04.6 base image. This particular choice of the base image and the corresponding version of the \ubuntu operating system was made to ensure compatibility with the European Space Agency (ESA) Datalabs platform \citep{navarro2021}. On top of this base, we build a \miniconda\footnote{\url{https://docs.anaconda.com/free/miniconda/index.html}} layer. Finally, we provide instructions for building the \cloe \conda environment (see Appendix~\ref{sec3p}).

The resulting \docker image is uploaded to the \gitlab Container Registry to facilitate the execution of CI/CD jobs. The use of this image minimizes the overhead on the dedicated \gitlab servers for \Euclid, as unit tests are run in a containerized environment with all of the required dependencies pre-installed. This reduces the overall runtime of CI/CD jobs as the environment can take a long time to build given the long list of dependencies.

For each version release of \cloe, we build an additional \docker image for users. This image is built on top of the CI/CD image base described above and simply installs the tagged version of \cloe into the pre-built \conda environment. This image is itself tagged with the corresponding release version and is also uploaded to the \gitlab Container Registry to be made available to users. This user image can be deployed on cloud computing platforms, such as ESA Datalabs, minimizing the effort required to install the software and guaranteeing a good level of reproducibility of the results obtained from a given release of \cloe.

\subsection{Code documentation}
\label{sec3s}

We provide details on how to install and run \cloe in the online documentation of the code. We also provide examples of how to use \cloe and the expected outputs in \jupyternotebooks, as detailed in Appendix~\ref{appnote}.

Our \sphinx documentation exists in the \texttt{docs} directory, which includes numerous restructured text (\texttt{rst}) files along with the build configuration file. We have created a comprehensive set of \python docstrings, which adhere to the \sphinx \numpydoc conventions and refer to comments inside the code that can be converted into \html. This includes details on what a given method or class does, inputs and outputs for methods and classes, and possible errors raised under certain conditions.

\subsection{Code validation}
\label{secval}

As noted earlier, \cloe has undergone rigorous benchmarking. This is detailed in \citet{Martinelli24} and briefly summarized here. 

The benchmarking procedure of the photometric 3$\times$2pt observables entails a comparison of the theoretical predictions in both Fourier and configuration space from \cloe against independent external codes (such as \ccl, itself previously benchmarked against \cosmolss in \citealt{Chisari19}). The benchmarking is extended to intermediate quantities such as comoving distances, growth rates, window functions, and auto- and cross-power spectra of matter, galaxies, and IAs. Analogously, the benchmarking of the spectroscopic observables, namely the multipole power spectra and correlation functions, is with respect to the codes \pbj \citep{Carrilho23, Moretti23} and \coffee\footnote{\coffeeaddress}\href{https://github.com/JCGoran/coffe}{\color{black}\faGithub} \citep{coffee2018}, respectively. As core parts of \pbj are incorporated in \cloe for the computation of the nonlinear redshift-space galaxy power spectrum, we are in this case partly focused on \pbj's accurate incorporation relative to the standalone version of the code.

The benchmarking in \citet{Martinelli24} exhibits sub-percent level agreement for the intermediate quantities, and most importantly, illustrates that all summary statistics agree with the benchmarks to better than our {\it a priori} requirement of $10\%$ of the \Euclid error bars over all relevant scales and for a wide range of cosmological and astrophysical modeling choices, including the photometric redshift distributions, IAs, galaxy bias, and magnification bias. In many cases, the agreement between the codes is significantly higher, down to the level of $\sim10^{-4}\sigma_{\rm Euclid}$. It should further be noted that the uncertainties $\sigma_{\rm Euclid}$ here refer to those of the final \Euclid data release (DR3), such that the agreement between the codes relative to the first \Euclid data release (DR1) uncertainties is even higher. The benchmarking has been performed for \cloe version 2.0.2, and subsequently repeated and confirmed to hold for the latest version.\footnote{In addition to the benchmarking of the photometric 3$\times$2pt and spectroscopic galaxy clustering probes in \citet{Martinelli24}, the latest version of \cloe includes the  benchmarking of the cluster counts and CMB cross-correlations, as discussed in Euclid Collaboration:~Sakr et al. (in prep) and Sect.~\ref{cmbsec}, respectively. This benchmarking exists in \texttt{DEMO\_Clusters\_of\_Galaxies.ipynb} and \texttt{CMBX\_probes.ipynb} in the \texttt{notebooks} directory of \cloe.}

In addition to the benchmarking of the theory predictions, \citet{Martinelli24} has benchmarked the calculation of the Gaussian log-likelihood (equivalently represented by the effective $\chi^2$), finding that it is well within the required probability threshold. We obtain additional confidence in the performance of \cloe through Monte Carlo runs using synthetic data, already achieved in a series of \Euclid publications (\citealt{GCH24,Blot24,Goh24}; Euclid Collaboration:~Carrilho et al., in prep.; Euclid Collaboration:~Moretti et al., in prep.; Euclid Collaboration:~Sciotti et al., in prep.). Other aspects of likelihood code validation consist in performing a comparison of the Monte Carlo results from \cloe and benchmarking codes, either using synthetic data or as part of a reanalysis of available (Stage III) datasets. The analysis of simulated data (e.g., Flagship in \citealt{castander24}) would moreover allow for stress tests of the analysis choices, such as the modeling assumptions and choice of scales. Such analyses are beyond the scope of the present work but represent a natural next step.

\section{Additional \cloe features}
\label{sec4}

In this section, we describe unique features in CLOE, such as the Bernardeau--Nishimichi--Taruya (BNT;~\citealt{bnt14}) transformation for the cosmic shear and galaxy-galaxy lensing observables, the \fftlog algorithm for the multipole correlation function integrals and non-Limber calculation, and the likelihood for the CMB cross-correlations.

\subsection{Transformations with \fftlog}
\label{fftlogsec}

A variety of operations, such as the transformation of the galaxy power spectrum multipoles to the corresponding multipole correlation functions, involve integrals that contain spherical Bessel functions, $j_\nu(x)$, in the form
\begin{equation}
    F(r)=\int_0^{\infty} \frac{\diff k}{k} \; f(k) \, j_{\ell}(k r) \; .
    \label{eq:fftlog}
\end{equation}
The oscillatory behavior of the Bessel function makes this integral expensive to compute with general purpose methods such as adaptive Gaussian quadrature. In order to address this, specialized algorithms have been developed, where \fftlog is the one most commonly used in cosmology. This algorithm was originally proposed in \cite{Talman78} and  applied to cosmology in \cite{hamilton2000}. It allows for the computation of the fast Fourier or Hankel transform of a periodic sequence of logarithmically spaced points and was recently generalized to include $n^{\rm th}$-order derivatives of the Bessel function, $j_{\ell}^{(n)}(k r)$~\citep{Fang2019}:
\begin{equation}
    F_n(r)=\int_0^{\infty} \frac{\diff k}{k} \; f(k) \, j_{\ell}^{(n)}(k r) \; .
\end{equation}
This is the version of the algorithm that we have implemented in the \texttt{fftlog} subpackage of \cloe. While we refer the reader to the original papers for a technical discussion, we outline the main ingredients of the algorithm here. The crucial observation is that the integral
\begin{equation}
    \tilde{g}_{\ell}(n, z)=4 \pi^{-1 / 2} \int_0^{\infty} \diff k \; k^{z-1} j_{\ell}^{(n)}(k) \;
\end{equation}
has an analytical solution:\footnote{As in~\cite{Fang2019}, our \fftlog implementation includes up to the second derivative of the Bessel function.}
\begin{equation}
\begin{aligned}
    \tilde{g}_{\ell}(0, z)=2^z \frac{\Gamma\left(\frac{\ell+z}{2}\right)}{\Gamma\left(\frac{3+\ell-z}{2}\right)} \; , \\
    \tilde{g}_{\ell}(1, z)=-2^{z-1}(z-1) \frac{\Gamma\left(\frac{\ell+z-1}{2}\right)}{\Gamma\left(\frac{4+\ell-z}{2}\right)} \; , \\
    \tilde{g}_{\ell}(2, z)=2^{z-2}(z-1)\,(z-2) \frac{\Gamma\left(\frac{\ell+z-2}{2}\right)}{\Gamma\left(\frac{5+\ell-z}{2}\right)} \; .
\end{aligned}
\end{equation}
The clear advantage is that the oscillatory functions are dealt  with by analytical integrals, which are exact. In order to use \fftlog, we thus have to decompose the $f(k)$ function as a sum of power laws,
\begin{equation}
    f\left(k_q\right)=\frac{1}{N} \sum_{m=-N / 2}^{N / 2} c_m \, k_0^\nu \, \left(\frac{k_q}{k_0}\right)^{\nu+\mathrm{i} \eta_m} \; ,
\end{equation}
where $N$ is the number of elements in the input function array, $\eta_m=2 \pi m /\left(N \Delta_{\ln k}\right)$, $m$ is the summation index, $\nu$ is the bias index used to stabilize numerical instabilities, $\Delta_{\ln k}$ is the linear spacing in $\ln(k)$, and $k_0$ is the smallest wavenumber, $k$, considered for the input array. 

Once the coefficients of the decomposition are computed by means of a fast Fourier transform (FFT), the original integral can be computed with an inverse fast Fourier transform (IFFT):
\begin{equation}    
F_n\left(r_p\right) = \frac{\sqrt{\pi}}{4 r_p^\nu} \operatorname{IFFT}\left[c_m^* \left(k_0 r_0\right)^{\mathrm{i} \eta_m} \tilde{g}_{\ell}\left(n, \nu-\mathrm{i} \eta_m\right)\right] \; .
\end{equation}
Hence, \fftlog trades an oscillatory integral with a couple of FFTs. In our case, this leads to a speed-up relative to standard brute force methods (e.g., the \texttt{scipy.integrate.quad} method) of three orders of magnitude. As our numerical implementation closely follows that of \cite{Fang2019}, we have confirmed that it agrees with this original implementation along with adaptive integration algorithms at a high precision (fractional difference of $10^{-10}$ in the former case and at the required accuracy of the \texttt{quad} algorithm in the latter case).

A natural extension of this implementation is to use \fftlog to perform the non-Limber integration in a similar way to \citet{Fang2019}, found to be favored relative to other methods in the N5K non-Limber integration challenge of the LSST Dark Energy Science Collaboration \citep{Leonard2023}.

\subsection{Bernardeau--Nishimichi--Taruya transformation}
\label{bntsec}

The BNT transform (\citealt{bnt14}, also see \citealt{Taylor18, Taylor21, gu24}) allows for an amelioration of the sensitivity of weak gravitational lensing to nonlinear scales, in that they can more safely be isolated. This is a subtle challenge for weak lensing, as the shear kernel is broad in redshift, extending from the actual redshift of the source galaxies down to redshift zero of the observer.

As reviewed in \citet{Cardone24}, the nulling technique proposed by \citet{bnt14} consists in constructing a linear combination of the binned shear kernels that effectively localizes them in redshift, under the constraint of leaving the lowest redshift bin unaltered. In other words, if \(\tens M=\{M_{ii'}\}\) is the matrix responsible for the aforementioned linear combination, with the indices $i$ and $i'$ running over all redshift bins, the BNT transform of the shear ($\gamma$) kernels is simply \(W^{\gamma}_i(z)\to M_{ii'}\,W^{\gamma}_{i'}(z)\), where summation over equal indices is assumed. This implies that the BNT-transformed lensing power spectrum, $C^{\rm LL}(\ell)$, can be obtained via
\begin{equation}
    C^{\rm LL} \rightarrow {\tens M} {\tens C}^{\rm LL} {\tens M}^{\sf T} \; ,
    \label{eq:BNT_transformed_Cij_LL}
\end{equation}
where \(\tens C^{\rm LL}(\ell)=\{\cl{\ell}[ij][\rm LL]\}\). Note that the transformation is here also applied to the IA terms despite the IA kernel differing from that of the shear \citep{Taylor21}. Under the assumption that the shear kernel dependence on the cosmological parameters is negligible compared to that of the matter power spectrum, the BNT transform can be computed once in some reference cosmology,\footnote{This is much in the same way as in spectroscopic galaxy clustering, where a reference cosmology is used to translate redshifts and angular positions into comoving coordinates.} and then applied to both the measured and theoretical tomographic shear power spectra at each step of the Monte Carlo analysis. However, in our case, we need to take into account that \cloe directly outputs the vectorized power spectrum due to our use of the \(4\)-point covariance matrix. As a result, Eq.~(\ref{eq:BNT_transformed_Cij_LL}) needs to be modified accordingly. The rest of this section will be devoted to the description of our implementation, found in the \texttt{matrix\_transforms} module of the \texttt{auxiliary} subpackage.

Following standard practice, let us render vectors with boldface letters and matrices with sans-serif letters. A vectorized matrix is a vector constructed by stacking matrix columns on top of each other, and is hence expressed as e.g.\ \(\vec M\equiv{\rm vec}(\tens M)\). However, bear in mind that the matrices we are dealing with are tomographic power spectra (at fixed multipole) and any auto-correlation \(C^{AB}_{ij}(\ell)\) with \(A=B\) is a symmetric matrix. In such cases, only the upper or lower half of the symmetric matrix \(\tens S\) carries independent information. Therefore, the operation required is that of half-vectorization, which we denote \(\slashed{\vec S}\equiv{\rm vech}(\tens S)\). In other words, only the elements above/below, and including, the main diagonal are stacked to create the vectorized matrix, so that if the original matrix has \(N_z\) redshift bins, and thus size \(N_z\times N_z\), the resulting half-vectorized matrix will be a column vector of size \(N_z\,(N_z+1)/2\).

Lastly, to work out the final expression in the case of the full data vector -- namely, vectorized not only in the \(N_z\) redshift-bin indices but also in the \(N_\ell\) multipole bins -- we need to stack an \(N_\ell\)-long array of \(N_z\times N_z\) tomographic power spectrum matrices. Focusing now on the tomographic lensing power spectrum, we write \(\slashed{\vec C}^{\rm LL}_\ell={\rm vech}[\tens C^{\rm LL}(\ell)]\) and \(\slashed{\vec C}^{\rm LL}={\rm vec}(\slashed{\vec C}^{\rm LL}_\ell)\). The former is the half-vectorization of the tomographic matrix at fixed \(\ell\), while the latter is the further stacking over all binned multipoles, i.e.\ the output of \cloe. The BNT transform of this quantity is then obtained through
\begin{equation}
    \slashed{\vec C}^{\rm LL}\to({\tens 1}_{N_\ell}\otimes{\tens E}_{N_z})\,[{\tens 1}_{N_\ell}\otimes({\tens M}\otimes{\sf M})]\,({\tens 1}_{N_\ell}\otimes{\tens D}_{N_z})\,\slashed{\vec C}^{\rm LL} \; , 
    \label{eq:BNT_LL}
\end{equation}
where \({\tens 1}_n\) is the \(n \times n\) identity matrix, \(\otimes\) is the Kronecker product, \({\tens E}_n\) is the elimination matrix that turns the vectorization of an \(n \times n\) matrix into its half-vectorization, and \({\tens D}_n\) is the unique duplication matrix that turns the half-vectorization of an \(n \times n\) symmetric matrix into its vectorization.

In the case of the lensing-clustering cross-correlation, the BNT transform acts on only one of the two tracers, that of the lensing, leaving the galaxy clustering unmodified. In the standard and vectorized cases this corresponds, respectively, to
\begin{align}
    \cl{\ell}[ij][{\rm L}{\rm G}]&\to M_{ii'}\,\cl{\ell}[i'j][{\rm L}{\rm G}]\;,\label{eq:BNT_transformed_Cij_LG}\\
    \slashed{\vec C}^{{\rm L}{\rm G}}&\to[{\tens 1}_{N_\ell}\otimes({\tens M}\otimes{\tens 1}_{N_\ell})]\,\slashed{\vec C}^{{\rm L}{\rm G}} \; , 
    \label{eq:BNT_LG}
\end{align}
where ${\rm G}$ refers to the galaxy tracer, while \(\slashed{\vec C}^{{\rm L}{\rm G}}_\ell={\rm vech}[\tens C^{{\rm L}{\rm G}}(\ell)]\) and \(\slashed{\vec C}^{{\rm L}{\rm G}}={\rm vec}(\slashed{\vec C}^{{\rm L}{\rm G}}_\ell)\). The BNT covariance follows an analogous approach, as described in \citet{Vazsonyi21}. Further details on the derivation of the BNT-transformed observables (Eqs.~\ref{eq:BNT_LL} and \ref{eq:BNT_LG}) and the elimination and duplication matrices are given in Appendix~\ref{eq:BNT_appendix}.

\subsection{CMB cross-correlations}
\label{cmbsec}

We have extended CLOE to include cross-correlations of the CMB temperature and  gravitational lensing with the \Euclid photometric tracers. Concretely, this includes the cross-correlation of photometric galaxy positions and CMB lensing potential, the cross-correlation of galaxy shear and CMB lensing potential, and the large-scale cross-correlation of photometric galaxy positions and CMB temperature anisotropies targeting the integrated Sachs-Wolfe (ISW) effect. In addition, this includes the CMB lensing potential power spectrum.

These auto- and cross-correlations are sensitive to the expansion and growth histories of the Universe, and thereby provide an increase in the statistical information content.  The additional observables allow for the breaking of degeneracies between parameters describing the underlying cosmology and systematic uncertainties when analyzing \Euclid data alone. 
We refer to \citet{Euclid:2021qvm} for a detailed \Euclid forecast with these cross-correlations in different cosmological models.

\cloe's \texttt{cmbx} module allows for the computation of the theoretical predictions of the aforementioned observables, and can be used together with both the \cobaya and \cosmosis interfaces and with both the \camb and \class Boltzmann solvers. As illustrated in \texttt{CMBX\_Probes.ipynb} in the \texttt{notebooks} directory of \cloe, these theoretical predictions have been benchmarked against \camb to a precision that is well below the requirement of $0.1$ standard deviations relative to the expected uncertainties on the observables in a prospective \Euclid$\times$~Simons Observatory analysis. 

The theoretical predictions of the CMB observables, along with the corresponding data vectors and covariance obtained from the \texttt{reader} module, enter a joint likelihood with the \Euclid primary probes in the \texttt{euclike} module. This builds on the photometric 3$\times$2pt analysis to reach a combined 7$\times$2pt analysis, where the vector of observables can be expressed as
\begin{equation}
    \mathcal{O}(\ell) = \left\{C_{ij}^{\rm{GG}}, C_{ij}^{\rm{GL}}, C_{ij}^{\rm{LL}}, C_i^{\rm{GT}},  C_i^{\rm{G}\phi}, C_i^{\rm{L}\phi}, C^{\phi\phi} \right\}(\ell) \, .
\label{cmbeqn}
\end{equation}
Here, ${\rm G}$ refers to photometric galaxy positions, ${\rm L}$ to galaxy lensing, ${\rm T}$ to CMB temperature, $\phi$ to CMB lensing, and $\{i, j\}$ to the tomographic bin indices. As a result, the terms on the right-hand side of the above equation correspond, respectively, to the photometric galaxy power spectrum, cosmic shear power spectrum, galaxy-galaxy lensing power spectrum, galaxy-CMB temperature cross-spectrum, galaxy-CMB lensing cross-spectrum, galaxy lensing-CMB lensing cross-spectrum, and CMB lensing power spectrum (see \citealt{Euclid:2021qvm} for their theoretical expressions).

The user is able to select the additional probes following the same triangular structure as described in Sect.~\ref{sec3b}. The file \texttt{observables\_selection.yaml} accepts the following added options:\\
\begin{figure}[h]
\vspace{-1.0em}
\includegraphics[width=0.685\linewidth, trim = {-1cm 1.7cm 9cm 1.7cm}]{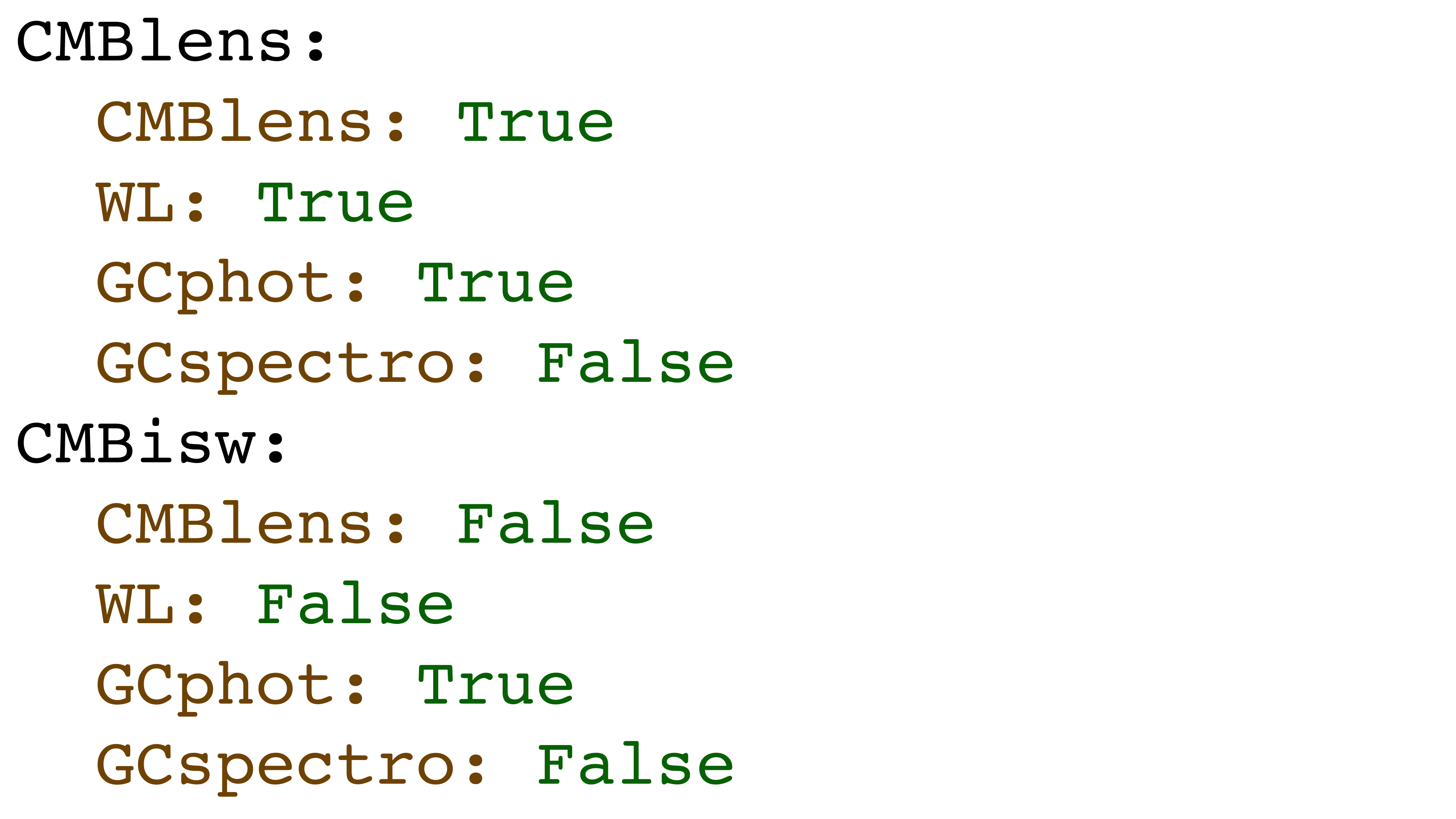}
\end{figure}
\vspace{-0.4em}
\\ \indent In this example, the user chooses to add the four latter angular power spectra in Eq.~(\ref{cmbeqn}) to a given analysis. In doing so, \cloe loads the 7$\times$2pt data vector and theoretical covariance matrix. The scale cuts are correspondingly specified via the files \texttt{CMBlensing.yaml}, \texttt{CMBlensing-GCphot.yaml}, \texttt{CMBlensing-WL.yaml}, and \texttt{iSW-GCphot.yaml}. For the chosen probes, \cloe then implements the same masking procedure for the measurements, covariance, and theory vector as described in Sect.~\ref{sec3i}. We note that the cross-correlations between the CMB and spectroscopic galaxy positions, between the CMB temperature and CMB lensing, and between the CMB temperature and galaxy lensing are not yet implemented in \cloe. 

The implementation of the CMB cross-correlations closely follows the baseline \cloe structure. To this end, the module \texttt{cmbx} contains the class \texttt{CMBX}, which inherits from the \texttt{Photo} class described in Sect.~\ref{sec3e}. This \texttt{CMBX} class contains the window functions of the CMB lensing and temperature anisotropies (restricted to the ISW effect), as described in \citet{Euclid:2021qvm}. The cross-spectra are then obtained by multiplying a given CMB window function with the corresponding window of the photometric tracer, and integrating the product following the same prescription as for the 3$\times$2pt observables in the \texttt{Photo} class. As a result, the 7$\times$2pt observables in Eq.~(\ref{cmbeqn}) are computed consistently relative to one another. This includes incorporating intrinsic galaxy alignments in the computation of the galaxy lensing - CMB lensing cross-spectrum, and magnification bias in the computation of the galaxy-CMB lensing and galaxy-temperature cross-spectra. We further note that the CMB cross-spectra employ the tomographic and multipole binning of the photometric probes.

As the user selects to include CMB lensing in the analysis, the \cobaya and \cosmosis interfaces request the matter power spectrum and expansion history from either \camb or \class across a broader range of redshifts up to the recombination epoch. Other CMB-related modifications to \cloe include the reading and formatting of the additional measurements and 7$\times$2pt covariance, along with the creation of the corresponding masking vector and CMB-related keys, in the \texttt{reader} and \texttt{data\_handler} modules. The \texttt{euclike} module then includes the methods \texttt{create\_photoxcmb\_data} and \texttt{create\_photoxcmb\_theory}, which create the masked data vector, covariance, and theory vector that enter the 7$\times$2pt likelihood calculation. 
To this end, the lowest multipole that enters the cosmological analysis with the CMB cross-correlations will need to be accounted for to avoid non-Gaussianity of the likelihood.

In \texttt{cloe/tests/test\_cmbx.py}, a unit test suite is included for the CMB probes. Moreover, in the \texttt{scripts} directory, \texttt{make\_cmbx\_data.py} generates a set of synthetic $C(\ell)$ to be used for testing purposes, and \texttt{make\_CMBX\_covmat.py} generates a Gaussian covariance matrix for \Euclid combined with the Simons Observatory \citep{SimonsObservatory:2018koc}\footnote{\url{https://github.com/simonsobs/so_noise_models}}. Moving forward, the covariance needs to account for realistic masks and non-Gaussianity, noting that it currently only includes a sky fraction ($f_{\rm sky}$) correction. In addition, as previously mentioned, in the \texttt{notebooks} directory, we provide the \jupyternotebook \texttt{CMBX\_probes.ipynb}, which computes and plots the theoretical predictions for the CMB probes and moreover illustrates the stringent benchmarking of \cloe for these calculations against \camb.

\subsection{Additional features}
\label{extendedsec}

There are additional features in \cloe that will be described in greater detail in forthcoming publications. This includes the emulators of the nonlinear matter power spectrum (\bacco and \eecode), the emulators of baryonic feedback (\bacco and \bcemu), the TATT model of intrinsic galaxy alignments, the one-loop perturbation theory model for the nonlinear photometric galaxy bias, and the \eftoflss implementation for the nonlinear redshift-space galaxy power spectrum, which will be described in Euclid Collaboration:~Carrilho et al. (in prep.) and Euclid Collaboration:~Moretti et al. (in prep.). 

\cloe further includes the computation of the theoretical predictions and likelihood for cluster number counts, which will be described in Euclid Collaboration:~Sakr et al. (in prep.).\footnote{In Euclid Collaboration:~Sakr et al. (in prep.), we will also describe the implementation of the likelihood for cluster weak lensing and cluster clustering, which currently exist in a separate (but not main) branch of the \cloe repository.} Moreover, \cloe includes the capability to account for the impact of magnification bias on the multipole correlation functions and the possibility to consider an effective Weyl power spectrum instead of the matter power spectrum, which are described in \citet{Goh24}. As modified gravity can both break the identity between the metric potentials and their relation to the density contrast, the ability of \cloe to directly consider the Weyl power spectrum paves the way for the propagation of modified gravity into the theoretical predictions.

\section{Computational speed}
\label{speedsec}

A core aspect of a parameter inference pipeline is not only its features and capabilities, but the speed with which it is able to compute the theoretical predictions and thereby likelihood. This likelihood-sampling speed is in turn strongly correlated with the speed with which a Monte Carlo run is completed. As future cosmological analyses will require the sampling of the likelihood over an ever-growing number of parameters, optimizing the computational speed of the code is therefore crucial for the Monte Carlo runs to reach convergence given a reasonable amount of time and computational resources. 

To this end, in the development of \cloe, we placed particular emphasis on optimizing its computational speed. This was further motivated by the early decision to write the code entirely in \python, which is intrinsically slower than low-level languages such as \ccode and \fortran.\footnote{Early in the development, we also opted against the use of \textsc{Cython} and \textsc{Numba} after finding marginal speed improvement in the former case and incompatibility with distinct \scipy functions in the latter case (noting that {numba-scipy} only extends to {scipy.special} functions).} The optimization of \cloe includes the consistent employment of broadcasting and vectorization of the code, along with efficient integration techniques as described in earlier Sects.~\ref{sec3e2} and \ref{sec3f2}. Meanwhile, due to \python's global interpreter lock (GIL), \cloe currently does not utilize open multi-processing (OpenMP) parallelization.\footnote{We note that \python 3.13 has an experimental option to disable the GIL and that multithreading should be straightforward in the future. A further caveat is that when \cloe is used together with \camb, the Boltzmann calculation and thereby overall likelihood evaluation does partially benefit from \openmp parallelization, as seen in Table~\ref{tabspeed}.} This can be contrasted against other codes such as \ccl, which parallelizes the computation of the angular power spectra at the level of the multipoles in \ccode. This allows likelihood calculations with \firecrown + \ccl to be sped up with an increasing number of threads.

In this regard, given a fixed amount of computational resources, we note that the usefulness of OpenMP parallelization depends on the choice of Monte Carlo sampler. While it is crucial for MCMC algorithms such as Metropolis--Hastings, it is largely unnecessary for a suite of other sampling algorithms such as the affine invariant ensemble MCMC sampler of \emcee and the nested samplers of \polychord and \nautilus. We moreover stress that OpenMP is distinct from  Message Passing Interface (MPI), where the former refers to thread-based parallelism and the latter to process-based parallelism. In the context of MCMC runs, MPI does not require any changes to the code and allows for the creation of multiple chains that are run in parallel, while OpenMP allows for the use of multiple threads in the computation of the likelihood for each chain. Hence, while \cloe is not OpenMP parallelized, it does allow for faster convergence of Monte Carlo runs via MPI parallelization. For these reasons among others, in \citet{GCH24}, \citet{Goh24}, and \citet{Blot24}, we have used \cloe together with the \nautilus sampler to perform the forecasted parameter inferences.

\subsection{The speed of \cloe}
\label{speedsecsub1}

In Table~\ref{tabspeed}, we provide the computational speed of \cloe for different probe combinations and analysis setups. Here, the computational speed refers to the amount of time \cloe uses to compute the likelihood at each point in parameter space. This is in turn strongly correlated with 
the amount of time required for a Monte Carlo run to reach convergence (which naturally also has a dependence on the sampling method). In quoting the likelihood-sampling speed, we neglect the amount of time spent to initialize the code, which occurs only once in a given Monte Carlo run and in the case of \cloe takes around one half of a minute (i.e., it is negligible as compared to the amount of time required for a full run).\footnote{As further clarification, the initialization of the code refers to parts of the likelihood calculation that are only executed once, such as the reading in of the measurements, covariances, mixing matrices, and redshift distributions. In contrast, the likelihood-sampling phase refers to the computation of the theoretical predictions and subsequently the likelihood at each point in parameter space. In other words, the theory predictions need to be recomputed at each step of the Monte Carlo run, while the data products remain fixed. In this regard, the impact of a cosmology-dependent covariance is commonly accounted for by iteratively performing Monte Carlo runs using the covariance recomputed at the best-fit cosmology of a previous run (e g. \citealt{Eifler09, Heymans21, Desy3}), 
and is demonstrated to be negligible in analyses of Stage IV photometric surveys \citep{kodwani19}.} 

In assessing the computational speed, we utilize a single core of an Ice Lake node with Intel Xeon Platinum 8368Q $2.60$ GHz processors. In addition, for the cases that support multithreading, we list the computational speed-up with increasing number of cores. Fiducially, the theory computations include our nonlinear galaxy power spectrum modeling (both photometric and spectroscopic), nonlinear matter power spectrum modeling with baryonic feedback using \hmcode, tomographic bin-dependent magnification bias, and the NLA model of intrinsic galaxy alignments. We additionally illustrate the impact of two representative changes in the fiducial setup, in the form of linear galaxy power spectrum modeling and the TATT model of IAs.

\begin{table*}
\begin{center}
\caption{Computational speed of \cloe for different probe combinations and analysis setups.}
\begin{tabularx}{\textwidth}{l|c|c|C}
\toprule
Code & Probe combination & No. cores & Time (s) \\
\midrule
\cloe &  3$\times$2pt & 1 & 0.80 \\
\cloe (linear galaxy bias) &  3$\times$2pt & 1 & 0.26 \\
\cloe (tidal alignment and tidal torquing) &  3$\times$2pt & 1 & 1.7 \\
\cloe (linear galaxy bias, $N_k = 200$, $N_z = 150$) &  3$\times$2pt & 1 & 0.23 \\
\cloe & 3$\times$2pt + Spectroscopic clustering & 1 & 1.7 \\
\cloe (linear galaxy bias) & 3$\times$2pt + Spectroscopic clustering & 1 & 0.77 \\
\cloe & 7$\times$2pt + Spectroscopic clustering & 1 & 1.8 \\
\midrule
\firecrown w/ \ccl (linear galaxy bias, $N_k = 200$, $N_z = 150$) &  3$\times$2pt & 1 & 6.4 \\
\firecrown w/ \ccl (linear galaxy bias, $N_k = 200$, $N_z = 150$) &  3$\times$2pt & 8 & 5.4 \\
\firecrown w/ \ccl (linear galaxy bias, $N_k = 200$, $N_z = 150$) &  3$\times$2pt & 16 & 4.6 \\
\midrule
\camb & Growth \& expansion histories & 1 &  6.5 \\
\camb & Growth \& expansion histories  & 8 & 1.4 \\
\camb & Growth \& expansion histories  & 16 & 1.1 \\
\class & Growth \& expansion histories & 1 & 8.7 \\
\bottomrule
\end{tabularx}
\label{tabspeed}
\end{center}
\vspace{-1.2em}
\tablefoot{Computational speed for a single likelihood evaluation on an Ice Lake node with two Intel Xeon Platinum 8368Q $2.60$ GHz processors, each with 38 cores. By default, in reporting the computational speed, we only utilize a single core of this node. However, for the cases where we found a noticeable change in speed from multithreading, we also list the computational speed-up with increasing number of cores. Note that we have separated out the time required for the computation of the growth and expansion histories by the Boltzmann code (i.e., quantities such as the linear matter power spectrum, expansion rate, and growth rate) from the rest of the likelihood calculation. In our nomenclature, the 3$\times$2pt probe combination refers to photometric galaxies alone (Sect.~\ref{sec3e}), while the 7$\times$2pt probe combination further includes the cross-correlations with the CMB (Sect.~\ref{cmbsec}). In the comparison of \cloe and \firecrown + \ccl, we have homogenized their analysis setups, which includes using the same synthetic data products and modeling of systematic uncertainties.}
\end{table*}

The likelihood is computed for a synthetic \Euclid survey setup. For the photometric 3$\times$2pt probes, we take this to entail $10$ tomographic bins across $z \in [0, 4]$ and $20$ multipole bins with $\ell \in [10, 5000]$. For the spectroscopic galaxy clustering, we take it to entail $4$ redshift bins across $z \in [0.9, 1.8]$ and $125$ bins in wavenumber with $k \in [0.004, 0.5] \, h \, {\rm Mpc}^{-1}$. Naturally, the computational time has some dependence on these exact configurations. For instance, for a \Euclid DR1 survey setup with only six tomographic bins, there would be a factor of $2.6$ fewer tomographic bin combinations for which the angular power spectra would need to be computed (however, note that this does not imply that the code would be a factor of $2.6$ faster, as elucidated below). We moreover note that distinct scale cuts will not meaningfully affect \cloe's computational time, as the masking procedure described in Sect.~\ref{sec3i} is currently applied after the unmasked theory vector has been computed. Avoiding the calculation of masked out terms of the theory vector would therefore constitute a natural speed optimization.

The representative cases that are considered include the photometric 3$\times$2pt probe combination on its own (Sect.~\ref{sec3e}) and together with spectroscopic galaxy clustering (Sect.~\ref{sec3f}) and CMB cross-correlations (Sect.~\ref{cmbsec}). In all cases, we restrict this assessment to the Fourier-space observables (i.e., angular power spectra and multipole power spectra), which form the basis of the other summary statistics derived from them. We moreover compare the performance of \cloe against \firecrown + \ccl, and in both cases separate out the time spent to compute the linear matter power spectrum and background expansion with either \camb or \class from the rest of the likelihood calculation. For the computational speed of the cluster probes, we refer to Euclid Collaboration:~Sakr et al. (in prep.). 

Given the analysis framework described, \cloe calculates the likelihood in $0.80$ seconds for the photometric 3$\times$2pt probe combination. Most of this computational time is spent inside the nonlinear module, which extends the linear galaxy power spectrum to nonlinear scales, rather than for instance the integrations to obtain the lensing kernels and angular power spectra. This is evident by the more than a factor of three decrease in the computational time to $0.26$ seconds when considering a linear galaxy bias instead. In addition, the computational time increases by more than a factor of two to $1.7$ seconds when considering the TATT model in lieu of the NLA model. 

Once spectroscopic galaxy clustering is further included, the computational time increases from $0.80$ seconds to $1.7$ seconds in the fiducial setup (and from $1.7$ to $2.6$ seconds when NLA $\rightarrow$ TATT). As before, when considering linear galaxy bias modeling for both the photometric and spectroscopic probes, this decreases down to $0.77$ seconds. These computational times identify the nonlinear modeling as the target for where the most significant speed improvements can be obtained, in particular on the photometric side (noting that the linear contribution to the computational time will be even lower in \Euclid's DR1 setup with six rather than ten tomographic bins). Lastly, when further including the CMB cross-correlations as part of a 7$\times$2pt + spectroscopic galaxy clustering analysis, the computational time only increases by $0.1$ seconds.

\subsection{Comparison with other Stage IV survey codes}
\label{speedsecsub2}
 
Next, we perform a comparison of \cloe's computational times with those obtained from \firecrown + \ccl. We begin by emphasizing that this comparison pertains to the specific code versions used in our analysis.\footnote{In this comparison, we are considering the most recent versions of the codes at the time of writing this manuscript (\cloe post-v2.1, \firecrown v1.10, and \ccl v3.2).} As \firecrown + \ccl is not designed for redshift-space galaxy clustering and is currently limited in terms of its nonlinear galaxy bias modeling, we restrict this comparison to the photometric 3$\times$2pt probe combination with linear galaxy bias. In the comparison of \cloe and \firecrown + \ccl, we have homogenized their analysis setups, which includes using the same synthetic measurements, covariance, and redshift distributions. This is achieved by creating a single file in \sacc\footnote{\saccaddress}\href{https://github.com/LSSTDESC/sacc}{\color{black}\faGithub} \citep{citesacc} format that contains these data products, which is then read in by \firecrown. We have also homogenized their treatment of systematic uncertainties, in particular as pertains to the nonlinear matter power spectrum with baryonic feedback from \hmcode, intrinsic galaxy alignments using the NLA model, and linear galaxy bias treatment.

In performing the comparison of \cloe and \firecrown + \ccl, one needs to be cognizant that there is generally a trade-off between computational speed and precision. In this regard, \cloe has performed its benchmarking of the theoretical predictions to better than $10\%$ of the \Euclid DR3 error bars \citep{Martinelli24} and \ccl has performed its benchmarking to better than $10\%$ of the cosmic variance limit \citep{Chisari19}. While the benchmarking of \cloe has been performed against \ccl itself, this does imply that there is potentially scope for further homogenization of their accuracy settings. Mindful of this caveat, we have employed \cloe and \firecrown + \ccl with their respective fiducial accuracy settings. Accordingly, we find that $\firecrown + \ccl$ performs the likelihood calculation in $6.4$ seconds, as compared to \cloe's $0.26$ seconds. As the \firecrown + \ccl calculation can be performed with multiple threads, its speed improves to $5.4$ seconds with $8$ threads and $4.6$ seconds with $16$ threads (utilizing one core for each thread). 

Beyond 16 threads, there is a saturation in the speed improvement of \firecrown + \ccl, such that for instance in the case of $76$ threads, the computational time is $4.5$ seconds (where the actual decrease in time is only $0.04$ seconds). This saturation is driven by the fact that the calculations of the window functions are not yet parallelized in \ccl, which introduces an effective floor to the computational time. In this regard, we note that for instance the Stage III survey code \cosmolss, written in \fortran, found it necessary to \openmp parallelize multiple levels of the angular power spectrum and correlation function calculations for a more optimal computational speed improvement with an increasing number of threads. This includes the calculation of the window and non-window components of the angular power spectrum integrands, the angular power spectrum integrations themselves, and the interpolation of the angular power spectra for the subsequent correlation function integrations. A corresponding multilayer parallelization of \ccl would allow it to more significantly benefit from multithreading.

One of the most important accuracy settings that can be tuned is the density of wavenumbers and redshifts that govern the calculation of the angular power spectra. In the case of \cloe, the matter power spectrum is generated for $N_k = 1000$ wavenumbers and $N_z = 100$ redshifts, and in the case of \firecrown + \ccl it is generated with a density of $N_k = 200$ and $N_z = 150$ (in both cases up to redshift $z=4.0$ and wavenumber $k=10.0 \, h \, {\rm Mpc}^{-1}$, with an extrapolation beyond the maximum wavenumber). When updating the density of redshifts and wavenumbers in \cloe to match that of \firecrown + \ccl, the likelihood calculation speeds up from $0.26$ seconds to $0.23$ seconds. We further note that the fiducial density of redshifts and wavenumbers is different in standalone \ccl (relative to \firecrown + \ccl), taking on the density $N_k = 167$ and $N_z = 250$. Employing \cloe with this updated setting would change the computational time from $0.26$ seconds to $0.25$ seconds. In other words, for the representative cases considered, \cloe currently seems to compute the photometric 3$\times$2pt likelihood between $18$ to $28$ times more rapidly than \firecrown + \ccl.\footnote{As an aside, we note that \cloe's default integration settings are intentionally conservative, with $400$ source redshifts and $1000$ lens redshifts for the photometric probes, $2001$ values of the cosine of the line-of-sight angle for the spectroscopic galaxy clustering, as well as $100$ redshifts and $1000$ wavenumbers of the matter power spectrum. If these settings are relaxed to less conservative values such as $100$ each (except for $N_k=200$), the runtime decreases to $0.06$ seconds for the 3$\times2$pt probes and $0.2$ seconds for 3$\times2$pt + spectroscopic clustering in the absence of nonlinear biasing, corresponding to a factor of $4$ to $5$ speed-up.}

We will end this section with two important caveats. The first caveat is that the comparison of \cloe with \firecrown + \ccl is for a configuration of systematic uncertainties that includes linear galaxy bias. In any future realistic Stage IV survey analysis, the nonlinearities in the galaxy bias will need to be incorporated. As previously noted and seen in Table~\ref{tabspeed}, the likelihood computation time increases from $0.26$ seconds to $0.80$ seconds when allowing for nonlinear galaxy bias in \cloe. If we therefore in an approximate way consider this additional $0.54$ seconds of computational time for both \cloe and \firecrown + \ccl, their computational speeds would only differ by a factor between $6$ to $9$ (as compared to a factor between $18$ to $28$ in the case of linear galaxy bias).\footnote{In addition, if we consider a further $0.9$ seconds of computational time for both \cloe and \firecrown + \ccl from using the TATT model in lieu of the NLA model, their computational speeds would only differ by a factor between $4$ to $5$.}

The second caveat is that, as noted earlier, we have compared the computational speeds of \cloe and \firecrown + \ccl after both codes have received the linear matter power spectrum and background expansion history from the Boltzmann solver (\camb/\class). In any Monte Carlo run, the additional computational time from the Boltzmann solver would need to be taken into account. Given the computational time of \camb of $6.5$ seconds and \class of $8.7$ seconds in the single-thread scenario, either of these codes act as the speed bottleneck relative to the rest of the likelihood calculation by \cloe or \firecrown + \ccl. In the single-thread scenario, this implies that \cloe + \camb is only $40$--$50\%$ faster than \firecrown + \ccl + \camb, while in a $16$-thread scenario this increases to 
between a factor of $3$ to $4$ depending on the specific linear/nonlinear configuration considered (likewise for $\class$). Note that this second caveat becomes largely immaterial when considering an emulator of the linear matter power spectrum for which the computational time decreases to below $0.1$ seconds (see Sect.~\ref{sec7a} for further discussion on emulators).

\section{Future developments}
\label{sec7}

\subsection{Additional features}
\label{sec7a}

While \cloe already includes a substantial set of features, as summarized in Table~\ref{tabfeat}, additional functionalities are relevant for analyses associated with the first \Euclid\ data release (DR1). This includes the non-Limber computation of the photometric 3$\times$2pt observables on large angular scales (e.g.,~\citealt{Campagne2017, Assassi2017, Fang2019, Chiarenza2024}) and an expansion in the number of photometric summary statistics, such that it also includes COSEBIs \citep{Schneider2010, Asgari2012} and angular power spectrum bandpowers \citep{Schneider02, Uitert18}. 

The impact of additional systematic uncertainties needs to be included, such as corrections due to the reduced shear \citep{deshpande20}, nonlocal contributions to the galaxy-galaxy lensing signal \citep{maccrann20}, and the mitigation of uncertainties in the widths of the lens redshift distributions (e.g.,~\citealt{Desy3}). Further methods for mitigating the existing systematic uncertainties are also needed. These include, for instance, the halo model prescription for the intrinsic galaxy alignments \citep{fortuna21} and the \flamingo emulator for baryonic feedback \citep{Schaye2023, Schaller2024}. They moreover include additional prescriptions for mitigating the photometric redshift uncertainties, such as the \textsc{Hyperrank} \citep{Cordero22} and resampling methods \citep{Hildebrandt17, Zhang23}, as well as the Gaussian mixture model used in past Stage III survey analyses \citep{Stolzner21, Johnston2024}.

On the spectroscopic galaxy clustering side, the impact of interloper contamination needs to be propagated (e.g.,~\citealt{pullen16, addison19, massara20, foroozan22}) and the likelihood of the baryon acoustic oscillations (BAOs) needs to be constructed. The BAOs can be analyzed both on their own and in combination with the existing full-shape galaxy clustering observables, enabling a unified treatment that incorporates their full covariance (e.g.,~\citealt{philcox20, gm22, desifs}).

In improving the computational speed of the theoretical calculations, the use of emulators is increasingly important. One of the most important aspects in this regard is the emulation of the linear matter power spectrum in lieu of the full calculation with \camb or \class (e.g.,~\bacco in \citealt{arico21}, \cosmopower in \citealt{mancini21}, and symbolic emulator in \citealt{bartlett24}). In the context of \cloe, this is considered together with the emulation of \hmcode in \citet{Blot24} and Euclid Collaboration:~Sciotti et al. (in prep.). When employing such emulators, particular attention must be paid to the size of the parameter space on which they are trained and to increasing the ranges of validity of the training sets.

One additional approach to improve the computational speed of the cosmological parameter inference is to perform an analytical marginalization of the nuisance parameters, which is exact for the spectroscopic galaxy clustering (e.g.,~\citealt{amico20, Ivanov20, mccann23}) and approximate for the photometric probes such as cosmic shear \citep{Ruiz-Zapatero23, Hadzhiyska23}. This is particularly important to reduce the size of the parameter space, which is otherwise substantially larger for Stage IV surveys as compared to past Stage III analyses (as previously highlighted in Sects.~\ref{sec3e1} and \ref{sec3f1}).

A range of additional features are relevant for \cloe in connection with work across multiple \Euclid science working groups. This primarily includes features pertaining to the galaxy clusters (in particular the cluster lensing profile and correlation function; Euclid Collaboration:~Sakr et al., in prep.), CMB cross-correlations, beyond-$\Lambda$CDM extensions, and higher-order weak lensing and galaxy clustering statistics. On a longer time scale, in particular as the combined analyses of \Euclid and overlapping spectroscopic surveys such as the Dark Energy Spectroscopic Instrument \citep{desi2024} and the 4-metre Multi-Object Spectroscopic Telescope \citep{4most19} are considered, it will be important to include the cross-clustering between the photometric and spectroscopic galaxy positions, along with galaxy-galaxy lensing with the spectroscopic galaxies acting as the lenses. These two additional correlations can also be considered for \Euclid alone, but are so far neglected due to the limited redshift overlap between the photometric and spectroscopic galaxy samples. Their inclusion would enable \Euclid to perform a popularly termed 6$\times$2pt analysis \citep{Johnston2024}, either on its own or together with the overlapping spectroscopic galaxy surveys.

\subsection{Code assumptions}
\label{sec7b}

As \Euclid and other Stage IV surveys prepare for their forthcoming cosmological inferences, it is important to safeguard against the impact of analysis assumptions. This includes safeguarding against a given treatment of a systematic uncertainty by having multiple competing prescriptions to choose from. \cloe has, for instance, multiple approaches to account for the impact of baryonic feedback on the matter power spectrum, via the prescriptions of \hmcode, \bacco, and \bcemu. However, due to the limitations in the accuracy of the hydrodynamical simulations or external astrophysical measurements, one will need to propagate the intrinsic uncertainty in these modeling approaches (see e.g.~Euclid Collaboration:~Carrilho et al., in prep.).

Another example pertains to the propagation of the photometric redshift distributions in the parameter inference code, where the existing and planned prescriptions include a histogram of values across redshift for each tomographic bin. Given the finite density of redshifts composing each histogram, an interpolation is created in \cloe. In this regard, the default choice to perform a cubic spline interpolation of the redshift distributions, as compared to other possible choices such as a linear interpolation or a non-interpolating approach (i.e.,~integrating over the histogram directly in the computation of the photometric observables), needs to be thoroughly assessed in terms of the impact on the cosmological parameter constraints.

\subsection{Differentiability and moving beyond classical sampling}
\label{sec7c}

One important distinction between the current version of \cloe and \firecrown with \ccl is that the theory and likelihood computations are more closely connected in \cloe than they are in \firecrown and \ccl. This is because \cloe's theory modules are entangled with the \cobaya and \cosmosis interfaces and the likelihood module, such that the theory predictions are generated as part of the likelihood computation, while \ccl returns the theory predictions to a user independently of the \firecrown likelihood framework. While the final likelihood stays the same, decoupling the theory modules from the rest of the code would simplify the use of samplers external to \cobaya and \cosmosis. It would moreover allow \cloe to be more suited for novel analysis approaches such as simulation based inference, which require the generation of a large number of realizations of the summary statistics, either at the level of theory predictions (e.g.,~\citealt{Lemos23, abellan24}) or at the level of mock catalogs (e.g.,~\citealt{Jeffrey24,vwk24}). 

The decoupling of the theory modules from the rest of the likelihood code (including the interfaces to \cobaya and \cosmosis) would further simplify the development of a differentiable version of \cloe, as gradient-based samplers require derivatives of the likelihood with respect to cosmological and nuisance parameters, which for a fixed covariance can be expressed in terms of derivatives of the theory vector. This can be achieved, for example, using Google's \jax library\footnote{\jaxaddress}\href{https://github.com/google/jax}{\color{black}\faGithub} \citep{bradbury18}, which contains features such as the automatic differentiation of native \python and \numpy methods, just-in-time compilation via \xla\footnote{\xlaaddress}\href{https://github.com/openxla/xla}{\color{black}\faGithub} \citep{leary17}, automatic vectorization capabilities, and the ability to run on not only central processing units (CPUs), but also graphics processing units (GPUs) and tensor processing units (TPUs). This is currently in progress within the \cloelib and \cloelike frameworks in the same \cloeorg \github organization. Creating a differentiable inference code would enable the use of gradient-based samplers such as stochastic variational inference \citep{hoffman12, zhang17}, Hamiltonian Monte Carlo (HMC; \citealt{duane87, Neal96, betancourt17}), and extensions such as the No-U-Turn Sampler (NUTS; \citealt{hoffman11}), Microcanonical HMC (MCHMC; \citealt{robnik}), Deep Learning HMC \citep{foreman21}, and Quantum Dynamical HMC \citep{lockwood24}.

These samplers can utilize automatic differentiation for computational efficiency and have the prospect to substantially speed up the cosmological parameter inference (e.g.,~\citealt{campagne23, zapatero24, piras24, Mootoovaloo24}, also see \citealt{nygaard23, balkenhol24, bonici24, Bonici25, Reeves25, Morawetz25}). This further holds when considered together with the latest developments of the linear matter power spectrum computation, such as a differentiable Einstein-Boltzmann solver \citep{hahn23} and JAX-based emulator \citep{piras23}. 
We note that the exact change in the Monte Carlo  runtime from gradient-based sampling varies with the size of the parameter space and shape of the posterior, and is sensitive to a trade-off between the increased efficiency of the sampler and the additional computational cost of the gradient calculation. The latest assessment, based on a comparison of the scaled effective sample size relative to the cost of the gradient calculation of NUTS with respect to the Metropolis--Hastings sampler, favors it to reduce the Monte Carlo runtime by a factor of two for an LSST-like survey configuration \citep{Mootoovaloo24}.\footnote{\citet{piras24} further reports more than two orders of magnitude improvement in the Monte Carlo runtime from the combined effects of emulating the matter power spectrum and using NUTS instead of the \polychord nested sampler for a future three-survey 3$\times$2pt setup. We note that this three-survey setup in \citet{piras24} has 157 free parameters as compared the single-survey setup in \citet{Mootoovaloo24} with 37 free parameters. In \citet{piras24}, the improvement in the Monte Carlo runtime is reported in terms of absolute time for a distinct configuration of GPUs and CPUs, rather than in terms of the scaled effective sample size, such that a comparison with the estimate in \citet{Mootoovaloo24} is not straightforward (however, also see \citealt{piras23}). For comparison, considering the CMB data from ACT, \citet{balkenhol24} finds a factor of four improvement in the effective sample size for NUTS relative to the Metropolis--Hastings sampler. Meanwhile, \citet{bonici24, Bonici25} find that the effective sample size of MCHMC is a factor of two and six higher than that of NUTS for this CMB dataset and BOSS, respectively.} Moving forward, it will be useful to expand this Stage IV survey assessment to gradient-based samplers other than NUTS.

\subsection{Blinding against confirmation bias}
\label{sec7d}

Another relevant capability is the ability to perform a blinded analysis of the \Euclid data, the overall strategy for which is described in \citet{Mellier24}. The \Euclid strategy involves a variety of validation and blinding layers by different groups in the consortium, and is executed to guard against possible confirmation biases in the cosmological analysis.

A range of blinding strategies have previously been considered in observational cosmology. This includes the strategy in the Kilo Degree Survey where the blinding is applied to either the galaxy ellipticities \citep{Kuijken15} or alternatively the photometric redshift distributions \citep{Hildebrandt20}. This further includes the approach in the Dark Energy Survey, where the data vector is modified through changes in the theory vector at two distinct points in parameter space \citep{Muir20}, and the approach in the Hyper Suprime-Cam survey where the blinding occurs at the level of the multiplicative shear bias \citep{Hikage19,dalal23}. Proposals further exist to perform the blinding at the level of the covariance matrix \citep{Sellentin20}, which resembles the data vector blinding in terms of the impact on the likelihood, and at the level of line-of-sight shifts in galaxy positions in the case of spectroscopic galaxy surveys \citep{Brieden20}. 

One of these existing approaches that is straightforward to incorporate in \cloe is that of \citet{Muir20}, as it will require the generation of the theory vectors for the considered probes at different points in parameter space. This approach can be self-consistently employed for the primary 3$\times$2pt and spectroscopic galaxy clustering analyses, along with the extended analyses involving cluster observables and CMB cross-correlations, given its flexibility to be applied to all of these probes.

\section{Conclusions}
\label{sec9}

We have described the code implementation and structure of the cosmological parameter inference pipeline \cloe. Developed within the EC, it provides a unified framework for analyzing a broad range of cosmological probes relevant to the \Euclid space mission, including the photometric 3$\times$2pt observables and spectroscopic galaxy clustering in both Fourier and configuration space, as well as cluster observables and CMB cross-correlations. The code is also sufficiently general to be used for other cosmological galaxy survey datasets by the broader community.

\cloe is a purely Pythonic code that has independent interfaces to the parameter sampling platforms of \cobaya and \cosmosis, and in turn the Boltzmann solvers of both \camb and \class and the full range of samplers (e.g.,~Metropolis--Hastings, \polychord, and \nautilus) and existing likelihoods (e.g., \Planck, ACT, and DESI) in these platforms. Once it receives the linear matter power spectrum, it is able to compute the likelihood for a \Euclid DR3 analysis setup in approximately one second, which is a performance that is superior to other existing Stage IV inference pipelines. The parameter inference speed can be further optimized by employing a range of strategies such as increased code vectorization, the use of emulators (in particular for the linear matter power spectrum), \jax just-in-time compilation, enhanced OpenMP parallelization, and GPU-accelerated gradient based sampling.

\cloe possesses the core features required for Stage IV cosmological analyses. This includes its capacity to take into account the forthcoming systematic uncertainties, such as baryonic feedback, intrinsic galaxy alignments, photometric redshift uncertainties of both source and lens galaxies, photometric and spectroscopic galaxy biases and redshift space distortions, multiplicative shear calibration uncertainties, magnification bias, spectroscopic redshift uncertainties, and spectroscopic sample impurities. It allows for both Gaussian and non-Gaussian likelihood forms (depending on whether the covariance is produced analytically or numerically), includes large-angle corrections, and has the capability to reweight the lensing kernels to be more compact in redshift through a BNT transformation.

\cloe can perform analyses in extended cosmologies that include the sum of neutrino masses, evolving dark energy, modified gravity, and nonzero curvature. It can also be interfaced with modified Boltzmann solvers that allow for a wider range of extended models. Additional \cloe features include a data reader for importing generic and \Euclid-specific data products, and a precise masking scheme that allows the user to select any combination of probes, scales, and redshifts to include in the analysis. Moreover, \cloe has a comprehensive suite of unit tests, extensive \sphinx documentation, and an early graphical user interface to ease the user experience. 

In highlighting the connection with the theoretical recipe, we have connected the key variables and methods in \cloe to the equations in \citet{Cardone24}. \cloe has undergone a rigorous benchmarking procedure in \citet{Martinelli24}, in which it is compared against external codes and verified to agree to high precision (broadly a $0.1\sigma$ criterion on the summary statistics). \cloe's ability to perform cosmological parameter inferences for Stage IV analysis setups is further demonstrated in \citet{GCH24}, \citet{Goh24}, and \citet{Blot24}.

The development of \cloe has been a significant collaborative effort by numerous groups in the EC over the past several years. While additional features and performance optimizations are valuable (as discussed in Sect.~\ref{sec7}), the code is already at an advanced stage where it can be used to inform us about the underlying nature of the Universe. To ensure long-term reproducibility, the \cloe repository is publicly available for use by the community at \url{https://github.com/cloe-org/cloe}.

\begin{acknowledgements}

\small{We thank the EC internal reviewer Martina Gerbino for their thoughtful and detailed feedback on the manuscript. We are also thankful for useful discussions with David Alonso, Nora Elisa Chisari, Francois Lanusse, Antony Lewis, Jacqueline Noder, Markus Michael Rau, Rafaela Gsponer, Jaime Ruiz Zapatero, Jesus Torrado, Tilman Tr\"{o}ster, and Angus H.~Wright. We are moreover grateful to the authors of external codes such as \camb \citep{lewis2000}, \class \citep{Lesgourgues11a, Lesgourgues11b, Blas11}, \cobaya \citep{Torrado21}, \cosmosis  \citep{Zuntz15}, \halofit  \citep{Smith03, Takahashi12}, \hmcode  \citep{Mead15, Mead16, Mead21}, \fastpt \citep{McEwen16, Fang17}, \bacco \citep{Angulo21}, \bcemu \citep{gs21}, and \eecode \citep{Knabenhans19, Knabenhans21} for making their codes public. 
\\
SJ acknowledges the Ram\'{o}n y Cajal Fellowship (RYC2022-036431-I) from the Spanish Ministry of Science and the Dennis Sciama Fellowship at the University of Portsmouth. StC acknowledges support from the Italian Ministry of University and Research (\textsc{mur}), PRIN 2022 `EXSKALIBUR – Euclid-Cross-SKA: Likelihood Inference Building for Universe's Research', Grant No.\ 20222BBYB9, CUP D53D2300252 0006, and from the European Union -- Next Generation EU. SD acknowledges support from the Italian Ministry of University and Research, PRIN 2022 ``LaScaLa – Large Scale Lab'' (grant no.\ PRIN\_20222JBEKN Macrosector PE - Physical Sciences and Engineering, sector PE2 ``Fundamental Constituents of Matter''), financed by the European Union -- Next Generation EU. AMCLB was supported by a Paris Observatory-PSL University Fellowship, hosted at the Paris Observatory, during part of this work. SP acknowledges support through the Conception Arenal Programme of the Universidad de Cantabria and funding from the proiect UC-LIME (PID2022-140670NA-I00), financed by MCIN AEI/ 10.13039/501100011033/FEDER, UE.
\\
This work used the DiRAC Data Intensive service (CSD3) at the University of Cambridge, managed by the University of Cambridge University Information Services on behalf of the STFC DiRAC HPC Facility (\url{www.dirac.ac.uk}). The DiRAC component of CSD3 at Cambridge was funded by BEIS, UKRI and STFC capital funding and STFC operations grants. This work also used the DiRAC Data Intensive service (DIaL3) at the University of Leicester, managed by the University of Leicester Research Computing Service on behalf of the STFC DiRAC HPC Facility. The DiRAC service at Leicester was funded by BEIS, UKRI and STFC capital funding and STFC operations grants. DiRAC is part of the UKRI Digital Research Infrastructure.
\\
The Euclid Consortium acknowledges the European Space Agency and a number of agencies and institutes that have supported the development of \Euclid, in particular the Agenzia Spaziale Italiana, the Austrian Forschungsf\"orderungsgesellschaft funded through BMIMI, the Belgian Science Policy, the Canadian Euclid Consortium, the Deutsches Zentrum f\"ur Luft- und Raumfahrt, the DTU Space and the Niels Bohr Institute in Denmark, the French Centre National d'Etudes Spatiales, the Funda\c{c}\~{a}o para a Ci\^{e}ncia e a Tecnologia, the Hungarian Academy of Sciences, the Ministerio de Ciencia, Innovaci\'{o}n y Universidades, the National Aeronautics and Space Administration, the National Astronomical Observatory of Japan, the Netherlandse Onderzoekschool Voor Astronomie, the Norwegian Space Agency, the Research Council of Finland, the Romanian Space Agency, the Swiss Space Office (SSO) at the State Secretariat for Education, Research, and Innovation (SERI), and the United Kingdom Space Agency. A complete and detailed list is available on the \Euclid web site (\url{www.euclid-ec.org/consortium/community}).}

\end{acknowledgements}

\bibliographystyle{aa}
\bibliography{biblio}

\appendix
\label{applabel}

\section{Details on BNT transform for vectorized spectra}
\label{eq:BNT_appendix}

In understanding the BNT transform for vectorized spectra, we begin by expressing the following well-known algebraic identity. Given a matrix \(\tens N\) that can be written as the product of three conformable matrices, namely \(\tens N=\tens O\,\tens P\,\tens Q\), it can be straightforwardly proven that
\begin{equation}
    \vec N=\left({\tens Q}^{\tens T}\otimes{\tens O}\right)\,\vec P\;,\label{eq:vectorisation_of_product}
\end{equation}
where the superscript `\(\tens T\)' denotes matrix transposition, \(\otimes\) the Kronecker product, and \(\vec N={\rm vec}(\tens N)\) and \(\vec P={\rm vec}(\tens P)\) the vectorization of two of the four matrices, as introduced in Sect.~\ref{bntsec}. We further note that we have defined vectors as columns here (rather than rows). The identity above can be straightforwardly related to our case, in which to obtain the BNT transform of the fully vectorized power spectrum, \(\slashed{\vec C}^{\rm LL}\), we need:
\begin{enumerate}
    \item To vectorize the BNT-transformed tomographic matrix at fixed multipole of Eq.\ (\ref{eq:BNT_transformed_Cij_LL}), \(\tens M\,\tens C^{\rm LL}(\ell)\,\tens M^{\tens T}\), over redshift bin pairs;
    \item And then to vectorize it further over multipole bins.
\end{enumerate}
Note that the BNT matrix \(\tens M\) is an \(N_z\times N_z\) matrix, whereas \(\tens C^{\rm LL}(\ell)\) is an \(N_\ell\)-long series of \(N_z\times N_z\) symmetric matrices.

The first step closely resembles Eq.\ (\ref{eq:vectorisation_of_product}), but for the fact that a half-vectorization is involved. This obstacle is bypassed by making use of the duplication matrix, \(\tens D_{N_z}\), of rank \(N_z\). In general, the duplication matrix is the unique linear transformation that turns half-vectorizations of matrices into vectorizations. In other words, given an \(n\times n\) matrix \(\tens A\), we have \(\tens D_n\,{\rm vech}(\tens A)={\rm vec}(\tens A)\). Then, to perform the second step, it is necessary to remove the duplicated entries due to the symmetric nature of the original matrix. To do so, we can now use an elimination matrix, \(\tens E_{N_z}\), of rank \(N_z\). Analogously to the duplication matrix, an elimination matrix is a linear transformation that turns vectorizations of matrices into half-vectorizations, i.e.\ in the general case \(\tens E_n\,{\rm vec}(\tens A)={\rm vech}(\tens A)\). Thus, starting from Eq.\ (\ref{eq:BNT_transformed_Cij_LL}) and using Eq.~(\ref{eq:vectorisation_of_product}), as well as the duplication and elimination matrices, we finally obtain Eq.\ (\ref{eq:BNT_LL}).

The situation for the lensing-clustering cross-correlation is far simpler. This is because the BNT-transformed tomographic power spectrum matrix at fixed \(\ell\) is only the product of two matrices, \(\tens M\,\tens C^{{\rm L}{\rm G}}(\ell)\) (see Eq.\ \ref{eq:BNT_transformed_Cij_LG}), where \(\tens C^{{\rm L}{\rm G}}(\ell)\) is not a symmetric matrix. As a result, to obtain Eq.\ (\ref{eq:BNT_LG}) neither an elimination nor duplication matrix is needed.

Lastly, to construct the duplication and elimination matrices, we follow \citet{doi:10.1137/0601049} and write
\begin{align}
    \tens E_n&=\sum_{i\ge j}\vec u_{ij}\,{\rm vec}\left(\vec e_i\,\vec e_j^{\tens T}\right)^{\tens T}\;,\\
    \tens D_n&=\sum_{i\ge j}\vec u_{ij}\,{\rm vec}\left(\tens T_{ij}\right)^{\tens T}\;.
\end{align}
Here, \(\vec u_{ij}\) is a vector of length \(n\,(n+1)/2\) having the value \(1\) in the position \((j-1)\,n+i-j\,(j-1)/2\) and \(0\) elsewhere; \(\vec e_i\) is a unit vector whose \(i\)-th element is \(1\) and \(0\) elsewhere; and \(\tens T_{ij}\) is an \(n\times n\) matrix with value \(1\) in position \((i,j)\) and \((j,i)\) and \(0\) elsewhere.

\section{Demonstration and validation notebooks}
\label{appnote}

\cloe contains \jupyternotebooks in the \texttt{notebooks} directory that demonstrate its use and perform distinct validations. This includes the main notebook \texttt{DEMO.ipynb} that demonstrates how to compute the theory predictions and likelihood for the primary probes given synthetic \Euclid data. It is executed as part of the continuous integration of \cloe (see Sect.~\ref{sec3q}), where it allows for an end-to-end verification test that covers a larger scope than individual code units and helps determine if \cloe as a whole is working as expected.

While \cobaya is considered as the backend in the main demo notebook, \cosmosis is additionally considered as the backend in \texttt{cosmosis\_validation.ipynb}. As the name suggests, this notebook illustrates the agreement of the log-likelihood and intermediate quantities between \cloe's \cosmosis and \cobaya backends. Another relevant notebook is \texttt{CMBX\_Probes.ipynb}, which demonstrates the computation and validation of the theory predictions for the CMB auto- and cross-correlations. It also includes a simple sampling of the sum of neutrino masses and plotting of the resulting log-posterior. Analogous demonstration notebooks exist for the clusters of galaxies probes (Euclid Collaboration:~Sakr et al., in prep.), spectroscopic magnification bias and Weyl potential \citep{Goh24}, and nonlinear implementations of galaxy clustering and IA modeling in \cloe (Euclid Collaboration:~Carrilho et al., in prep., Euclid Collaboration:~Moretti et al., in prep.). The benchmarking of the primary probes is also illustrated in \jupyternotebooks, albeit in a separate repository, as described in \citet{Martinelli24}.

\section{\conda environment and code dependencies}
\label{sec3p}

We allow for a straightforward \cloe installation via \conda, as described in the online documentation of the code. The file \texttt{environment.yml} contains \cloe's dependencies on other codes. This currently includes \python and \pip. It also includes standard third party \python packages in the form of \astropy, \jupyter, \matplotlib, \mpiforpy, \numpy, \scikit, \scipy, \seaborn, and \tensorflow. It includes unit testing and documentation dependencies in the form of \numpydoc, \pytest, \pytestcov, \pytestcode, \sphinx, and \sphinxrtd. It moreover includes domain-specific third party \python packages in the form of \fastpt, along with 
additional non-\python dependencies in the form of \camb, \gsl, \cxxcomp, \gfortran and \fitsio. Lastly, the environment file contains installations of \cobaya, \deepdish, \class, \bacco, \bcemu, \euclidemu, \pyhmcode, \euclidlib, \pytestdoc, \cosmosis, and \cosmosislib. We further note that \cosmosis uses the version of \camb that is included in the \conda environment.

The \cloe \conda environment does not include the likelihood of external datasets. For instance, in the case of the \Planck CMB, the user will need to separately install the likelihood to consider a combined analysis of \Euclid and \Planck. Once the \Planck likelihood is installed, the \texttt{likelihood} block for \cloe with \cobaya and the \texttt{pipeline} block for \cloe with \cosmosis will need to be updated in the standard way 
for the use of this dataset on the \cobaya/\cosmosis platforms, as further described in \cloe's online documentation.

\section{Repository structure}
\label{apprepo}

We provide here a summary of the directory structure of \cloe:
\begin{itemize}
    \item[$\blacksquare$] \texttt{cloe}: Main source directory, containing the following subpackages for the calculation of theoretical predictions and likelihood (as well as the interface to \cobaya).
    \begin{itemize}
        \item[$\square$] \texttt{user\_interface},  \texttt{photometric\_survey}, \texttt{gui}, \texttt{spectroscopic\_survey},  \texttt{non\_linear}, \texttt{tests},  \texttt{data\_reader}, \texttt{like\_calc}, \texttt{masking}, \texttt{cmbx\_p}, \texttt{clusters\_of\_galaxies}, \texttt{auxiliary}, \texttt{fftlog},   \texttt{cosmo}.
    \end{itemize}
    \item[$\blacksquare$] \texttt{configs}: Configuration \yaml files with user specifications, such as probe selection, systematics treatment, scale cuts, and parameter priors (e.g. \texttt{config\_default.yaml}, \texttt{model\_default.yaml}, and the \texttt{models} subdirectory).
    \item[$\blacksquare$] \texttt{data}: Data files, including measurements, covariance matrices, mixing matrices, and redshift distributions.
    \item[$\blacksquare$] \texttt{chains}: Monte Carlo chains, directory automatically created when performing a run.
    \item[$\blacksquare$] \texttt{cosmosis}: Interface to \cosmosis.
    \item[$\blacksquare$] \texttt{docs}: \sphinx\ documentation.
    \item[$\blacksquare$] \texttt{gui}: Graphical user interface.
    \item[$\blacksquare$] \texttt{mcmc\_scripts}: Example scripts to run MCMC chains.
    \item[$\blacksquare$] \texttt{scripts}: Example scripts to create synthetic data.
    \item[$\blacksquare$] \texttt{notebooks}: Demonstration and validation \jupyter notebooks.
    \item[$\blacksquare$] Top-level files:
    \begin{itemize}
        \item[$\square$] \texttt{run\_cloe.py}: Main script for running \cloe.
        \item[$\square$] \texttt{example\_mcmc\_script\_for\_cluster.sh}: Example submission script on a computing cluster.
        \item[$\square$] \texttt{setup.py}: Installation script.
        \item[$\square$] \texttt{environment.yml}: \conda\ environment file.
        \item[$\square$] \texttt{Dockerfile}: Reproducible environment build.
        \item[$\square$] \texttt{README.md}: Overview and usage guide.
        \item[$\square$] \texttt{CONTRIBUTING.md}: Contribution guidelines.
        \item[$\square$] \texttt{LICENSE}: LGPL license for \cloe.
    \end{itemize}
\end{itemize}
This structure underpins the modular architecture of \cloe.

\section{Development history}
\label{secdev}

There have been multiple internal releases of \cloe within the EC. Here, we provide a brief overview of the development history, which has thus far consisted in over 300 code mergers.
\begin{itemize}
\item[$\bullet$] \texttt{v1.0}: This version of the code was internally released to the EC in April 2021. It includes the basic aspects of \cloe, such as an interface to \cobaya (and through it to \camb), along with modules for reading in the measurements and covariances, computing the theory predictions for the photometric and spectroscopic probes, and calculating the likelihood. This version of the code is thereby capable of performing an end-to-end inference that produces parameter constraints. However, it also has major limitations, such as no masking of the measurements and covariance, limited treatment of the systematic uncertainties, absence of nonlinear corrections to the theory predictions (aside from \halofit and \hmcode for the matter power spectrum and the NLA model for the IAs), and a single summary statistics for the core observables (i.e., 3$\times$2pt angular power spectra and redshift-space multipole power spectra). 
\item[$\bullet$] \texttt{v2.0}: This version of the code was internally released to the EC in June 2023. There were two subsequent minor patches, \texttt{v2.0.1} and \texttt{v2.0.2}, internally released in August and November 2023, respectively. This version of the code has major upgrades in a variety of areas. This includes the creation of the \cloe overlayer, the interface to the linear matter power spectrum and the background expansion and growth histories of \class, the masking prescription, an expanded treatment of systematic uncertainties (such as magnification bias, galaxy bias, photometric redshift uncertainties, and sample impurities), modified gravity corrections to the growth history, and the capability to consider summary statistics in configuration space (i.e., 3$\times$2pt correlation functions and multipole correlation functions). This version of \cloe also includes new features such as the BNT transformation, GUI, and redshift-space distortions for the photometric galaxy clustering and galaxy-galaxy lensing. As further discussed in Euclid Collaboration:~Carrilho et al. (in prep.) and Euclid Collaboration:~Moretti et al. (in prep.), this version of \cloe moreover includes emulators for the nonlinear matter power spectrum and the \eftoflss model for the nonlinear galaxy power spectrum. The patches \texttt{v2.0.1} and \texttt{v2.0.2} primarily incorporate a third-order polynomial in redshift for the magnification bias, spectroscopic redshift errors, and the \cosmosis backend. We note that \texttt{v2.0.2} has been benchmarked in \citet{Martinelli24} and used to perform the parameter inference in \citet{GCH24}, \citet{Blot24}, and Euclid Collaboration:~Sciotti et al. (in prep.). In the latter two cases, \cloe was further interfaced with an emulator of the matter power spectrum to increase the computational speed of the theory predictions.
\item[$\bullet$] \texttt{v2.1}: This version of \cloe includes code developed until August 2024, which is when the Euclid Inter-Science Taskforce: Likelihood completed its tenure. It includes the \fftlog implementation, a unified treatment of \cobaya and \cosmosis via the \cloe overlayer, and the ability to read in data files in the \Euclid OU-LE3 format via \euclidlib. This version of \cloe moreover includes an optimization of the integration strategy for the weak lensing and magnification kernels (resulting in a factor of six speed improvement), along with bug fixes pertaining to the magnification bias, \conda environment, and measurement units. This version also includes the masking of the configuration space probes and the window convolution for the spectroscopic galaxy clustering, i.e. summary statistics in the form of window-convolved multipole power spectra. In addition, as discussed in Euclid Collaboration:~Carrilho et al. (in prep.) and Euclid Collaboration:~Moretti et al. (in prep.), this \cloe version includes the baryonic feedback emulators of \bacco and \bcemu, the 1-loop perturbation theory prescription for the nonlinear photometric galaxy clustering, the TATT model for IAs, and an extension of the spectroscopic galaxy clustering modeling for compatibility with massive neutrinos. As further discussed in \citet{Goh24}, this version of \cloe includes the spectroscopic magnification bias and interface to the Weyl potential power spectrum. This version moreover includes additional probes, in the form of CMB cross-correlations and cluster counts (the latter described in Euclid Collaboration:~Sakr et al., in prep.). These additional probes have the same functionalities as the primary probes in that they can be run using either \cobaya or \cosmosis, and using either \camb or \class. The minor patch \texttt{v2.1.1} includes a bug fix for the ISW calculation in November 2024, and we release an additional distinct branch of the code used in \citet{Goh24}. We note that \texttt{v2.1} of \cloe passes the same benchmarking tests as \texttt{v2.0}, alongside the added benchmarking tests of the new probes. We have also confirmed that there is backward compatibility, as Monte Carlo runs yield results that are consistent with earlier versions of the code when considering the same analysis setups, and that Monte Carlo runs are consistent between the choice of either \cobaya or \cosmosis as the backend.
\item[$\bullet$] \texttt{Further developments}: The main branch of \cloe includes further developments beyond the latest tagged release. This includes the feature to write the theory predictions to file, enhancements to the \sphinx documentation, and bug fixes pertaining to the data handler module and checkpointing of Monte Carlo runs. This further includes the window convolution for the photometric probes, i.e. summary statistics in the form of 3$\times$2pt pseudo-$C_\ell$, as well as the vectorization of the matter, IA, and galaxy auto- and cross-power spectra. In addition, this includes the removal of deprecated code, refinements to notebooks and scripts for compatibility with the latest version of \cloe, and minor technical updates required for the migration from \gitlab to \github. We have confirmed that the benchmarking holds for the same set of summary statistics as earlier versions of the code.
\end{itemize}

\section{Hard-coded variables}
\label{appe}

\cloe inevitably contains a number of hard-coded variables that can be generalized. This primarily includes the extrapolation mode of \scipy's \texttt{InterpolatedUnivariateSpline} method, where it is set to \texttt{ext=2} in most cases. In this mode, the interpolator raises an error when requested outside of the predefined interval. Likewise, for \scipy's 2D interpolator \texttt{RectBivariateSpline}, the degree of the bivariate spline is set to unity with \texttt{kx=1} and \texttt{ky=1} in nearly all cases. For the cases where there are deviations in these settings, we set \texttt{ext=1} and \texttt{kx=ky=3}.

Other hard-coded variables include the range of scales for the linear matter power spectrum in \texttt{cobaya\_interface.py} and \texttt{cosmosis\_interface.py}, along with the range of angular scales for the photometric 3$\times$2pt correlation functions in the constructor of \texttt{Euclike}. Moreover, the range of redshifts for the integrations to obtain the 3$\times$2pt angular power spectra and the range of scales for the subsequent integration to obtain the 3$\times$2pt correlation functions are hard-coded within the constructor of the \texttt{Photo} class.
The density of $\mu_k$ values to include in the integration of the multipole power spectra is further hard-coded in the constructor of the \texttt{Spectro} class.

\end{document}